\documentclass[11pt]{cernrep}

\usepackage{epsfig}
\usepackage{graphicx}
\usepackage{cite,./mcite}

\newcommand{\leqsim}{\,\raisebox{-0.6ex}{$\buildrel < \over \sim$}\,}
\newcommand{\geqsim}{\,\raisebox{-0.6ex}{$\buildrel > \over \sim$}\,}

\newcommand{\asmz}{\alpha_s(M_Z^2)}
\newcommand{\msbar}{\mbox{$\overline{\rm{MS}}$}\ }
\newcommand{\bs}{\overline{s}}
\newcommand{\bc}{\overline{c}}
\newcommand{\bu}{\overline{u}}
\newcommand{\bd}{\overline{d}}

\newcommand{\bU}{\overline{U}}
\newcommand{\bD}{\overline{D}}

\def\totalave{ 1402}
\def\uniqueave{  741}

\def\chiave{636.5}
\def\dofave{656}

\def\asq{\alpha_s(Q^2)}
\def\as{\alpha_s}

\begin{document}
\setcounter{page}{1}
\title{What did HERA teach us about the structure of the proton? }
\author{Amanda Cooper-Sarkar}
\institute{Oxford University}
\maketitle
\begin{abstract}
Starting in 2008 the H1 and ZEUS experiments have been combining their data
in order to provide the most complete and 
accurate set of deep-inelastic data as the legacy of HERA. The present review 
presents these combinations, both published and preliminary, and explores how 
they have been used to give information on the structure of the proton.
The HERAPDF parton distribution functions (PDFs) are presented and compared 
with other current PDFs and with data from the Tevatron and LHC colliders.
\end{abstract}
\pagenumbering{roman}
\section{Introduction}	
\label{sec:xsecns}
\vspace*{-0.5pt}
\noindent

HERA was an electron(positron)-proton collider located at DESY, Hamburg. 
It ran in two phases HERA-I from 1992-2001 and HERA-II 2003-2007. 
Two similar experiments, H1 and ZEUS, took data. In HERA-1 running each 
experiment collected  $\sim 100$pb$^{-1}$ of $e^+p$ data 
and $\sim15$pb$^{-1}$ of $e^-p$ data with electron beam energy $27.5$GeV and 
proton beam energies $820,920~$GeV. In HERA-II running each experiment took 
$\sim 140$pb$^{-1}$ of $e^+p$ data 
and $\sim180$pb$^{-1}$ of $e^-p$ data with the same electron beam energy and 
proton beam energies $920~$GeV. In addition to this, before the shut-down in 
2007, each experiment took $\sim 30$pb$^{-1}$of data with reduced proton 
beam energies $460,575~$GeV. 

Deep inelastic lepton-hadron 
scattering data has been used both to investigate the theory of 
the strong interaction and to determine the momentum distributions of the 
partons within the nucleon. The data from the HERA collider now dominate 
the world data on deep inelastic scattering since they cover an unprecedented 
kinematic range: in $Q^2$, the (negative of the) 
invariant mass squared of the virtual exchanged boson, 
$0.045 < Q^2 < 3\times 10^{-5}$; in Bjorken $x$, $6\times10^{-7} < x < 0.65$.
Futhermore, because the HERA experiments investigated
 $e^+p$ and $e^-p$, charge current(CC) and neutral current (NC) scattering, 
information can be gained on flavour separated up- and down-type quarks and 
antiquarks and on the gluon- from its role in the scaling violations of 
perturbative quantum-chromo-dynamics. 
From 2008, the H1 and ZEUS experiments 
began to combine their data in order to provide the most complete and 
accurate set of deep-inelastic data as the legacy of HERA. Data on 
inclusive cross-sections have been combined for the HERA-I phase of running 
and a preliminary combination has been made also using the HERA-II data. 
This latter exersize also includes the data run at lower proton beam 
energies in 2007. Combination of  $F_2^{c\bar{c}}$ data is also underway, 
and combination of  $F_2^{b\bar{b}}$ data and of jet data is foreseen.
The HERA collaborations have used these combined data to 
determine parton distribution functions (PDFs). These analyses had resulted
in the HERAPDF sets. The present review 
concentrates on the information on proton structure which has been 
gained from these HERA data.

\section{Formalism}
\label{formalism}

\begin{figure}[ht]
\centerline{\psfig{figure=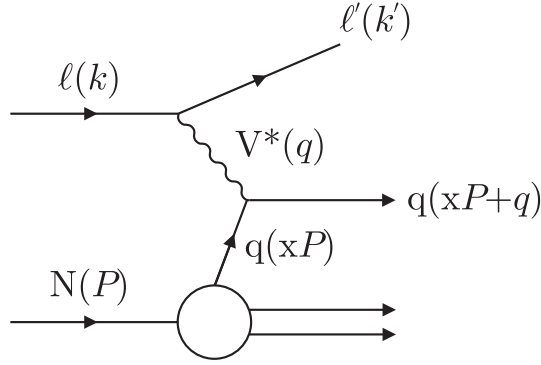,height=0.2\textheight}}
\caption{Schematic diagram of lepton-hadron scattering in the 
quark-parton model}
\label{fig:qpm}
\end{figure}

In the quark parton model deep inelastic  lepton-hadron scattering is pictured
as in Fig.~\ref{fig:qpm}. 
\noindent
where $l,l'$ represent leptons (lepton is taken to include 
antileptons, unless it is necessary to distinguish them), and $N$ represents 
the nucleon. The associated four vectors are $k,k'$ for the
incoming and outgoing leptons respectively, and $p$ for the target 
(or incoming) nucleon. The process is mediated by the exchange of a 
virtual vector boson, $V^*$($\gamma, W$ or $Z$), with four 
momentum given by
\[
               q = k - k'.
\]
Various Lorentz invariants are useful in the description of the kinematics of 
the process:
\[
                   s = ( p + k )^2,
\]
the centre of mass energy squared for the $lp$ interaction,
\[
                   Q^2 = -q^2,
\]
the (negative of) the invariant mass squared of the virtual exchanged boson,
\[
                   x = Q^2/2p.q,
\]
the Bjorken $x$ variable, 
which is interpreted in the quark-parton model as the fraction 
of the momentum of the incoming nucleon taken by the struck quark, 
and
\[
                   y = p.q/p.k,
\]
which gives
a measure of the amount of energy transferred between the lepton and the hadron
systems. 
Note that (ignoring masses),
\[
                   Q^2 = s x y,
\]
so that only two of these quantities are independent. 
Finally the centre of mass of the $V^*p$ system (or equivalently the invariant 
mass of the final state hadronic system) is often denote by $W$
\[ 
                   W^2=(q+p)^2.
\]

Neutral current (NC) deep inelastic scattering is mediated by $\gamma$ and $Z$ 
exchange and the NC deep inelastic $e^{\pm}p$ scattering cross sections can %%@
be expressed as
\begin{eqnarray} \label{eq:ncsi}     
 \tilde{\sigma}^{\pm}_{NC} = \frac{Q^4 x}{2\pi \alpha^2 Y_+} %%@
\sigma^{\pm}_{NC}                                                     
  =            F_2 \mp \frac{Y_-}{Y_+} xF_3 -\frac{y^2}{Y_+} F_L~,
\end{eqnarray}                                                                  
where the electromagnetic coupling constant $\alpha$, the photon              
propagator and a helicity factor are absorbed 
in the definition of a reduced cross section $\tilde{\sigma}$, and %%@
$Y_{\pm}=1 \pm (1-y)^2$.                                                                              
The structure functions $F_2$, $F_L$  and $xF_3$
 are given by
\begin{eqnarray} \label{strf}                                                   
 F_2 &=& F_2^{\gamma} - \kappa_Z v_e  \cdot F_2^{\gamma Z} +                      
  \kappa_Z^2 (v_e^2 + a_e^2 ) \cdot F_2^Z~, \nonumber \\   
 F_L &=& F_L^{\gamma} - \kappa_Z v_e  \cdot F_L^{\gamma Z} +                      
  \kappa_Z^2 (v_e^2 + a_e^2 ) \cdot F_L^Z~, \nonumber \\                     
 xF_3 &=&  \kappa_Z a_e  \cdot xF_3^{\gamma Z} -                     
  \kappa_Z^2 \cdot 2 v_e a_e  \cdot xF_3^Z~.                                   
\end{eqnarray} 
where $v_e$ and $a_e$ are the vector and axial-vector weak couplings of 
the electron 
and $\kappa_Z(Q^2) =   Q^2 /[(Q^2+M_Z^2)(4\sin^2 \theta_W \cos^2 %%@
\theta_W)]$. 
At low $Q^2$, the contribution of $Z$
 exchange is negligible and $xF_3 = 0, F_2 =  F_2^{\gamma}, F_L =  F_L^{\gamma}$ and  $\tilde{\sigma} = F_2  - y^2 F_L/Y_+$.
The contribution of the term containing the structure function $F_L$ is %%@
only significant for large values of $y$.

In the Quark Parton Model (QPM),  
$F_L=0$, and the 
other strcuture functions are given by
\begin{eqnarray} \label{ncfu}                                                   
  (F_2^{\gamma}, F_2^{\gamma Z}, F_2^Z) &=&  [(e_u^2, 2e_uv_u, %%@
v_u^2+a_u^2)(xU+ x\bar{U})
  +  (e_d^2, 2e_dv_d, v_d^2+a_d^2)(xD+ x\bar{D})]~,            
                                 \nonumber \\                                   
  (xF_3^{\gamma Z}, xF_3^Z) &=& 2  [(e_ua_u, v_ua_u) (xU-x\bar{U})
  +  (e_da_d, v_da_d) (xD-x\bar{D})]~,                        
\end{eqnarray} 
such that at low $Q^2$
\begin{equation}
F_2^{\gamma} =  [e_u^2(xU+ x\bar{U})  +  e_d^2(xD+ x\bar{D})]~,
\end{equation}
where  $e_u,e_d$ denote the electric charge of up- or
down-type quarks while $v_{u,d}$ and $a_{u,d}$ are 
the vector and axial-vector weak couplings of the up- or 
down-type quarks.
Here   $xU$, $xD$, $x\bU$ and $x\bD$ denote
the sums of up-type, of down-type and of their 
anti-quark momentum distributions, respectively. 
In the QPM these ditributions are 
functions of Bjorken $x$ only, and not also of $Q^2$ as they would be in full 
generality- this is what is meant by Bjorken scaling.
Below the $b$ quark mass threshold,
these sums are related to the quark distributions as follows
\begin{equation}  \label{ud}
  xU  = xu + xc\,,    ~~~~~~~~
 x\bU = x\bu + x\bc\,, ~~~~~~~~
  xD  = xd + xs\,,    ~~~~~~~~
 x\bD = x\bd + x\bs\,, 
\end{equation}
where $xs$ and $xc$ are the strange and charm quark distributions.
Assuming symmetry between sea quarks and anti-quarks, 
the valence quark distributions result from 
\begin{equation} \label{valq}
xu_v = xU -x\bU\,, ~~~~~~~~~~~~~ xd_v = xD -x\bD\,.
\end{equation}

Charge current (CC) deep inelastic scattering is mediated by $W^+$ and $W^-$ 
exchange and the CC deep inelastic $e^{\pm}p$ scattering cross sections can %%@
be expressed as
\begin{equation}
 \label{Rnc}
 \tilde{\sigma}^{\pm}_{CC} =  
  \frac{2 \pi  x}{G_F^2}
 \left[ \frac {M_W^2+Q^2} {M_W^2} \right]^2 \sigma^{\pm}_{CC}      
\end{equation}
where analogously to Eq~\ref{eq:ncsi}, 
\begin{equation}
 \label{ccsi}
 \tilde{\sigma}^{\pm}_{CC}=
  \frac{Y_+}{2}W_2^\pm   \mp \frac{Y_-}{2} xW_3^\pm - \frac{y^2}{2} W_L^\pm. 
\end{equation}
In the QPM, $W_L^\pm = 0$, 
and the CC structure functions represent sums and differences
of quark and anti-quark-type distributions depending on the 
charge of the lepton beam: 
\begin{eqnarray}
 \label{ccstf}
    W_2^{+}  =  x\bU+xD\,,\hspace{0.05cm} ~~~~~~~
  xW_3^{+}  =   xD-x\bU\,,\hspace{0.05cm}  ~~~~~~~ 
    W_2^{-}  =  xU+x\bD\,,\hspace{0.05cm} ~~~~~~~
 xW_3^{-}  =  xU-x\bD\,.
\end{eqnarray}
From these equations it follows that
\begin{equation}
\label{ccupdo}
 \tilde{\sigma}^+ = x\bU+ (1-y)^2xD\,, ~~~~~~~
 \tilde{\sigma}^- = xU +(1-y)^2 x\bD\,. 
\end{equation}
Therefore the NC and CC measurements may be used
to determine the combined sea quark distribution functions, $x\bU$ and %%@
$x\bD$,
and the valence quark distributions, $xu_v$ and $xd_v$. 

Perturbative QCD extends the formalism of the QPM such that the parton 
momentum distributions (PDFs) become functions of $Q^2$ as well as $x$. However
this scaling violation induces only a logarithmic dependence on $Q^2$, 
as described by the DGLAP equations
~\cite{Gribov:1972ri,Lipatov:1974qm,Dokshitzer:1977sg,Altarelli:1977zs}.
The DGLAP equations are coupled equations for the change of the quark,
antiquark and gluon densities as $\ln Q^2$ changes
\begin{eqnarray}
{\partial\over \partial\ln Q^2}\left(\matrix{q_i(x,Q^2)\cr g(x,Q^2)}\right) &=&
{\asq\over 2\pi}\sum_j\int_x^1{d\xi\over \xi} \nonumber \\
&&\left(
\matrix{P_{q_iq_j}({x\over \xi},\asq)&P_{q_ig}({x\over \xi},\asq)\cr
        P_{gq_j}({x\over \xi},\asq)&P_{gg}({x\over \xi},\asq)\cr}
\right)
\left(\matrix{q_j(\xi,Q^2)\cr g(\xi,Q^2)}\right),\nonumber \\
\label{eqn:ap_gen}
\end{eqnarray}
where the $q_i,q_j$ are taken to include both quarks and antiquark 
distributions. The splitting functions are expanded as power series
in the strong coupling $\as$,
\begin{eqnarray}
P_{q_iq_j}(z,\as)&=&\delta_{ij}P^{(0)}_{qq}(z)+
{\as\over 2\pi}P^{(1)}_{qq}(z)+\dots \nonumber \\
P_{qg}(z,\as)&=&P^{(0)}_{qg}(z)+
{\as\over 2\pi}P^{(1)}_{qg}(z)+\dots \nonumber \\
P_{gq}(z,\as)&=&P^{(0)}_{gq}(z)+
{\as\over 2\pi}P^{(1)}_{gq}(z)+\dots \nonumber \\
P_{gg}(z,\as)&=&P^{(0)}_{gg}(z)+
{\as\over 2\pi}P^{(1)}_{gg}(z)+\dots \nonumber 
\end{eqnarray}
and are calculable within pQCD. Thus the gluon momemtum distribution 
influences the quark distributions through its contribution to their scaling 
violations, and the gluon PDF is determined by analysing the $Q^2$ dependence 
of the data.

To leading order in pQCD the equations for the structure functions in terms 
of the PDFs are still given by the QPM expressions. However beyond leading 
order a convolution of parton distributions and QCD-calculable coefficient 
functions is necessary.
\begin{eqnarray}
{F_2(x,Q^2)\over x}&=&\int{d\xi\over \xi}\left[\sum_ie^2_iq_i(\xi,Q^2)
C_q\left({x\over \xi},\as\right)+
\bar{e}^2g(\xi,Q^2)C_g\left({x\over \xi},\as\right)\right],\nonumber \\
&&
\label{eqn:dglap_f2}
\end{eqnarray}
where, $\bar{e}^2=\sum_ie^2_i$, and the sums run over all active quark 
and antiquark flavours. $C_q$ an $C_g$ are the coefficient functions, 
which may also be expanded as power series in $\as$,
\begin{eqnarray}
C_q(z,\as)&=&\delta(1-z)+{\as\over 2\pi}C^1_q(z)+\dots \nonumber \\
C_g(z,\as)&=&{\as\over 2\pi}C^1_g(z)+\dots \nonumber.
\end{eqnarray}
In the QPM the transverse momentum of the partons is assumed to be zero and
one of the consequences of this for spin ${1\over 2}$ quarks is that the
longitudinal structure function ($F_L=F_2-2xF_1$) is zero. However this is
no longer true beyond leading order, and $F_L$ is given by.
\begin{equation}
{F_L(x,Q^2)\over x}={\as\over 2\pi}\int_x^1{d\xi\over \xi}\left[
\sum_ie^2_i{8\over 3}\left({x\over \xi}\right)q_i(\xi,Q^2)+
\bar{e}^2 4\left({x\over \xi}\right)^2\left(1-{x\over \xi}\right)g(\xi,Q^2)
\right]
\label{eqn:fl_qg}
\end{equation}
Thus the gluon distribution also influences the longitudinal structure function
particularly at low $x$.

\section{Data sets}

The deep inelastic $ep$ scattering cross sections
depend on the centre-of-mass energy $s$ and on two other independent 
kinematic variables, usually taken to be $Q^2$ and $x$.  The 
salient feature of the  HERA collider experiments is the possibility
to determine the $x$ and $Q^2$ from the scattered
electron, or from the hadronic final state, or using a combination %%@
of the
two.  The choice of the most appropriate kinematic reconstruction method
for a given phase space region is 
based on resolution, measurement accuracy and radiative correction effects %%@
and has
been optimised differently for the two HERA experiments H1 and ZEUS, as %%@
described in the original publications.
The use of different  reconstruction techniques by the two experiments
contributes to improved accuracy when the data sets are combined, since %%@
although
the detectors were built following similar physics considerations they
opted for different technical solutions, both for the calorimetric
and the tracking measurements. Thus the experiments can calibrate each %%@
other.

\subsection{The combined inclusive HERA-I data set}
\label{sec:datacomb}
The inclusive cross-sections data collected by each experiment in the %%@
HERA-I running period have been combined~\cite{h1zeuscomb}.
A summary of the data used in this analysis is given in %%@
Table\,\ref{tab:data}.
The NC data cover a wide range in $x$ and $Q^2$.
The lowest $Q^2 \ge 0.045$~GeV$^2$ data come from the measurements of ZEUS %%@
using
the BPC and BPT~\cite{Breitweg:1997hz,Breitweg:2000yn}. The $Q^2$ range %%@
from $0.2$~GeV$^2$ to $1.5$~GeV$^2$
is covered using special HERA runs, in which the interaction vertex
position was shifted forward allowing for larger angles 
of the backward scattered electron
to be accepted~\cite{Breitweg:1998dz,Collaboration:2009bp}.
The lowest $Q^2$ for the shifted vertex data 
was reached using events in which the effective electron
beam energy was reduced by initial state %%@
radiation~\cite{Collaboration:2009bp}. 
Values of  $Q^2\ge 1.5$~GeV$^2$ are measured using  the nominal vertex %%@
settings.
For $Q^2 \le 10$~GeV$^2$, the cross section is very high 
and the data were  collected using dedicated
trigger setups~\cite{Chekanov:2001qu,Collaboration:2009bp}. 
The highest accuracy of the cross-section measurement  is achieved
for  $10 \le Q^2 \le 100$~GeV$^2$  %%@
\cite{Chekanov:2001qu,Collaboration:2009kv}.
For $Q^2\ge 100$~GeV$^2$, the statistical uncertainty of the data becomes
relatively large.
The high $Q^2$ data included here were collected with  
positron~\cite{Adloff:1999ah,Chekanov:2001qu,Adloff:2003uh,Chekanov:2003yv} %%@
and with 
electron~\cite{Adloff:2000qj,Chekanov:2002ej} beams.  
The CC data for $e^+p$ and $e^-p$ scattering cover the range $300\le Q^2\le %%@
30000$~GeV$^2$~\cite{Adloff:1999ah,zeuscc97,Chekanov:2002zs,Adloff:2003uh,C%%@
hekanov:2
003vw}.

\begin{table}
\begin{center}
\begin{scriptsize}
\begin{tabular}{|lr|lr|lr|c|c|c|c|c|}
\hline
\multicolumn{2}{|c|}{Data Set} &
\multicolumn{2}{|c|}{$x$ Range} &
\multicolumn{2}{|c|}{$Q^2$ Range} &
${\cal L}$ & $e^+/e^-$ & $\sqrt{s}$ & Reference \\
\multicolumn{2}{|c|}{ } &
\multicolumn{2}{|c|}{ } &
\multicolumn{2}{|c|}{GeV$^2$} &
pb$^{-1}$ &   \\
\hline
H1~svx-mb  & $95$-$00$ & $5\times 10^{-6}$  & $0.02$  & $0.2$ & $12$  & %%@
$2.1$ &$e^+p$   & $301$-$319$    & \cite{Collaboration:2009bp} \\
H1~low~$Q^2$      & $96$-$00$ & $2\times 10^{-4}$  & $0.1$   & $12$  & %%@
$150$ & $22$ &$e^+p$   & $301$-$319$      &\cite{Collaboration:2009kv}\\
H1~NC             & $94$-$97$ & $0.0032$  &$0.65$   &$150$  &$30000$   & %%@
$35.6$  & $e^+p$ & $301$           &   \cite{Adloff:1999ah}\\ 
H1~CC             & $94$-$97$ & $0.013$   &$0.40$   &$300$  &$15000$   & %%@
$35.6$  & $e^+p$ & $301$            &   \cite{Adloff:1999ah}\\ 
H1~NC             & $98$-$99$ & $0.0032$  &$0.65$   &$150$  &$30000$   & %%@
$16.4$  & $e^-p$ & $319$            &    \cite{Adloff:2000qj}\\
H1~CC             & $98$-$99$ & $0.013$   &$0.40$   &$300$  &$15000$   & %%@
$16.4$  & $e^-p$ & $319$             &  \cite{Adloff:2000qj}\\
H1~NC HY          & $98$-$99$ & $0.0013$  &$0.01$   &$100$  &$800$   & %%@
$16.4$  & $e^-p$ & $319$               &  \cite{Adloff:2003uh}\\
H1~NC             & $99$-$00$ & $0.0013$ &$0.65$   &$100$  &$30000$   & %%@
$65.2$  & $e^+p$ & $319$              &  \cite{Adloff:2003uh}\\
H1~CC             & $99$-$00$ & $0.013$   &$0.40$   &$300$  &$15000$   & %%@
$65.2$  & $e^+p$ & $319$              &   \cite{Adloff:2003uh} \\
\hline
ZEUS~BPC            & $95$ & $2\times 10^{-6}$    & $6\times 10^{-5}$  %%@
&$0.11$  & $0.65$   & $1.65$  &  $e^+p$  & $301$ &       %%@
\cite{Breitweg:1997hz} \\
ZEUS~BPT            & $97$ & $6\times 10^{-7}$    & $0.001$   & $0.045$ & %%@
$0.65$  & $3.9$  & $e^+p$   &  $301$  &     \cite{Breitweg:2000yn}\\
ZEUS~SVX            & $95$ & $1.2\times 10^{-5}$  & $0.0019$  & $0.6$   & %%@
$17$  & $0.2$  & $e^+p$   &  $301$    &      \cite{Breitweg:1998dz}\\
ZEUS~NC             & $96$-$97$ & $6\times10^{-5}$  &$0.65$& $2.7$ & %%@
$30000$  & $30.0$  & $e^+p$    & $301$        \cite{Chekanov:2001qu}\\   
ZEUS~CC             & $94$-$97$ & $0.015$  & $0.42$  & $280$  & $17000$   %%@
&$47.7$   &  $e^+p$   & $301$         &      \cite{zeuscc97}\\
ZEUS~NC             & $98$-$99$ & $0.005$  & $0.65$  & $200$  & $30000$  %%@
&$15.9$   & $e^-p$    & $319$          &     \cite{Chekanov:2002ej} \\   
ZEUS~CC             & $98$-$99$ & $0.015$  & $0.42$  & $280$  & $30000$  %%@
&$16.4$   & $e^-p$    & $319$          &      \cite{Chekanov:2002zs} \\
ZEUS~NC             & $99$-$00$ & $0.005$  & $0.65$  & $200$  & $30000$  %%@
&$63.2$   & $e^+p$   & $319$           &      \cite{Chekanov:2003yv} \\   
ZEUS~CC             & $99$-$00$ & $0.008$  & $0.42$  & $280$  & $17000$  %%@
&$60.9$   &  $e^+p$   &$319$           &       \cite{Chekanov:2003vw}\\
\hline
\end{tabular}
\end{scriptsize}
\end{center}
\caption{H1 and ZEUS data sets used for the combination. }
\label{tab:data}
\end{table} 
The full details of the combination procedure are given in %%@
ref~\cite{h1zeuscomb}. 
The combination of the data sets uses the
$\chi^2$ minimisation method described in~\cite{Collaboration:2009bp}.
The $\chi^2$ function takes into account the correlated systematic %%@
uncertainties
for the H1 and ZEUS cross-section measurements.
Global normalisations of the data sets 
are split into an overall normalisation uncertainty of $0.5\%$, common to %%@
all data sets, due to uncertainties of higher order corrections 
to the Bethe-Heitler process used for  the luminosity calculation,
and experimental uncertainties which are treated as correlated systematic %%@
sources.
Some sources of point-to-point correlated uncertainties
are common for CC and NC data and for several data sets of the same %%@
experiment.  The systematic uncertainties were treated as independent %%@
between H1 and ZEUS apart from the $0.5\%$ overall normalisation %%@
uncertainty. 
All the NC and CC cross-section data from H1 and ZEUS are combined in one %%@
simultaneous minimisation. 
Therefore resulting shifts of 
the correlated systematic uncertainties propagate coherently to both CC and %%@
NC data. 

There are in total $110$ sources of correlated systematic uncertainty, %%@
including global normalisations, characterising 
the separate data sets. None of these systematic sources
shifts by more than $2\, \sigma$ of the nominal value in the averaging %%@
procedure. 
The absolute normalisation of the combined data set
is to a large extent defined by the most precise measurements of
NC $e^+p$ cross-section in the $10\le Q^2\le 100$~GeV$^2$ kinematic range. %%@
Here
the H1~\cite{Collaboration:2009kv} and ZEUS~\cite{Chekanov:2001qu} results %%@
move towards each other and 
the other data sets follow this adjustment.

The influence of several correlated systematic uncertainties is reduced
significantly for the averaged result. 
For example, the uncertainty due to the  H1 LAr calorimeter energy scale  %%@
is halved  while the uncertaintydue to the
ZEUS photoproduction background is reduced by a factor of 3. 
There are two main reasons for thess significant reductions. 
Since H1 and ZEUS use different reconstruction methods
similar systematic sources influence 
the measured cross section differently as a function of $x$ and $Q^2$. 
Therefore, requiring the cross sections to agree at all $x$ and $Q^2$ %%@
constrains 
the systematics efficiently. In addition, for some regions of the phase %%@
space, one of the two experiments has superior precision compared to the %%@
other.
For these regions, the less precise measurement is fitted to the more %%@
precise one, with a simultaneous reduction of the correlated systematic %%@
uncertainty.
This reduction propagates to the other average points, including those %%@
which are based solely on the measurement from the less precise experiment.
 
In addition to the 110 sources of systematic uncertainty which result from 
the separate data sets there are three sources of procedural uncertainty deriving
 from the choices made in the combination. Firstly all systematic 
uncertainties were treated as multiplicative, this has been varied by treating 
all sources bar the normalisations as additive, and the difference is used 
to estimate a procedural systematic error. Secondly, the correlated 
systematics from H1 and ZEUS were treated as uncorrelated between the 
experiments, but this may not be 
completely true due to some similarity of methods. An alternative combination 
procedure treats 12 sources of similar systematics as correlated. 
This results in some differences in the result for the photo-producton background and the 
hadronic energy scale and these differences are use to estimate two further 
procedural systematic errors.

The data averaging procedure results in a set of measurements for each %%@
process:
the average cross section value at a point $i$, its relative
correlated systematic, relative statistical and relative uncorrelated %%@
systematic uncertainties, respectively. 
The number of degrees of
freedom, $ndf$, is calculated as the difference between the total 
number of measurements and 
the number of combined data points. The value of $\chi^2_{\rm min}/ndf$ is %%@
a measure of the consistency of the data sets.

Tabulated results for the average NC and CC cross sections and the %%@
structure function $F_2$
together with statistical, uncorrelated systematic and procedural
uncertainties are given in ref~\cite{h1zeuscomb}. 
The total integrated luminosity of the combined data set corresponds to %%@
about $200$~pb$^{-1}$ for $e^+p$ and
 $30$~pb$^{-1}$ for $e^-p$.
In total $\totalave$ data points are combined to $\uniqueave$ cross-section %%@
measurements. 
The data show good consistency, with $\chi^2/ndf = \chiave/\dofave$, and %%@
there are no tensions between the input data sets.

For $Q^2\ge 100$~GeV$^2$ the precision of the H1 and
ZEUS measurements is about equal and thus the systematic uncertainties are %%@
reduced uniformly.
For $2.5\le Q^2 < 100$~GeV$^2$ and $Q^2<1$~GeV$^2$ the precision is %%@
dominated by the H1~\cite{Collaboration:2009bp,Collaboration:2009kv}
and ZEUS~\cite{Breitweg:2000yn} measurements, respectively.
Therefore the overall reduction of the uncertainties is smaller, 
and it is essentially obtained from the reduction of the correlated %%@
systematic uncertainty. 
The total uncertainty of the combined measurement is typically smaller than %%@
$2\%$ for $3 < Q^2 < 500$~GeV$^2$ and reaches $1\%$ for $20 <  Q^2 < %%@
100$~GeV$^2$. 
The uncertainties are larger for high inelasticity $y>0.6$ 
due to the photoproduction background.
%\begin{figure}
%\centerline{\epsfig{file=uncertainty.eps,width=\linewidth}}
%\caption{Total experimental uncertainties $\delta_{\rm ave,exp~tot}$ 
%given in per mill of the combined NC $e^+p$ data set.
%The colour coding, given in the legend, groups the measurements
%based on the size of the uncertainty.\label{fig:uncertainty}}
%\end{figure}

\begin{figure}[tbp]
\vspace{-0.5cm} 
%\vspace*{5pt}
\centerline{
\epsfig{figure=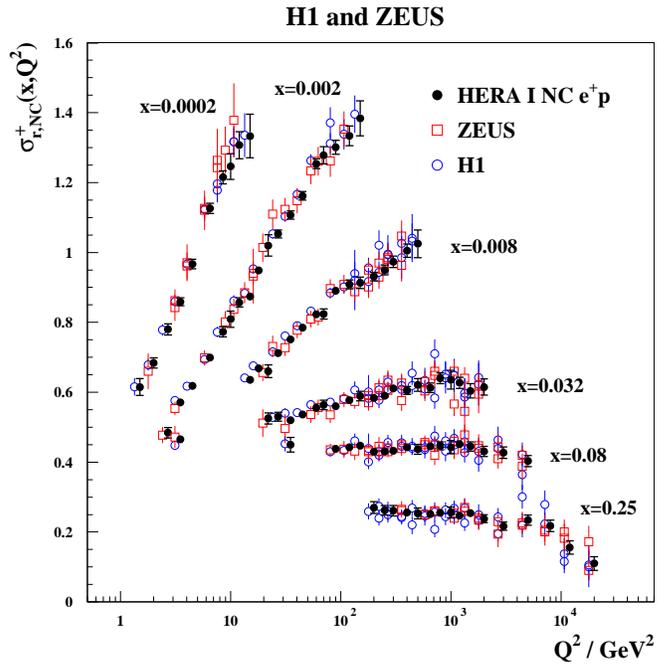,height=0.4\textheight}}
\caption {HERA combined NC $e^+p$ reduced 
cross section as a function of 
$Q^2$ for six $x$-bins compared to the separate 
H1 and ZEUS data input to the averaging procedure.
The individual measurements are displaced horizontally for a better %%@
visibility.
}
\label{fig:quality}
\end{figure}

\begin{figure}[tbp]
\vspace{-0.5cm} 
%\vspace*{5pt}
\centerline{
\epsfig{figure=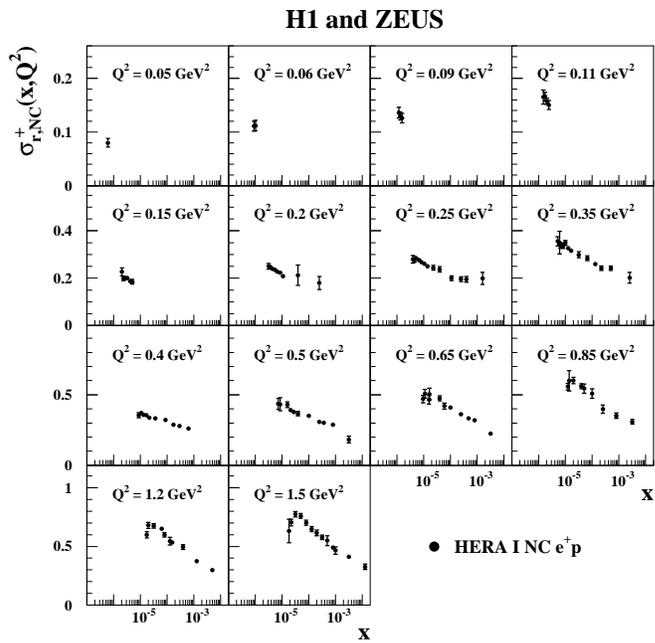,height=0.4\textheight}
}
\caption {HERA combined NC $e^+p$ reduced cross section at very low $Q^2$. 
}
\label{fig:vsQ2l}
\end{figure}

\begin{figure}[tbp]
\vspace{-0.5cm} 
%\vspace*{5pt}
\centerline{
\epsfig{figure=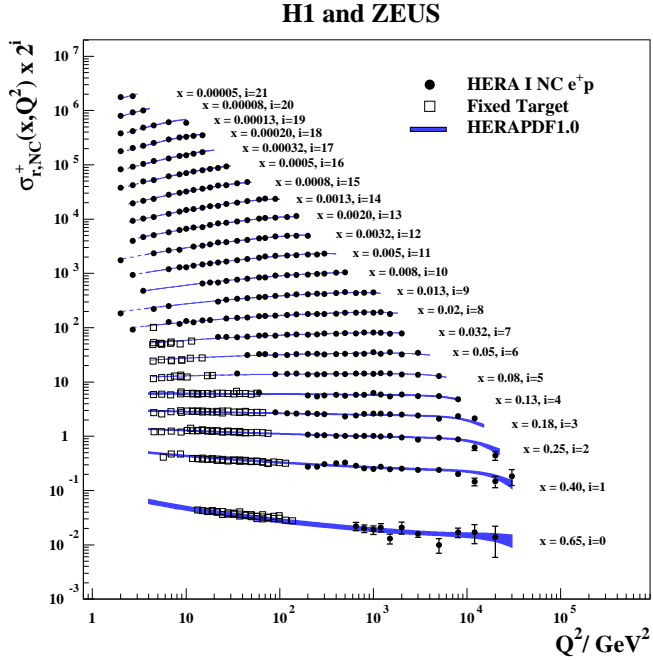,height=0.4\textheight}}
\caption {HERA combined NC $e^+p$ reduced cross section
 and fixed-target data as a function of $Q^2$.
The  HERAPDF1.0 fit is superimposed.
 The  bands represent the total uncertainty of the fit.
 The dashed lines are shown for $Q^2$ values not included in the QCD %%@
analysis. 
}
\label{fig:scal}
\end{figure}

\begin{figure}[tbp]
\vspace{-0.5cm} 
%\vspace*{5pt}
\centerline{
\epsfig{figure=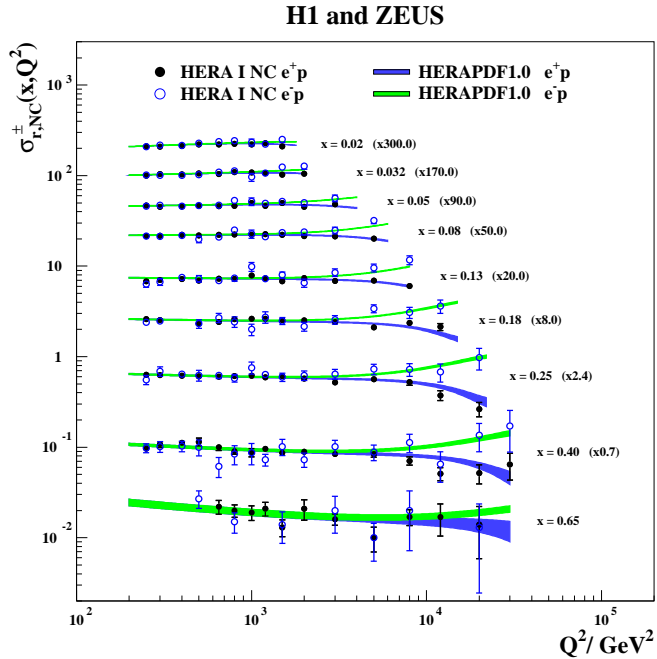,height=0.4\textheight}}
\caption {HERA combined NC $e^{\pm}p$ reduced cross sections
at high $Q^2$. 
The  HERAPDF1.0 fit is superimposed.
 The  bands represent the total uncertainty of the fit.
}
\label{fig:ncepem}
\end{figure}

\begin{figure}[tbp]
%\vspace{-1.0cm} 
%\vspace*{5pt}
\centerline{
\epsfig{figure=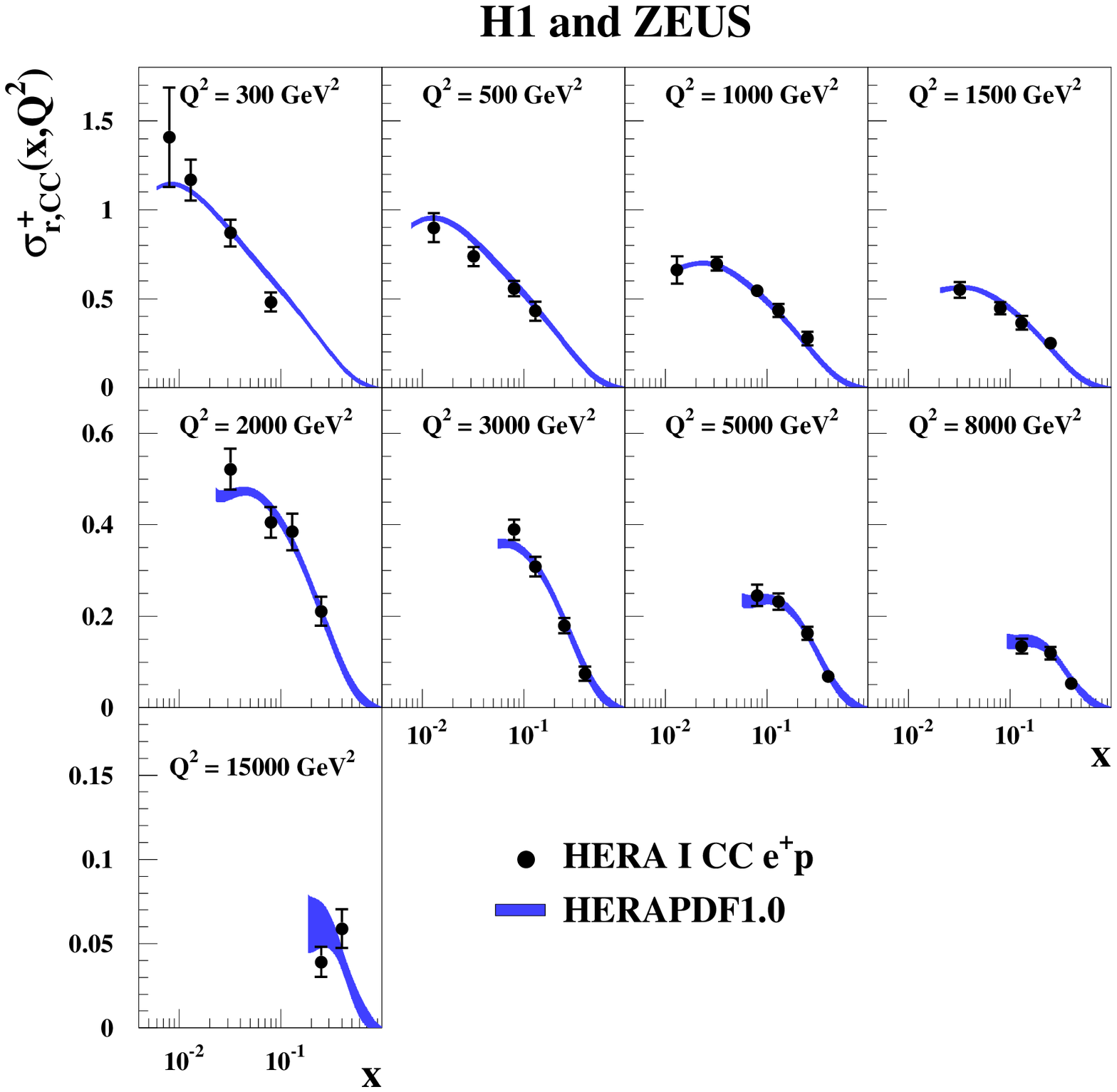,height=0.4\textheight}
}
\caption {HERA combined CC $e^+p$ reduced cross section. 
The  HERAPDF1.0 fit is superimposed.
 The  bands represent the total uncertainty of the fit.
}
\label{fig:dataCCp}
\end{figure}

\begin{figure}[tbp]
%\vspace{-1.0cm} 
%\vspace*{5pt}
\centerline{
\epsfig{figure=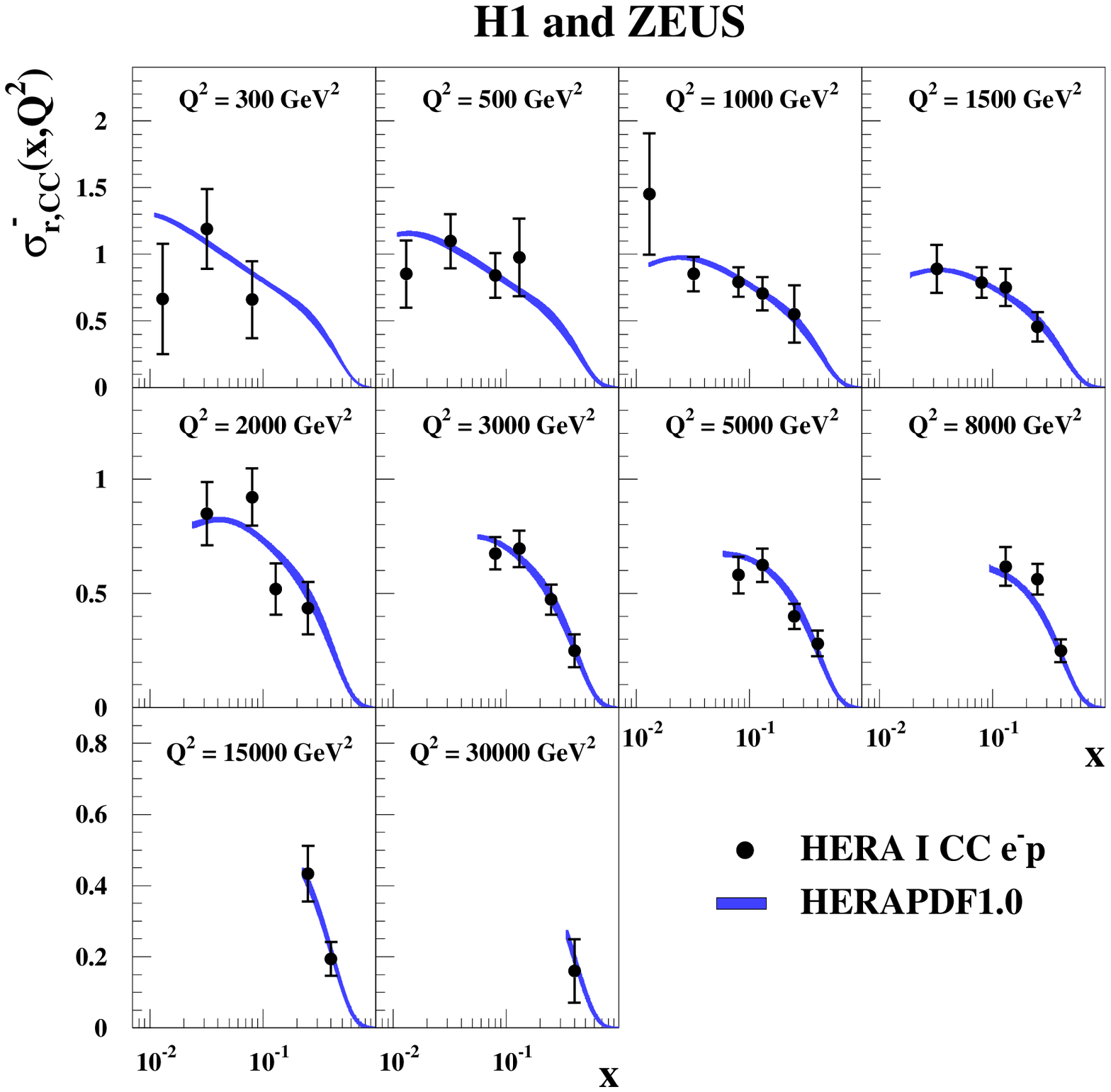,height=0.4\textheight}}
\caption {HERA combined  CC $e^-p$ reduced cross section. 
The  HERAPDF1.0 fit is superimposed.
 The  bands represent the total uncertainty of the fit.
}
\label{fig:dataCCm}
\end{figure}

In Fig~\ref{fig:quality} averaged data are compared to the 
input H1 and ZEUS data, illustrating the improvement in precision. Because %%@
of the reduction in size of the systematic error this improvement is far %%@
better than would be expected simply from the rough doubling of statistics %%@
which combining the two experiments represents. 
In Fig~\ref{fig:vsQ2l}, the combined NC $e^+p$ data at very low $Q^2$ are %%@
shown.
In Fig~\ref{fig:scal} the NC reduced cross section, for $Q^2 > 1$\,GeV$^2$, %%@
is shown
as a function of $Q^2$ for the 
HERA combined $e^+p$ data and for fixed-target data~\cite{bcdms,nmc} across 
the whole of the measured kinematic plane.
The combined NC $e^{\pm}p$ reduced cross sections are compared in the %%@
high-$Q^2$
region in Fig~\ref{fig:ncepem}.
In Figs~\ref{fig:dataCCp} and \ref{fig:dataCCm}
 the combined data set is shown for CC scattering at high $Q^2$.
The HERAPDF1.0 fit, described in Sec.~\ref{sec:pdfchap}, used these data as input. 
It is superimposed on the data 
 in the kinematic region suitable for the application of perturbative QCD. 

\subsection{ HERA-II inclusive data sets}

The published inclusive data combination does not included the data 
from the HERA-II running period 2003-2007. These data were collected 
 for both electron and positron beams, polarised both positively %%@
and negatively, at  $\surd{s}$ = 318 GeV. 
 The polarised data can be used to measure electroweak %%@
parameters~\cite{ewpapers}. This is beyond the scope of the present review. 
For investigation of the parton distribution functions 
these data have been combined into unpolarised cross-sections.
 Details of the luminosities collected are given in %%@
Table~\ref{tab:lumi} 
\begin{table}[tbp]
\centerline{
\begin{tabular}{|l|r|r|}
%\vspace{-1.0cm}
\hline
 New Data Set& Luminosity in pb$^{-1}$ & Reference \\
\hline
% CC $e^-p$ 04/06 P = -0.27 & $104$    \\
% CC $e^-p$ 04/06 P = +0.30  & $71$   \\
% NC $e^-p$ 05/06 P = -0.27  & $99$    \\
% NC $e^-p$ 05/06 P = +0.29 & $71$   \\
% CC $e^+p$ 06/07 P = -0.36  & $56$  \\
% CC $e^+p$ 06/07 P = +0.33  & $76$    \\
% NC $e^+p$ 07 $\surd{s}$ = 318  & $44$    \\
% NC $e^+p$ 07 $\surd{s}$ = 251 & $7.1$   \\
% NC $e^+p$ 07 $\surd{s}$ = 225 & $14$   \\
 CC $e^-p$ 04/06 ZEUS & $175$  &  \cite{zeusccem}\\
 CC $e^-p$ 04/06 H1  & $180$  & \cite{h1cc}\\
 NC $e^-p$ 05/06 ZEUS & $170$ & \cite{zeusncem} \\
 NC $e^-p$ 05/06 H1 & $180$   & \cite{h1nc}\\
 CC $e^+p$ 06/07 ZEUS  & $132$ & \cite{zeusccep} \\
 CC $e^+p$ 06/07 H1  & $149$   & \cite{h1cc}\\
 NC $e^+p$ 06/07 ZEUS & $ 135$ & \cite{zeusncep} \\
 NC $e^+p$ 06/07 H1 & $149$   & \cite{h1nc}\\
\hline
\end{tabular}}
\caption{Luminosities of the HERA-II data sets.
}
\label{tab:lumi}
\end{table}

\begin{figure}[tbp]
\vspace{-0.5cm} 
%\vspace*{5pt}
\centerline{
\epsfig{figure=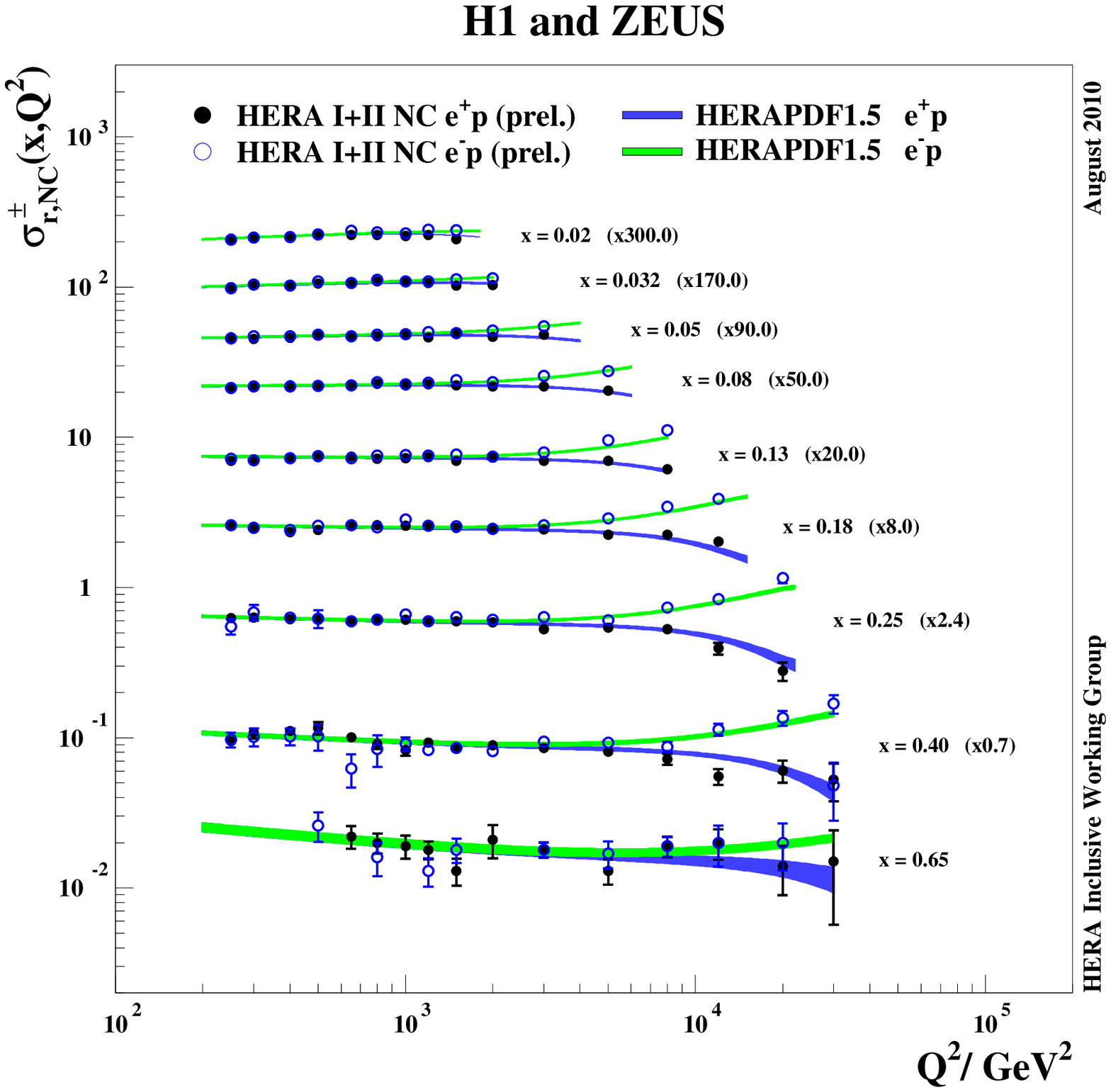,height=0.4\textheight}}
\caption {HERA I+II preliminarycombined NC $e^{\pm}p$ reduced cross sections
at high $Q^2$. 
The  HERAPDF1.5 fit is superimposed.
 The  bands represent the total uncertainty of the fit.
}
\label{fig:ncepem2}
\end{figure}

\begin{figure}[tbp]
%\vspace{-1.0cm} 
%\vspace*{5pt}
\centerline{
\epsfig{figure=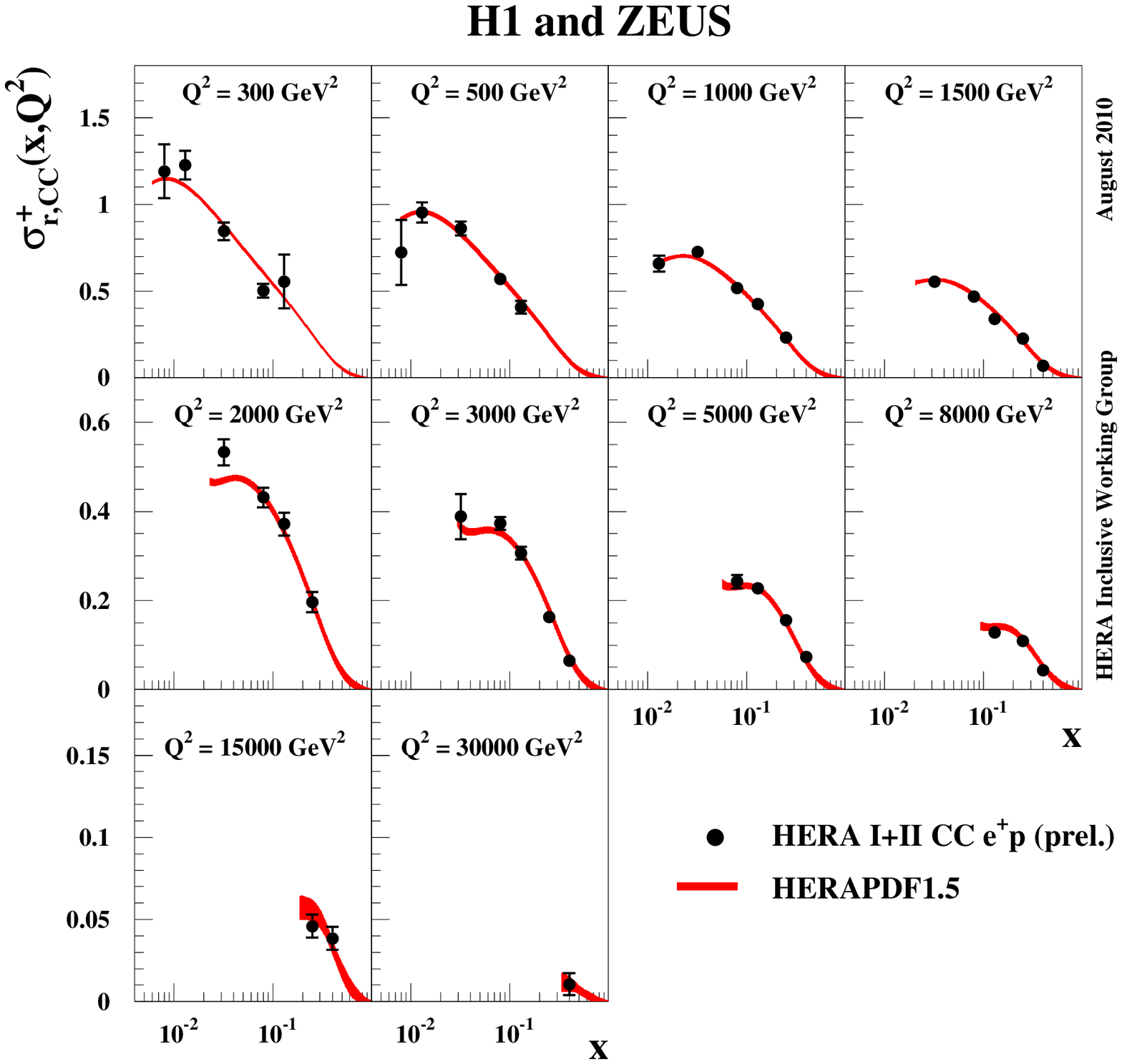,height=0.4\textheight}
}
\caption {HERA I+II preliminary combined CC $e^+p$ reduced cross section. 
The  HERAPDF1.5 fit is superimposed.
 The  bands represent the total uncertainty of the fit.
}
\label{fig:dataCCp2}
\end{figure}

\begin{figure}[tbp]
%\vspace{-1.0cm} 
%\vspace*{5pt}
\centerline{
\epsfig{figure=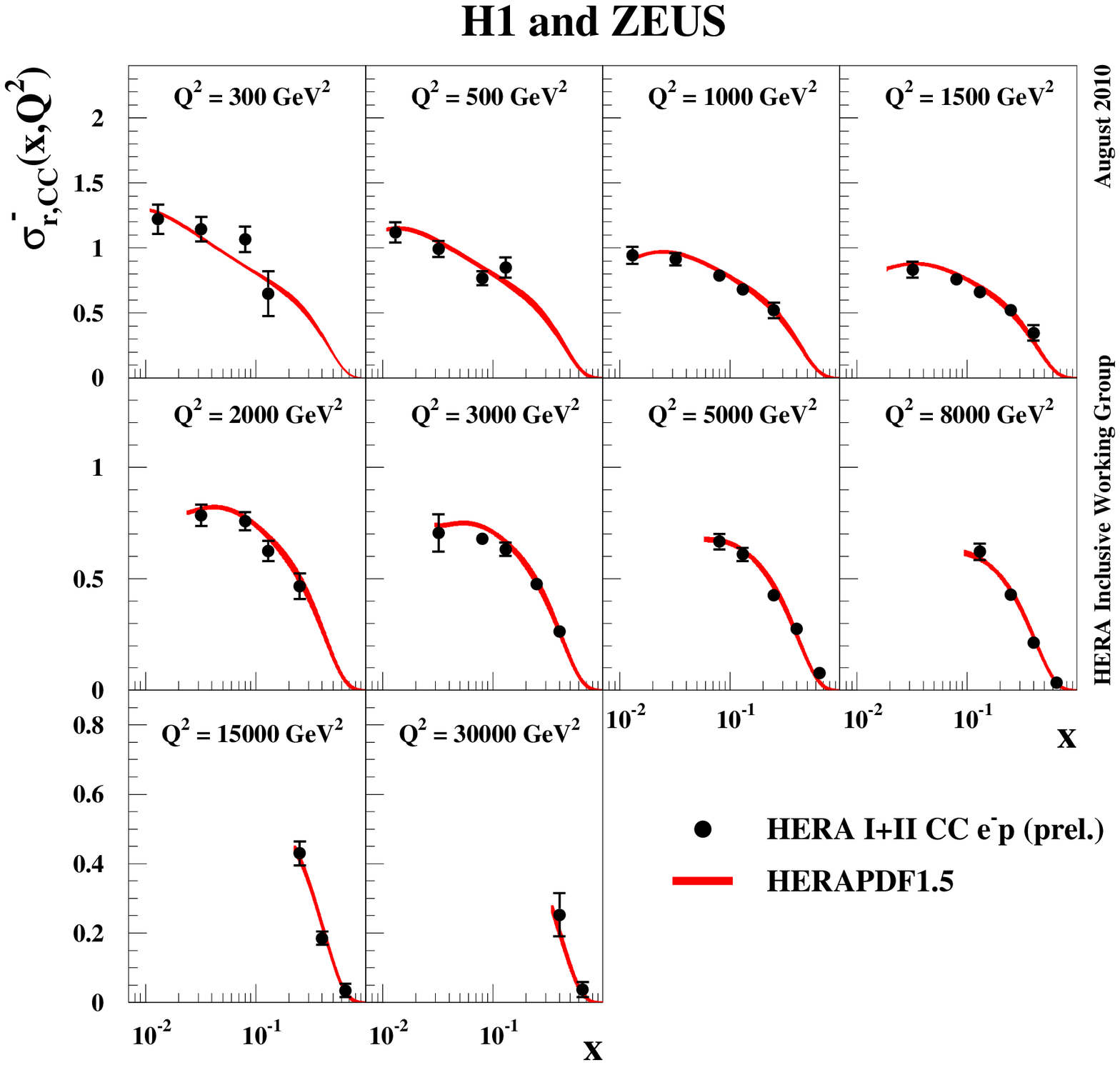,height=0.4\textheight}}
\caption {HERA I+II preliminary combined  CC $e^-p$ reduced cross section. 
The  HERAPDF1.5 fit is superimposed.
 The  bands represent the total uncertainty of the fit.
}
\label{fig:dataCCm2}
\end{figure}

A preliminary combination of these data with the HERA-I data has been made~\cite{hera2comb} 
using the same $\chi^2$ minimisation method, such that a new a set of 
measurements for each process, NC and CC $e^+p$ and $e^-p$, results.
There are in total $131$ sources of correlated systematic uncertainty, 
characterising the separate data sets, and three sources of procedural 
uncertainty, plus the overall normalisation uncertainty of $0.5\%$, as before.
These preliminary combined data are shown in 
Figs.~\ref{fig:ncepem2}-~\ref{fig:dataCCm2}. Comparison of these figures 
with Figs.~\ref{fig:ncepem}-~\ref{fig:dataCCm} shows how much the addition
of the HERA-II data improves precision at high $x$ and $Q^2$. 
The HERAPDF1.5 fit~\cite{hera2fit}, described in Sec.~\ref{sec:pdfchap}, used these data as input. 
It is superimposed on the data in the figures.  


\subsection{$F_L$ data}
\label{sec:lowE}
During the final running period the proton beam ran at three different %%@
proton beam energies ($920, 575, 460~$GeV) and NC $e^+p$ data were collected. 
These data access high-$y$ and have been used to measure 
the longitudinal structure function $F_L$~\cite{h1fl,zeusfl}. 
The luminosities for the input data for the combination are specfied in 
Table~\ref{tab:fl}
\begin{table}[tbp]
\centerline{
\begin{tabular}{|l|r|r|}
%\vspace{-1.0cm}
\hline
 New Data Set& Luminosity in pb$^{-1}$ & Reference \\
\hline
 ZEUS $E_p=460~$GeV& $13.9$  &  \cite{zeusfl}\\
 ZEUS $E_p=575~$GeV& $7.1$  &  \cite{zeusfl}\\
 ZEUS $E_p=920~$GeV& $45$  &  \cite{zeusfl}\\
 H1 $E_p=460~$GeV& $12.4$  &  \cite{h1fl}\\
 H1 $E_p=575~$GeV& $6.2$  &  \cite{h1fl}\\
 H1 $E_p=920~$GeV& $21.6$  &  \cite{h1fl}\\
\hline
 H1 $E_p=460~$GeV& $12.2$  &  \cite{h1newfl}\\
 H1 $E_p=575~$GeV& $5.9$  &  \cite{h1newfl}\\
 H1 $E_p=920~$GeV& $97.6,5.9$  &  \cite{h1newfl}\\
\hline
\end{tabular}}
\caption{Luminosities of the data sets for the low energy running. 
The first 6 data sets have already been combined. The final 3 will be combined.
}
\label{tab:fl}
\end{table}

\begin{figure}[tbp]
\vspace{-0.5cm} 
%\vspace*{5pt}
\centerline{
\epsfig{figure=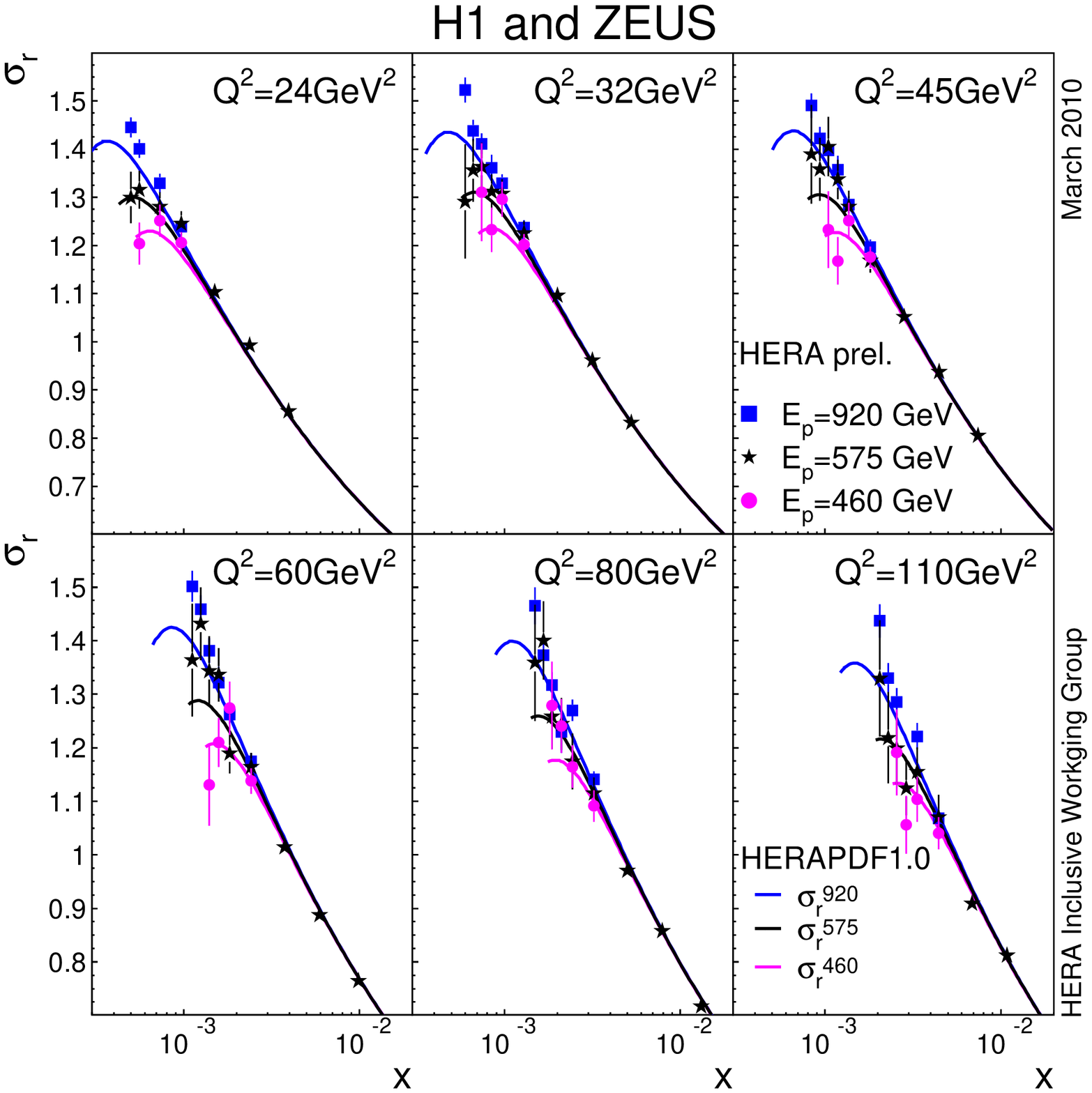,height=0.4\textheight}}
\caption {HERA combined NC $e^{+}p$ reduced cross sections from running at
three different proton beam energies. the predictions of HERAPDF1.0 are 
superimposed. 
}
\label{fig:lowE}
\end{figure}

\begin{figure}[tbp]
\vspace{-0.5cm} 
%\vspace*{5pt}
\centerline{
\epsfig{figure=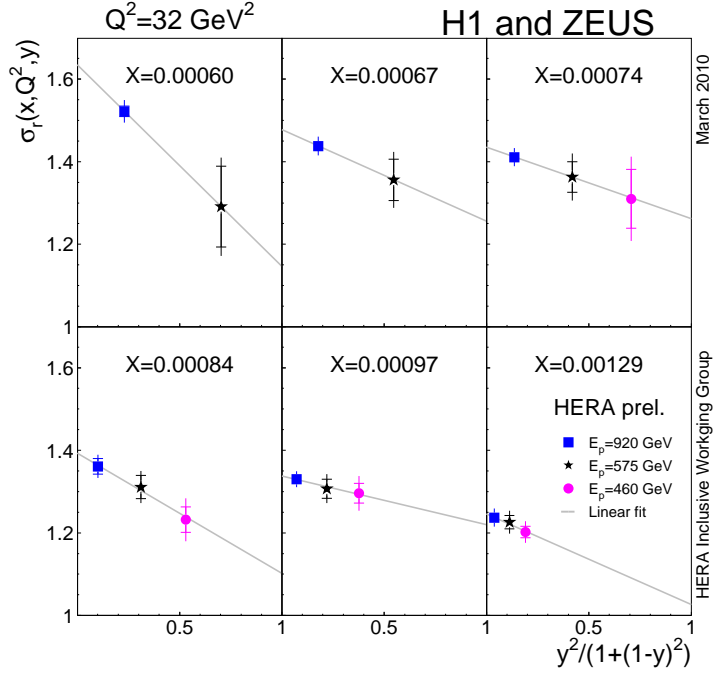,height=0.4\textheight}}
\caption {The slope of $\tilde{\sigma}$ vs $f(y)= y^2/Y_+$ for various $x$ 
bins at $Q^2=32$GeV$^2$.
}
\label{fig:lowEslope}
\end{figure}

\begin{figure}[tbp]
\vspace{-0.5cm} 
%\vspace*{5pt}
\centerline{
\epsfig{figure=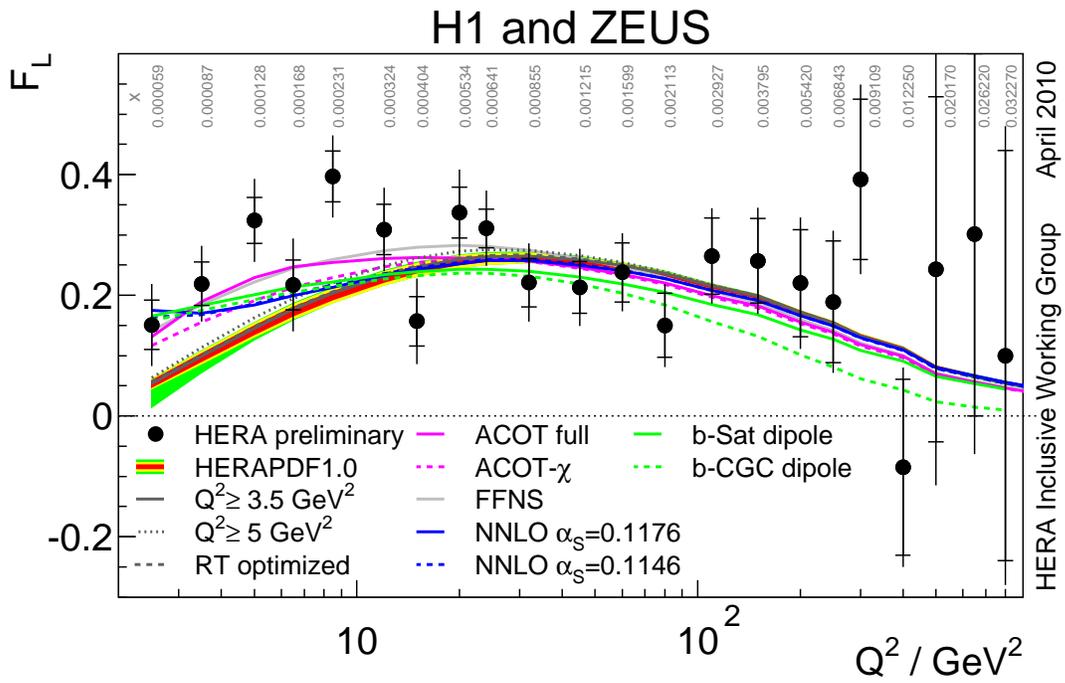,height=0.4\textheight}}
\caption {The HERA combined measurement of $F_L$ avberaged in $x$ at a 
given value of $Q^2$.
}
\label{fig:fl}
\end{figure}

The reduced cross-section data from these runs have been combined~\cite{lowEcomb}
and the combination is shown in Fig~\ref{fig:lowE}. 
Using these data a combined measurement of $F_L$ can be made~\cite{lowEcomb}.
Recall that the NC $e^+p$ reduced cross section is given by, 
$\tilde{\sigma} = F_2  - y^2 F_L/Y_+$, for $Q^2 \ll M_Z^2$. 
Since $Q^2=s x y$ one needs measurements at different $s$ values in order 
to access different 
$y$ values for the same $x,Q^2$ point.
The structure function $F_L$ is measured as a slope of a linear fit of 
$\tilde{\sigma}$ versus $f(y)= y^2/Y_+$, in $x,Q^2$ bins. 
Fig~\ref{fig:lowEslope} shows an 
example of such a fit, for various $x$ values, at $Q^2 = 32~$GeV$^2$. 
The measured $F_L$ is shown, averaged in $x$ as a function of $Q^2$, in 
Fig~\ref{fig:fl}, with various theoretical predictions superimposed.
At low-$x$, NLO QCD 
in the DGLAP formalism predicts that 
this structure function is strongly related to the gluon PDF, 
see Eqn.~\ref{eqn:fl_qg}. 

\begin{figure}[tbp]
\vspace{-0.5cm} 
%\vspace*{5pt}
\centerline{
\epsfig{figure=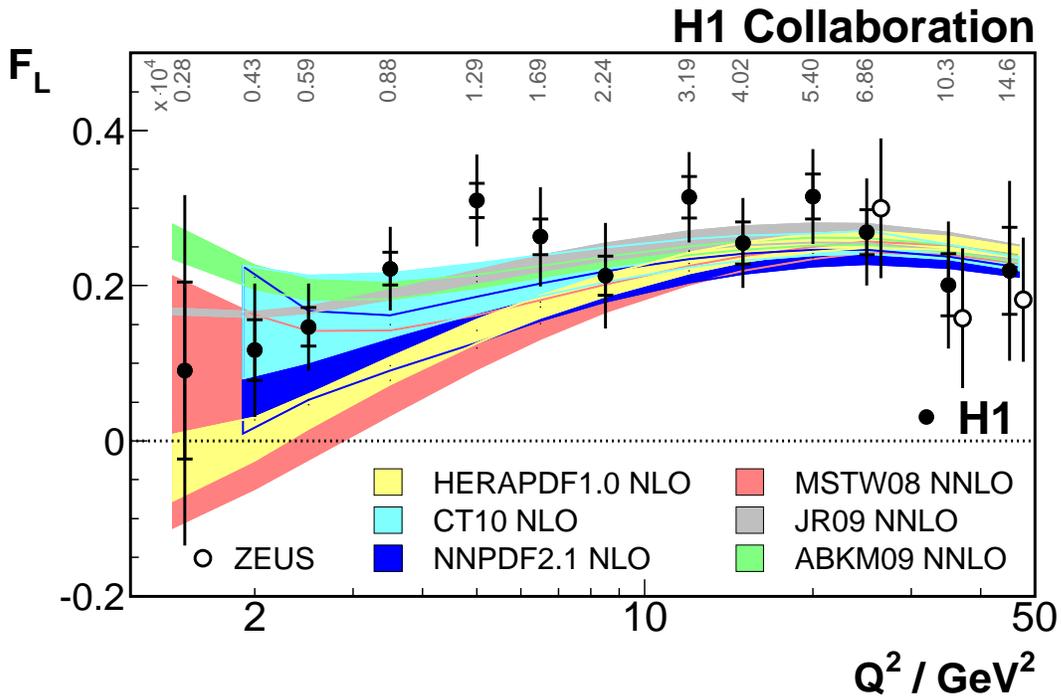,height=0.4\textheight}}
\caption {The H1 2011 measurement of $F_L$ averaged in $x$ at a 
given value of $Q^2$.
}
\label{fig:h1newfl}
\end{figure}

This  combination will be updated to include the recently
published
H1 data, which extend to lower $Q^2$~\cite{h1newfl}. 
The luminosities used for this extension are given in Table~\ref{tab:fl}
and the $F_L$ 
measurement from these H1 data is shown in Fig~\ref{fig:h1newfl}.

\subsection{$F_2^{c\bar c}$ and $F_2^{b\bar b}$ data sets}
\label{sec:charmdata}
A preliminary combination has been made of data on $F_2^{c\bar c}$~\cite{f2ccomb} from 
various different methods of tagging charm: using the $D^*$, using 
the vertex detectors to see the displaced decay vertex, using direct 
$D_0, D^+$ production identified using the vertex detectors, and indentifying 
semi-leptonic charm decays via muons, also using the vertex detectors.
The details of the data sets used in the combination are given in Table~\ref{tab:f2c}.

\begin{table}[tbp]
\centerline{
\begin{tabular}{|l|r|r|}
%\vspace{-1.0cm}
\hline
 Data Set& Luminosity in pb$^{-1}$ & Reference \\
\hline
% NC $e^+p$ 07 $\surd{s}$ = 225 & $14$   \\
 $D^*$ 99/00 H1 & $47$  &  \cite{h1d99}\\
 $D^*$ 04/07 H1  & $340$  & \cite{h1d07}\\
 Vtx. 99/00 H1 & $57.4$ & \cite{h1vtx99}  \\
 Vtx. 06/07 H1 & $189$   & \cite{h1vtx07}\\
 $D^*$ 98/00 ZEUS  & $82$ & \cite{zeusd00} \\
 $D^*$ 96/97 ZEUS  & $37$   & \cite{zeusd97}\\
 $D^0,D^{\pm}$ 04/05 ZEUS & $ 134$ & \cite{zeusd0dp} \\
 muons 05 ZEUS & $126$   & \cite{zeusmuon}\\
\hline
\end{tabular}}
\caption{Luminosities of the $F_2^{c\bar c}$ data sets.
}
\label{tab:f2c}
\end{table}

\begin{figure}[tbp]
\vspace{-0.5cm} 
%\vspace*{5pt}
\centerline{
\epsfig{figure=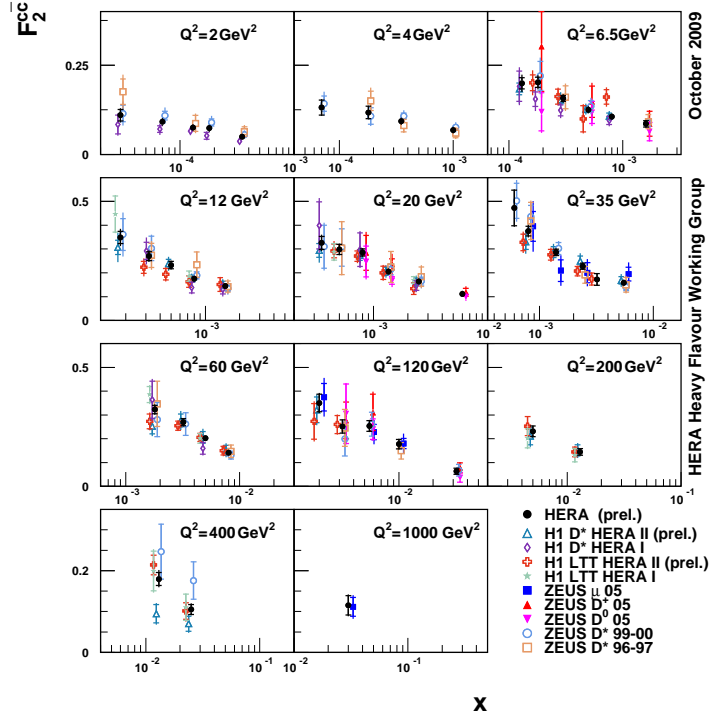,height=0.4\textheight}}
\caption {The HERA combined measurement of $F_2^{c\bar c}$ compared to the 
data sets of H1 and ZEUS used for the combination. these data sets are 
slightly displaced in $x$ for visibility.
}
\label{fig:f2ccomb}
\end{figure}

\begin{figure}[tbp]
\vspace{-0.5cm} 
%\vspace*{5pt}
\centerline{
\epsfig{figure=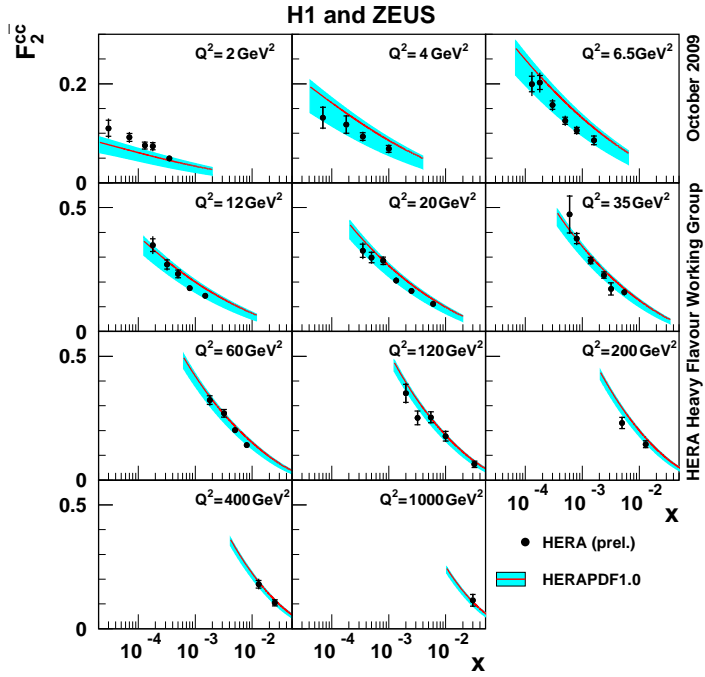,height=0.4\textheight}}
\caption {The HERA combined measurement of $F_2^{c\bar c}$ compared to the 
predictions of HERAPDF1.0
}
\label{fig:f2cfit}
\end{figure}

The results of the $F_2^{c\bar c}$ combination compared to the 
separate measurements which 
went into it are shown in Fig~\ref{fig:f2ccomb}.
The  $F_2^{c\bar c}$ combination is shown compared to the predictions of 
HERAPDF1.0 in Fig.~\ref{fig:f2cfit}

\begin{figure}[tbp]
\vspace{-0.5cm} 
%\vspace*{5pt}
\centerline{
\epsfig{figure=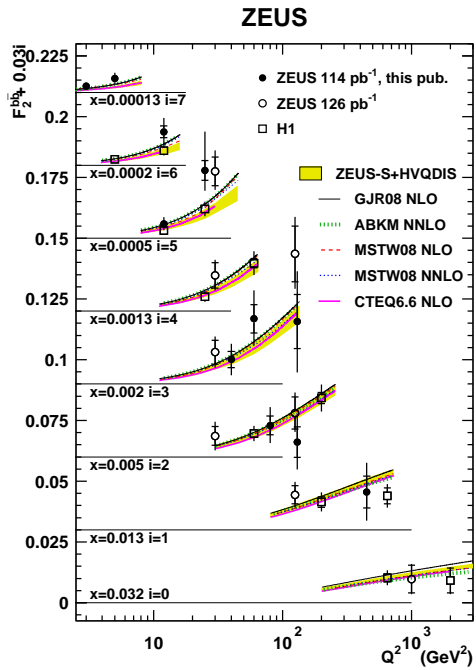,height=0.4\textheight}}
\caption {The H1 and ZEUS measurements of $F_2^{b\bar b}$.
}
\label{fig:f2b}
\end{figure}

Data on $F_2^{b\bar b}$ have not yet been combined. A recent comparison of 
H1~\cite{h1f2b} 
and ZEUS~\cite{zeusf2b} separate results is shown in Fig.~\ref{fig:f2b}.

\subsection{ Jet data sets}
\label{sec:jetdata}
\begin{figure}[tbp]
\vspace{-0.5cm} 
%\vspace*{5pt}
\centerline{
\epsfig{figure=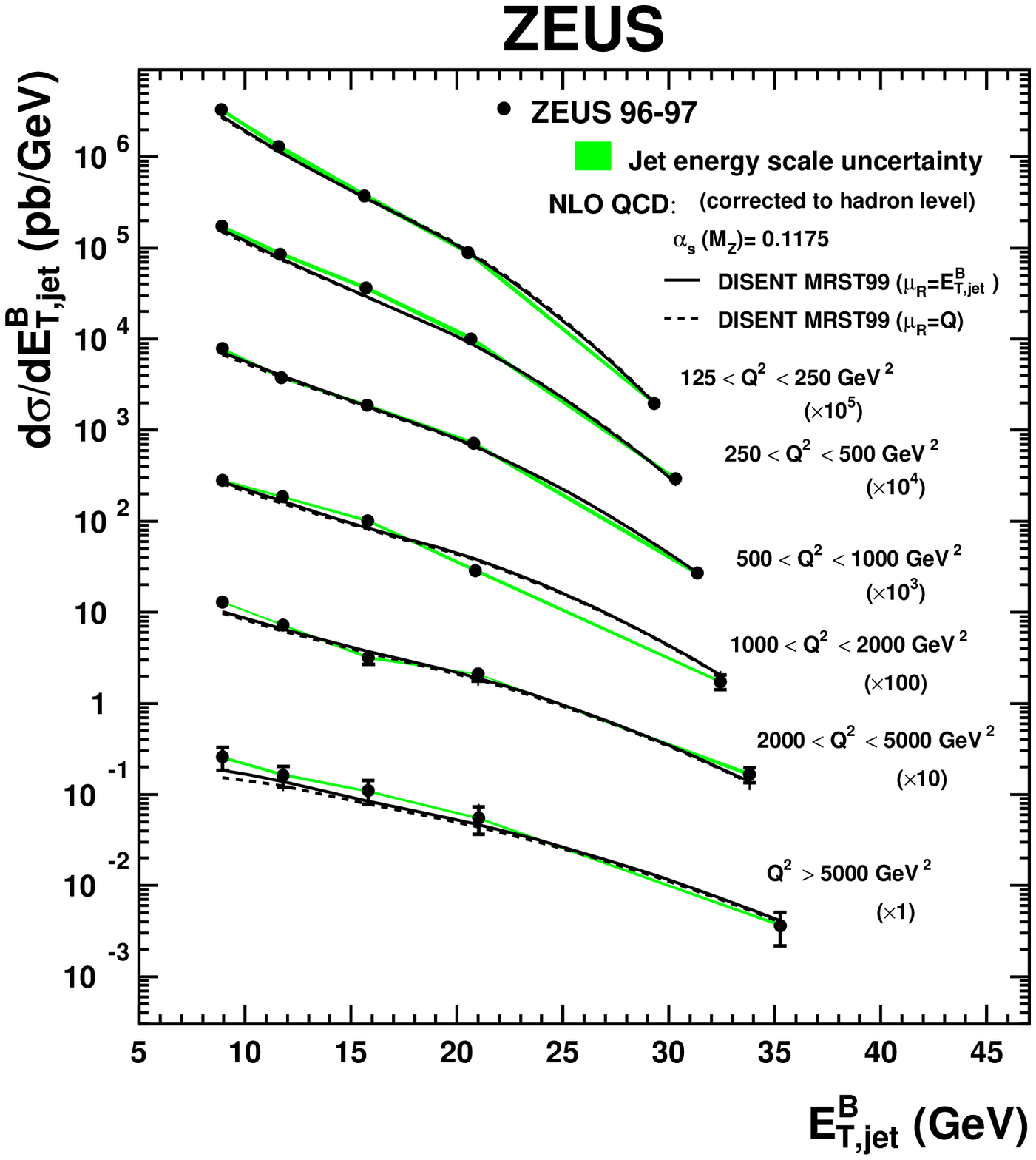,height=0.4\textheight}}
\caption {ZEUS 96/97 measurements of tne inclusive jet cross section, as a function of $E_T$ jet in the Breit frame for various $Q^2$ bins. 
}
\label{fig:jetzeus1}
\end{figure}

\begin{figure}[tbp]
\vspace{-0.5cm} 
%\vspace*{5pt}
\centerline{
\epsfig{figure=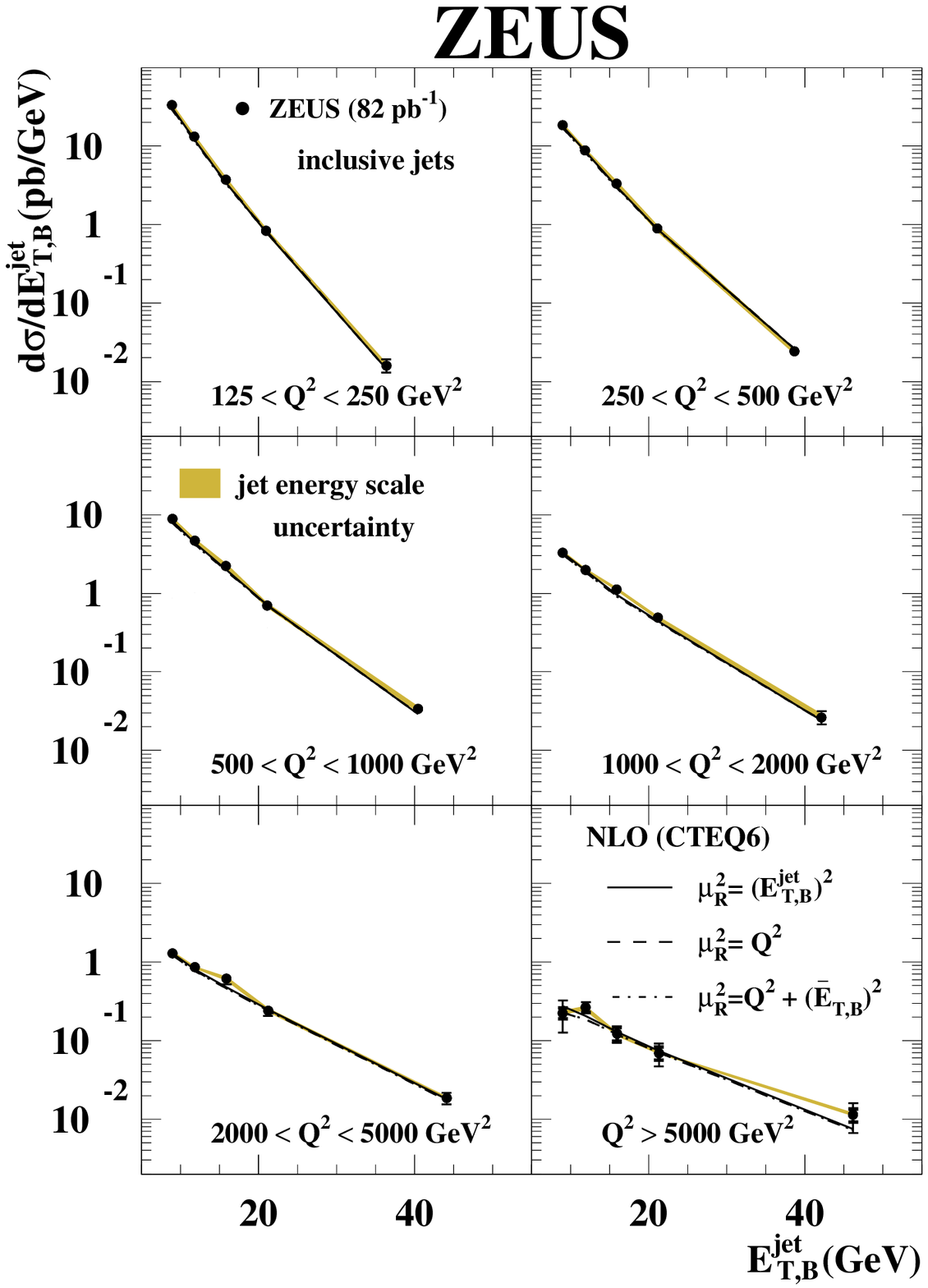,height=0.4\textheight}}
\caption {ZEUS 98/00 measurements of the inclusive jet cross section, as a function of $E_T$ jet in the Breit frame for various $Q^2$ bins. .
}
\label{fig:jetzeus2}
\end{figure}

\begin{figure}[tbp]
\vspace{-0.5cm} 
%\vspace*{5pt}
\centerline{
\epsfig{figure=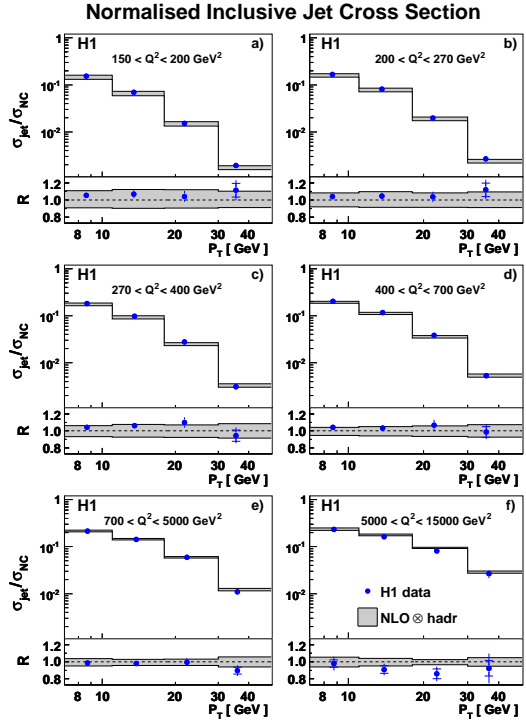,height=0.4\textheight}}
\caption {H1 HERA-I+II  measurements of tne normalised inclusive jet cross section, as a function of $p_T$  for various $Q^2$ bins. 
}
\label{fig:jeth11}
\end{figure}

\begin{figure}[tbp]
\vspace{-0.5cm} 
%\vspace*{5pt}
\centerline{
\epsfig{figure=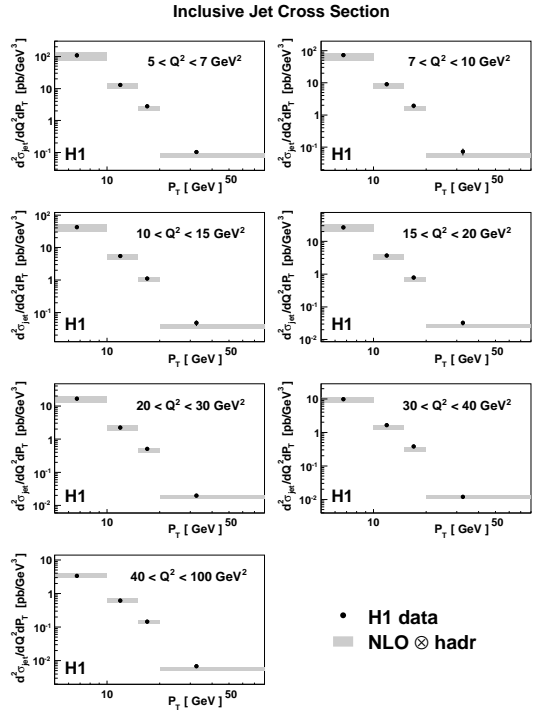,height=0.4\textheight}}
\caption {H1 HERA-I  measurements of tne low-$Q^2$ inclusive jet cross section, as a function of $p_T$ jet for various $Q^2$ bins. 
}
\label{fig:jeth12}
\end{figure}

Jet data may also be used to constrain the PDFs. 
So far H1 and ZEUS jet data have not 
been combined but some separate H1 and ZEUS jet data sets have been input to 
the HERAPDF fits in order to exploit their ability to constrain the gluon PDF 
and to make a determination of the value of $\alpha_s(M_Z)$ simultaneously 
with the PDF determination~\cite{herapdf16}. The jet data which have been used are summarised in 
Table~\ref{tab:jetdata}
\begin{table}[tbp]
\centerline{
\begin{tabular}{|l|r|r|}
%\vspace{-1.0cm}
\hline
 Data Set& Luminosity in pb$^{-1}$ & Reference \\
\hline
 High $Q^2$ norm. inc. jets 96/07 H1 & $395$  &  \cite{h1hq2}\\
 Low $Q^2$  inc. jets 96/00 H1 & $43.5$  &  \cite{h1lq2}\\
 High $Q^2$  inc. jets 96/97 ZEUS & $38.6$  &  \cite{zeus97}\\
 High $Q^2$  inc. jets 98/00 ZEUS & $82$  &  \cite{zeus00}\\
\hline
\end{tabular}}
\caption{Luminosities of the jet data sets.
}
\label{tab:jetdata}
\end{table}
These data are illustrated in Figs~\ref{fig:jetzeus1},~\ref{fig:jetzeus2},~\ref{fig:jeth11},~\ref{fig:jeth12}.

\section{Extraction of parton densities}
\label{sec:pdfchap}

The section discusses how parton momentum densities are extracted from the %%@
HERA data\footnote{Open access code for the HERA PDF fits, 
and many other useful utilities, are available from the HERAFitter website 
http://herafitter.hepforge.org}. There are several PDF fits to different 
HERA data sets. 
The HERAPDF1.0~NLO set used only the HERA-I combined data~\cite{h1zeuscomb}. 
However there are 
studies, using the same fit formalism, including the preliminary low-energy 
combined data~\cite{lowEpdf} and using the preliminary combined $
F_2^{c\bar{c}}$ data~\cite{charmpdf}. The
HERAPDF1.5~NLO set~\cite{hera2fit} used the HERA-I+II preliminary combined 
data~\cite{hera2comb} and the same 
fit formalism. Studies were also made using the same data but
extending the parametrisation HERAPDF1.5f~NLO~\cite{herapdf16}. 
This extended parametrisation 
was then used for the HERAPDF1.5~NNLO fit~\cite{herapdf15nnlo}. 
Subsequent fits also used the 
extended parametrisation: HERAPDF1.6~NLO fit~\cite{herapdf16} 
which included the HERA-I+II 
inclusive data and separate H1 and ZEUS jet data; 
HERAPDF1.7~NLO fit~\cite{herapdf17} which used all the data 
described in the previous sections (HERA-I+II 
inclusive data and low energy inclusive data, $F_2^{c\bar{c}}$ data and H1 and 
ZEUS jet data sets).
   
The relationship of the measured cross-sections to the parton 
distributions, presented in Sec.~\ref{sec:xsecns}, is not so %%@
straightforward beyond LO since the evolved parton 
distributions must be convoluted with coefficient functions and all types %%@
of
parton may contribute to a particular structure function through the %%@
evolution.
However, the simple LO formulae still give a good guide to the major %%@
contributions.
The cross-sections for NC and CC, $e^+$ and $e^-$ scattering on protons
provide enough information to extract the $u$ and $d$ valence PDFs and the 
$\bar{U}$ and $\bar{D}$ PDFs, as well as the gluon PDF from scaling %%@
violation. Briefly: 
\begin{itemize}
\item 
HERA NC $e^+$ reduced cross-section data at low $Q^2$ give information on %%@
the shape of the sea 
distribution at low $x$ region, whereas the high $Q^2$ NC $e^+$ and $e^-$ 
cross-sections not only extend the coverage to high $x$ ($x < 0.65$) but %%@
also provide information on the valence combination 
$xF_3 = x(B^0_u u_v + B^0_d d_v)$, which is extracted from the 
difference between the high-$Q^2$ NC $e^+$ and $e^-$ cross sections. 
The $x$ range of the valence measurement is, $0.01 < x < 0.65$.
\item
HERA CC data gives further information on flavour separation. The 
$e^-$ cross-section at high-$x$ is $u$-valence dominated and the $e^+$ 
cross-section is $d$-valence dominated, giving unique information on the %%@
$d$ quark. Fixed proton target data are 
$u$-quark dominated, and so historically information on the $d$-quark has 
been extracted from deuterium target %%@
data or from neutrino scattering. However, each of these methods has %%@
difficulties. The neutrino data uses heavy isoscalar targets such that %%@
uncertain nuclear corrections~\cite{cjstuff} are necessary. 
The deuterium target data also %%@
needs some nuclear binding corrections~\cite{jlabstuff} and extraction of 
the $d$-quark is %%@
dependent on the assumption of strong isospin invariance.
\item
NC data on $F_2$ have also been used to constrain the gluon distribution.
Since the gluon does not couple to the 
photon it does not enter the expressions for the structure functions at all 
in the QPM. However, it is constrained by the momentum sum rule, and by the %%@
way that gluon to quark-antiquark splitting feeds into the sea distributions
(from the $P_{qg}$ term in the DGLAP equations). 
The shape of the gluon distribution
extracted from a DGLAP QCD fit will be correlated with
the value of $\alpha_s$, since
an increase in $\alpha_s$ increases the negative contribution from the
$P_{qq}$ term but this may be compensated by a positive contribution from
the $P_{qg}$ term if the gluon is made harder.
Hence, a fixed value of $\alpha_s$,
as determined from independent data, has  often been assumed in PDF fits.
HERA data are invaluable in constraining the low-$x$gluon distribution, 
since at small $x$ QCD evolution becomes gluon
dominated and the uncertainties referred to above are reduced. 
This is because $F_2$ is essentially given
by the singlet sea quark distribution for $x \leqsim 0.01$, and this in %%@
turn is driven by the gluon through the $P_{qg}$ term in %%@
Eqn.~\ref{eqn:ap_gen}. The approximate LO relationship
\begin{equation}
 xg(x)\simeq \frac{3\pi}{\bar{e}^2\alpha_s}\frac{\partial F_2(x/2,Q^2)}{\partial lnQ^2}
\end{equation}
illustrates how the gluon distribution depends on the scaling violation of 
$F_2$ at low $x$.
Hence in this kinematic region the gluon distribution may be obtained %%@
almost directly from the $F_2$ scaling violation data. 
\item
Jet production data from HERA can give more direct information on the %%@
gluon since vector-boson gluon fusion (BGF) to quark-anti-quark pairs makes a 
significant 
contribution to final state jet production. Such data has also been input to
 the PDF fits to
constrain the gluon distribution in the $x$ range, $0.01 < x < 0.1$, 
and to simultaneously determine %%@
$\alpha_s(M_Z)$, see Sec.~\ref{sec:jets}.
\item
The longitudinal structure function $F_L$ 
can also give information on the gluon as can be seen from 
Equation~\ref{eqn:fl_qg}. At low $x$ the dominant contribution comes from the 
gluon and the integral over the gluon distribution approximates to a $\delta$ 
function such that a measurement of $F_L(x,Q^2)$ is almost a direct 
measurement of the gluon distribution $yg(y,Q^2)$ at $y=2.5x$~\cite{amcs1987}.
The heavy quark structure functions $F_2^{c\bar{c}}$ and $F_2^{b\bar{b}}$ 
may also yield information on 
the gluon since heavy quarks are generated by the BGF process. However, 
currently such data are most useful for distinguishing between different 
schemes for heavy quark production and fixing the value of the heavy quark mass
parameters that enter into these schemes, see Sec.~\ref{sec:charm}
\end{itemize}

Perturbative QCD predicts the $Q^2$ evolution of the parton distributions, %%@
but
not the $x$ dependence.
The parton distributions are extracted by 
performing a direct numerical integration of the DGLAP equations at NLO and 
NNLO~\cite{nnloevol}.
For most PDF extractions (the notable exception is the NNPDF analysis) 
a parametrised analytic shape for the parton distributions (valence, sea %%@
and gluon) is assumed to be valid  at some starting value of $Q^2 = Q^2_0$. %%@
This starting
value is arbitrary, but should be large enough to ensure that 
$\alpha_s(Q^2_0)$ is small enough for perturbative calculations to be
applicable. For the HERAPDF the value $Q^2_0 = 1.9$GeV$^2$ is chosen such %%@
that the starting scale is below the charm mass threshold, $Q_0^2 < m_c^2$. 
Then the DGLAP equations
are used to evolve the parton distributions up to higher $Q^2$ values, 
where they  are convoluted with coefficient functions to make predictions %%@
for the structure functions and cross sections. 
These predictions are then fitted to data to 
determine the PDF parameters, and thus the shapes of the parton distributions 
at the starting scale and, through evoution, at any other value of $Q^2$. 

The QCD evolution is performed using the programme 
QCDNUM~\cite{qcdnum}. The HERADF uses the \msbar renormalisaton %%@
scheme, with the renormalisation and factorisation scales chosen to be %%@
$Q^2$. The light quark coefficient functions~\cite{nlo,nnlo}
are calculated using the programme QCDNUM. The heavy quark 
coefficient functions are calculated in the general-mass
variable-flavour-number scheme of \cite{Thorne:1997ga}, with recent 
modifications and extension to NNLO~\cite{Thorne:2006qt,ThornePrivComm}. 
(This scheme will be called the RT-VFN scheme).
The heavy quark masses  for the central fit were chosen to be $m_c=1.4~$GeV %%@
and $m_b=4.75~$GeV and the strong coupling constant was fixed to 
$\asmz =  0.1176$. These choices are varied to evaluate model 
uncertainties. The predictions are then fitted to the %%@
combined HERA data sets on differential cross sections 
for NC and CC $e^+p$ and $e^-p$ scattering. 
A minimum $Q^2$ cut, $Q^2_{min} = 3.5$~GeV$^2$, was imposed 
to remain in the kinematic region where
perturbative QCD should be applicable. This choice is also varied when 
evaluating model uncertainties. It is also conventional to apply a 
minimum cut 
on $W$, invariant mass of the hadronic system, to avoid sensitivity to 
target mass and large-$x$ higher-twist contributions. However the HERA data 
have $W>15$\,GeV and $x< 0.65$, so that no further cuts are necessary.
 
PDFs were parametrised at the input scale by the generic form 
\begin{equation}
 xf(x) = A x^{B} (1-x)^{C} (1 + \epsilon\surd x + D x + E x^2).
\label{eqn:pdf}
\end{equation}
The parametrised PDFs  are the gluon distribution $xg$, the valence quark %%@
distributions $xu_v$, $xd_v$, and the $u$-type and $d$-type anti-quark %%@
distributions
$x\bar{U}$, $x\bar{D}$. Here $x\bar{U} = x\bar{u}$, 
$x\bar{D} = x\bar{d} +x\bar{s}$, at the chosen starting scale. 
The normalisation parameters, $A_g, A_{u_v}, A_{d_v}$, are constrained 
by the quark number sum-rules and momentum sum-rule. 
The $B$ parameters  $B_{\bar{U}}$ and $B_{\bar{D}}$ are set equal,
 $B_{\bar{U}}=B_{\bar{D}}$, such that 
there is a single $B$ parameter for the sea distributions. 
The strange quark distribution is expressed 
as $x$-independent fraction, $f_s$, of the $d$-type sea, 
$x\bar{s}= f_s x\bar{D}$ at $Q^2_0$. For $f_s=0.5$ the %%@
$s$ and $d$ quark densities would be the same, but the value $f_s=0.31$  
is chosen to be consistent with determinations 
of this fraction using neutrino-induced di-muon 
production~\cite{Martin:2009iq,Nadolsky:2008zw}. 
This choice is varied when evaluating model uncertainties. 
The further constraint 
$A_{\bar{U}}=A_{\bar{D}} (1-f_s)$, together with the requirement  
$B_{\bar{U}}=B_{\bar{D}}$,  ensures that 
$x\bar{u} \rightarrow x\bar{d}$ as $x \rightarrow 0$.
For the HERAPDF1.0 and 1.5 NLO central fits, the valence $B$ parameters, 
$B_{u_v}$ and $B_{d_v}$  are also set equal, 
but this assumption is dropped for fits using the extended paramterisation.
%and when parametrisation variations are considered.
The form of the gluon parametrisation is also extended for these latter
fits such that a term of the form $A_g'x^{B_g'}(1-x)^{C_g'}$ is subtracted 
from the standard parametrisation, where $C_g' = 25$ is 
fixed and $A_g'$ and $B_g'$ are fitted. This allows for the gluon 
distribution to become negative at low $x,Q^2$,
 although it does not do so within the kinematic range of the fitted data.
The central %%@
fit is found by first setting the $\epsilon$, $D$ and $E$ parameters to %%@
zero and then varying them, one at a time, the best fit is achieved for 
$E_{u_v}\not= 0$. This is then adopted as standard and the other $\epsilon$, $D$ and $E$ parameters are then varied, one at a %%@
time, However these fits do not 
represent a significant improvement in fit quality for the HERAPDF1.0, 1.5 and 
1.7 NLO fits, and thus
a central fit with just $E_{u_v}\not= 0$ is chosen.
For the HERAPDF1.5f, 1.6 and HERAPDF1.5NNLO fit an extra parameter, $D_{u_v}\not= 0$ is used.
The HERAPDF1.0 and 1.5 NLO fits have $10$ parameters, 
and the 1.5f, 1.6 NLO and 1.5NNLO fits have $14$ parameters and the 
HERAPDF1.7NLO fit has $13$ parameters.

The assumptions made in setting the parameters for this central fit are now %%@
discussed:
\begin{itemize}
\item In common with most PDF fits it is assumed that $q_{sea} = \bar{q}$. 
\item
 The HERAPDF parametrizes $\bar U$ and $\bar D$ separately to allow for the %%@
fact that $\bar u \ne \bar d$ at high $x$, but the restriction 
$x\bar u \to x\bar d$ as $x \to 0$ is imposed. 
\item
The strange sea is suppressed. However determinations of the degree of 
suppression are not very accurate and hence model uncertainty on this 
fraction is evaluated by allowing the variation, $0.23 < f_s < 0.38$.
\item
The $u$-valence and $d$-valence shapes are parametrized separately, but the %%@
form of the parametrization imposes $d_v/u_v = (1-x)^p$ as $x \to 1$.
\item The heavy quarks are treated using a General-Mass-Variable-Flavour %%@
Number-Scheme. There is some model uncertainty in the choice of the heavy %%@
quark masses. The ranges  $1.35 < m_c < 1.65~$GeV and $4.3 < m_b < 5.0~$GeV
are considered as model variations. There are also different heavy 
quark schemes. The ACOT %%@
scheme~\cite{acot} has been used as a cross-check to the Thorne-Roberts scheme.
\item All PDF extractions make choices concerning the fitted kinematic %%@
region, i.e the minimum values of $Q^2$, $W^2$, $x$. 
These choices can have small systematic effects on the PDF shapes %%@
extracted. 
The choice of $Q^2_{min}$ is varied in the range $2.5 < Q^2_{min} < 5.0$.
\item The PDFs
extracted for $Q^2 \gg Q^2_0$ lose sensitivity to the exact form of the
parametrisation at $Q^2_0$. However the choices of $Q^2_0$ and of the form %%@
of parametrisation represent a parametrisation uncertainty. The HERAPDF %%@
uses the technique of saturation of the $\chi^2$, increasing the number of %%@
parameters systematically until the $\chi^2/ndf$ no longer decreases %%@
significantly. However, a number of variations on the central fit %%@
parametrisation, which have similar fit quality, are considered in order %%@
to give an estimate of parametrization uncertainty. The value %%@
of $Q^2_0$ is also varied in the range $1.5 < Q^2_0 < 2.5~$GeV$^2$ 
for the same %%@
purpose.
\end{itemize}

Table~\ref{tab:model} summarizes the variations in numerical values %%@
considered when evaluating model uncertainties on the HERAPDF. 
\begin{table}[tbp]
\centerline{
\begin{tabular}{|l|l|l|r|}
%\vspace{-1.0cm}
\hline
  Variation& Standard Value & Lower Limit & Upper Limit  \\
\hline
$f_s$ & $0.31$  & $0.23$ & $0.38$   \\
$m_c$ [GeV] & $1.4$  & $1.35\, (Q^2_0=1.8)$ & $1.65$   \\
$m_b$ [GeV] & $4.75$  & $4.3$ & $5.0$  \\
$Q^2_{min}$ [GeV$^2$] & $3.5$  & $2.5$ & $5.0$   \\
$Q^2_0$ [GeV$^2$] & $1.9$  & $1.5\,(f_s=0.29)$ & $2.5\,(m_c=1.6,f_s=0.34)$  %%@
\\
\hline
\end{tabular}}
\caption{Standard values of input parameters  and the variations 
considered.  
}
\label{tab:model}
\end{table}
Note that the variations of $Q^2_0$ and $f_s$ are not independent, since %%@
QCD evolution 
will ensure that the strangeness fraction increases as $Q^2_0$ increases. 
The value $f_s=0.29$ is used for $Q^2_0=1.5~$GeV$^2$ and the value %%@
$f_s=0.34$ 
is used for $Q^2_0=2.5~$GeV$^2$ in order 
to be consistent with the choice $f_s=0.31$ at $Q^2_0=1.9~$GeV$^2$.
The variations of $Q^2_0$ and $m_c$ are also not independent, 
since $Q_0 < m_c$ is required in the fit programme. Thus when $m_c = %%@
1.35~$\,GeV, 
the starting scale used is 
$Q^2_0=1.8~$\,GeV$^2$. Similarly, when $Q^2_0 = 2.5~$GeV$^2$ the 
charm mass used is $m_c=1.6~$GeV. In practice, the variations of $f_s$, %%@
$m_c$,
$m_b$, mostly affect the model uncertainty of the $x\bar{s}$, $x\bar{c}$, 
$x\bar{b}$, quark distributions, respectively, and have little effect on %%@
other parton flavours. 
The difference between the central fit and the fits corresponding to model %%@
variations of $m_c$, $m_b$, $f_s$, $Q^2_{min}$ are 
added in quadrature, separately for positive and negative deviations, to
represent the model uncertainty of the HERAPDF sets. 

The variation in $Q^2_0$ is regarded as a parametrisation uncertainty, 
rather than a model uncertainty. The variations of $Q^2_0$ mostly increase 
the PDF uncertainties of the sea and gluon  at small $x$.
At the starting scale the gluon shape is valence-like, so
for the downward variation of the starting scale, $Q^2_0 = 1.5~$GeV$^2$, 
a gluon parametrisation 
which explicitly allows for a negative gluon contribution at low $x$ 
is considered for the 1.0 and 1.5 NLO fits- in all other HERAPDF fits it 
is already a standard part of the parametrisation. 
Similarly a parametrisation variation, $B_{u_v}\not= B_{d_v}$, 
which is standard for the 1.5f, 1.6 and 1.7 NLO and the 1.5NNLO fits, 
is also allowed for the 1.0 and 1.5 NLO fits. This increases the 
uncertainties on the valence quarks at low $x$. Finally, variation of 
the number of terms in
the polynomial $(1 + \epsilon\surd x + D x + E x^2)$ is considered for each 
fitted parton distribution. In practice only a small number of these 
variations have significantly different PDF shapes from the central fit, 
notably: $D_{u_v}\not= 0$ (standard for 1.5f, 1.6 NLO and 1.5NNLO), $D_{\bar{U}}\not= 0$
 and $D_{\bar{D}}\not=0$. 
These variations mostly increase the PDF uncertainty at high $x$, 
but the valence PDFs at low $x$ are also affected 
because of the constraints of the quark number sum rules. 
The difference between all these parametrisation 
variations and the central fit is stored 
and an envelope representing the maximal deviation at each $x$ value
is constructed  
to represent the parametrisation uncertainty.

The HERAPDF uses a form of the $\chi^2$ specified in ref~\cite{h1zeuscomb} 
to perform the fit of the predictions to the HERA data. 
The consistency of the input data  
justifies the use of the conventional 
$\chi^2$ tolerance, $\Delta\chi^2=1$, when determining the $68\%$C.L. %%@
experimental uncertainties on the HERAPDF1.0 fit.
Modern deep inelastic scattering experiments
have very small statistical uncertainties, so that the contribution of %%@
correlated
systematic uncertainties has become dominant for individual data sets and  %%@
consideration of the treatment of such errors is essential.
However, the HERA data combination has changed this situation. The %%@
combination of the H1 and ZEUS data sets has resulted in a data set for NC %%@
and CC $e^+p$ and $e^-p$ scattering with correlated systematic 
uncertainties which are smaller or comparable to the statistical and %%@
uncorrelated uncertainties. 
Thus the central values and experimental uncertainties on the
PDFs which are extracted from the combined data are not much dependent 
on the method of treatment of correlated systematic uncertainties in the 
fitting procedure. For the HERAPDF1.0(1.5) NLO central fit, 
the $110(131)$ systematic uncertainties 
which result from the ZEUS and H1 data sets are combined 
in quadrature, and the three sources of uncertainty which result from 
the combination procedure are treated as correlated 
by the Offset method~\cite{offhesse}.
The resulting experimental uncertainties on the PDFs are small. 
For the HERAPDF1.5f, 1.6, 1.7 NLO fits and the HERAPDF1.5 NNLO fit it 
was decided
to treat the three procedural errors as correlated by the Hessian 
method~\cite{offhesse}. 
This has a negligible effect on the size of the experimental uncertainties 
and a small effect on the resulting $\chi^2$ value, 
see Sec.~\ref{sec:herapdf1.5}. 
The total PDF uncertainty is obtained by adding in quadrature experimental, %%@
model and parameterisation uncertainties.

\subsection{Results from the HERAPDF fit}
\label{sec:results}
\subsubsection{HERAPDF1.0}
 
\begin{figure}[tbp]
\vspace{-0.5cm} 
%\vspace*{5pt}
\centerline{
\epsfig{figure=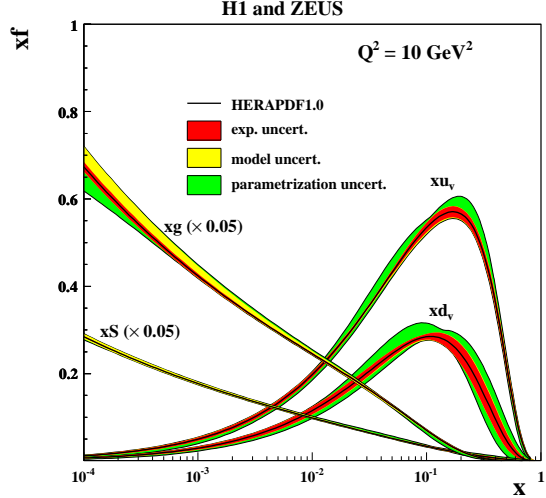 ,width=0.5\textwidth}}
\vspace*{-0.6cm}
%\centerline{
%\epsfig{figure=herapdf1.0_summary_10_logy.eps ,width=0.5\textwidth}}
\caption { 
The parton distribution functions from 
HERAPDF1.0, $xu_v,xd_v,xS=2x(\bar{U}+\bar{D}),xg$, at
$Q^2 = 10$~GeV$^2$.
The experimental, model and parametrisation 
uncertainties are shown separately. The gluon and sea %%@
distributions are scaled down by a factor $20$. 
%For the bottom plot, 
%a logarithmic scale is used to emphasize regions where PDFs are small.
}
\label{fig:summary}
\end{figure}

We first discuss results from the published HERAPDF1.0 fit. This fit has a 
$\chi^2$ per degree of freedom of $574/582$.
Fig~\ref{fig:summary} shows summary plots of the HERAPDF1.0 PDFs at %%@
$Q^2=10~$GeV$^2$. 
%A logarithmic scale is used in the bottom plot to %%@
%emphasize the regions where PDFs are small.

\begin{figure}[tbp]
\vspace{-0.3cm} 
%\vspace*{5pt}
\centerline{
\epsfig{figure=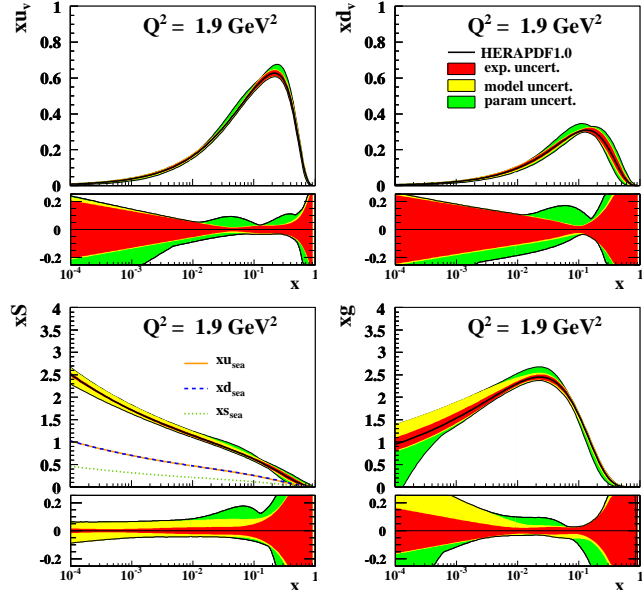 ,width=0.4\textheight}}
\caption {
The parton distribution functions from 
HERAPDF1.0, $xu_v,xd_v,xS=2x(\bar{U}+\bar{D}),xg$, at 
$Q^2=1.9~$GeV$^2$. The break-up of the Sea PDF, $xS$, into the flavours,
$xu_{sea}=2x\bar{u}$, $xd_{sea}=2x\bar{d}$, $xs_{sea}=2x\bar{s}$, 
$xc_{sea}=2x\bar{c}$ is illustrated. Fractional uncertainty bands are shown 
below each PDF. 
The experimental, model and parametrisation 
uncertainties are shown 
separately. 
\label{fig:pdfs19}}
\end{figure}

\begin{figure}[tbp]
\vspace{-0.3cm} 
%\vspace*{5pt}
\centerline{
\epsfig{figure=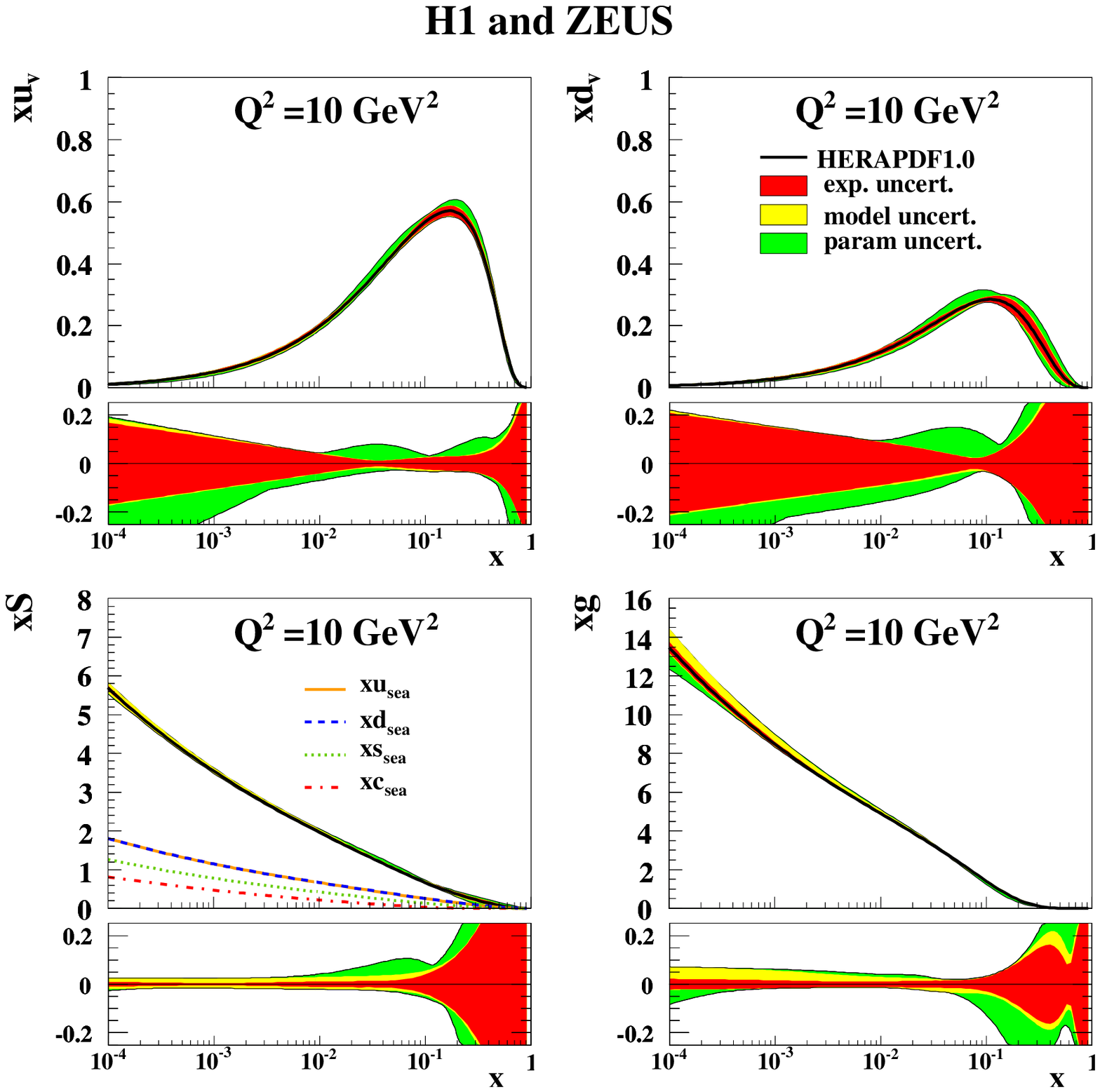 ,width=0.4\textheight}}
\caption {
The parton distribution functions from 
HERAPDF1.0, $xu_v,xd_v,xS=2x(\bar{U}+\bar{D}),xg$, at 
$Q^2=10~$GeV$^2$. The break-up of the Sea PDF, $xS$, into the flavours,
$xu_{sea}=2x\bar{u}$, $xd_{sea}=2x\bar{d}$, $xs_{sea}=2x\bar{s}$, 
$xc_{sea}=2x\bar{c}$ is illustrated. Fractional uncertainty bands are shown 
below each PDF. 
The experimental, model and parametrisation 
uncertainties are shown 
separately. 
\label{fig:pdfs10}}
\end{figure}

\begin{figure}[tbp]
\vspace{-0.3cm} 
%\vspace*{5pt}
\centerline{
\epsfig{figure=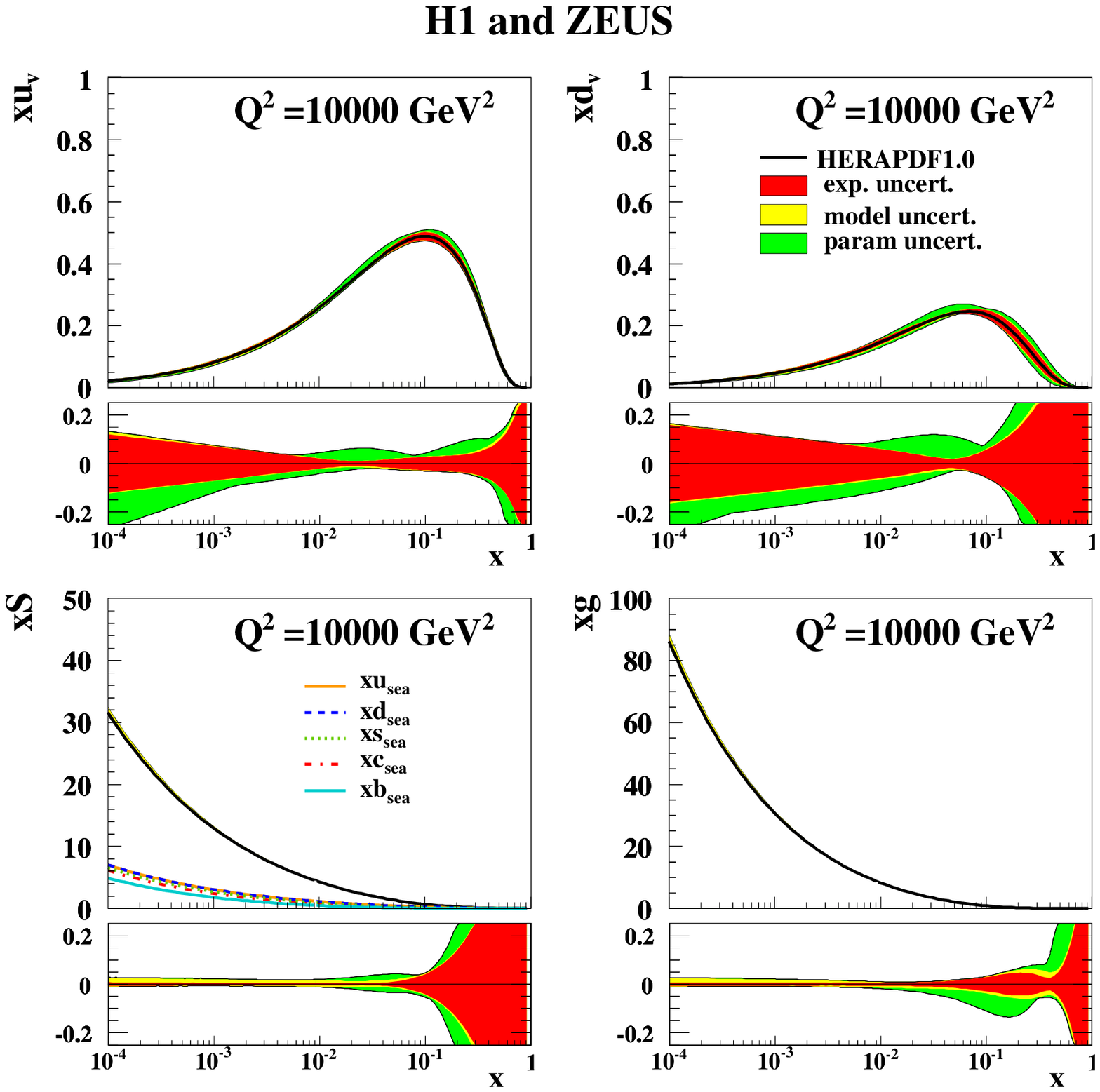 ,width=0.4\textheight}}
\caption {
The parton distribution functions from 
HERAPDF1.0, $xu_v,xd_v,xS=2x(\bar{U}+\bar{D}),xg$, at 
$Q^2=10,000~$GeV$^2$. The break-up of the Sea PDF, $xS$, into the flavours,
$xu_{sea}=2x\bar{u}$, $xd_{sea}=2x\bar{d}$, $xs_{sea}=2x\bar{s}$, 
$xc_{sea}=2x\bar{c}$, $xb_{sea}=2x\bar{b}$ is illustrated. 
Fractional uncertainty bands are shown 
below each PDF. The experimental, model and parametrisation 
uncertainties are shown 
separately.}
\label{fig:pdfs10000}
\end{figure}

Figs~\ref{fig:pdfs19}-\ref{fig:pdfs10000} show the HERAPDF1.0 %%@
distributions, 
 $xu_v,xd_v,xS,xg$, as a function of $x$ at 
 $Q^2=10, 10000$~GeV$^2$,
where $xS =2x (\bar{U} + \bar{D})$ is the sea PDF. Note that for $Q^2 > %%@
m_c^2$,  
$x\bar{U} = x\bar{u} + x\bar{c}$, and for $Q^2 > m_b^2$, 
$x\bar{D} = x\bar{d} +x\bar{s} +x\bar{b}$, so that the heavy quarks are 
included in the sea distributions. The break-up of $xS$ into the flavours
$xu_{sea}=2x\bar{u}$, $xd_{sea}=2x\bar{d}$, $xs_{sea}=2x\bar{s}$, 
$xc_{sea}=2x\bar{c}$, $xb_{sea}=2x\bar{b}$ is illustrated so that the %%@
relative 
importance of each flavour at different $Q^2$ may be assessed. 
Fractional uncertainty bands are shown below each PDF. The experimental, 
model and parametrisation uncertainties are shown separately. 
The model and parametrisation uncertainties are asymmetric. For the sea 
and gluon distributions, the variations in parametrisation which have 
non-zero $\epsilon$, $D$ and $E$ affect the large-$x$ 
region, and the uncertainties arising from the 
variation of $Q^2_0$ and $Q^2_{min}$ affect the small-$x$ 
region. For the valence distributions the non-zero $\epsilon$, $D$ and $E$ 
parametrisation uncertainty is important 
for all $x$, and is their dominant uncertainty. 
The total uncertainties at low $x$ decrease with increasing $Q^2$ due to %%@
QCD evolution resulting, for instance,
in $2\%$ uncertainties for $xg$ at $Q^2=10000$~GeV$^2$ for $x<0.01$.

\begin{figure}[tbp]
\vspace{-0.3cm} 
%\vspace*{5pt}
\centerline{
\epsfig{figure=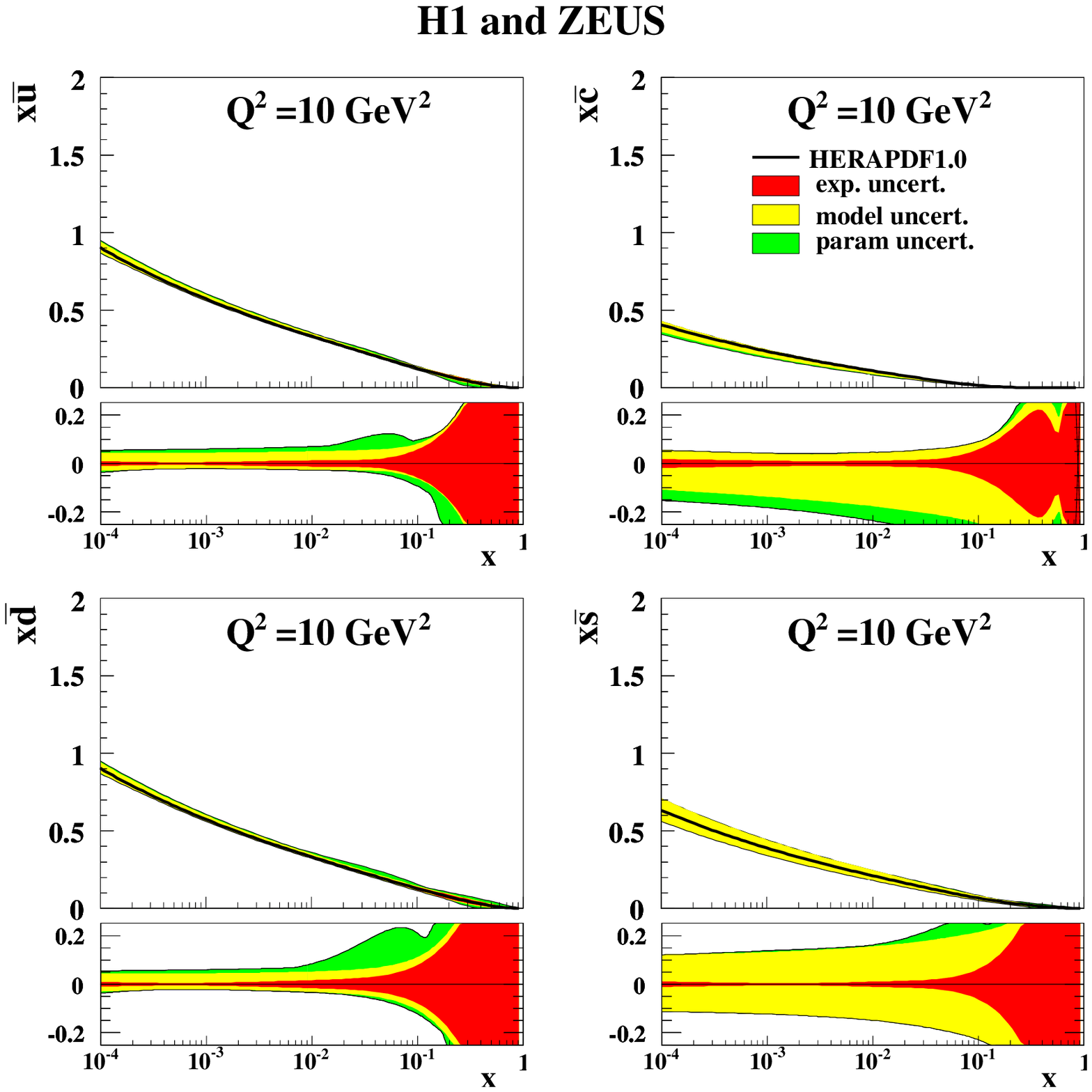 ,width=0.4\textheight}}
\vspace{0.5cm}
\caption {
The parton distribution functions from 
HERAPDF1.0, $x\bar{u}, x\bar{d}, x\bar{c}, x\bar{s}$ at 
$Q^2=10~$GeV$^2$.
Fractional uncertainty bands are shown 
below each PDF. The experimental, model and parametrisation 
uncertainties are shown 
separately.}
\label{fig:pdfsudcs}
\end{figure}

The break-up of the PDFs into different flavours is further illustrated in 
Fig~\ref{fig:pdfsudcs}, where the quark distributions  
$x\bar{u}, x\bar{d}, x\bar{c}, x\bar{s}$ are shown at $Q^2=10$~GeV$^2$. 
The $u$ flavour is better constrained than the $d$ flavour because of 
the dominance of this flavour in all interactions except $e^+p$ CC %%@
scattering.
The quark distribution $x\bar{s}$ is derived from $x\bar{D}$ 
through the assumption on the value of $f_s$, and the uncertainty on %%@
$x\bar{s}$
directly reflects the uncertainty on this fraction. 
The charm PDF, $x\bar{c}$, is strongly related to the 
gluon density such that it is affected by the same variations which 
affect the gluon PDF (variation of $Q^2_0$ and
 $Q^2_{min}$) as well as by the variation of $m_c$. 
  The uncertainty on the
bottom PDF, $x\bar{b}$ (not shown), is dominated by the variation of $m_b$.

The shapes of the gluon and the sea distributions can be compared by 
considering Figs~\ref{fig:pdfs19}-\ref{fig:pdfs10000}.
For $Q^2 \geqsim 10\,$GeV$^2$, the gluon density rises dramatically 
towards low $x$ 
and this rise increases with increasing $Q^2$. This rise is one of the most 
striking discoveries of HERA. 
However, at low $Q^2$ the gluon shape flattens at low $x$. At 
$Q^2 = 1.9\,$GeV$^2$, the gluon shape
becomes valence like and the parametrisation variation which includes a 
negative
gluon term increases the uncertainty on the gluon at low $x$. 
However the gluon distribution itself is not negative in the fitted 
kinematic region.  

 The uncertainty in the sea distribution 
is considerably less that that of the gluon distribution. 
For $Q^2 > 5$~GeV$^2$, the gluon density becomes much 
larger than the sea density, but for lower $Q^2$ the sea density continues %%@
to
rise at low $x$, whereas the gluon density is suppressed. This may be a %%@
signal
that the application of the DGLAP NLO formalism for $Q^2 \leqsim 5$~GeV$^2$
is questionable. Kinematically low $Q^2$ HERA 
data is also at low $x$ and the DGLAP formalism may be indequate at 
low $x$ since it is missing 
$ln(1/x)$ resummation terms and possible non-linear effects
- see Ref.~\cite{lowx}. 
Discussion of this topic is beyond the scope of the 
present review. PDF fits within the DGLAP formalism are successful down to 
$Q^2\sim 2$~GeV$^2$ and $x\sim 10^{-4}$ and this is the kinematic region 
considered in the present review.

\subsubsection{Including Heavy Quark data in PDF fits}
\label{sec:charm}
\begin{figure}[tbp]
\vspace{-0.5cm} 
%\vspace*{5pt}
\centerline{
\epsfig{figure=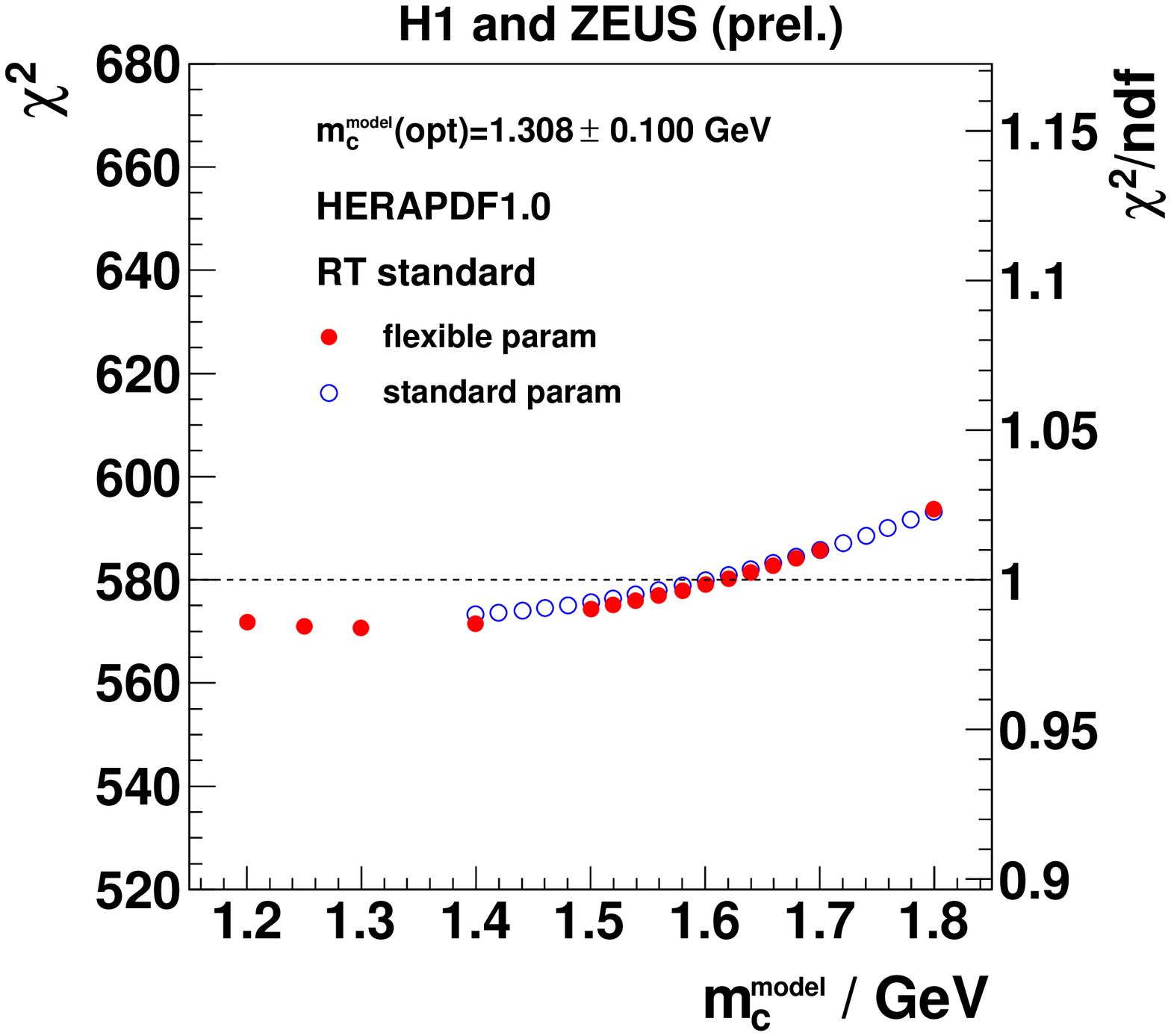 ,width=0.4\textwidth}
\epsfig{figure=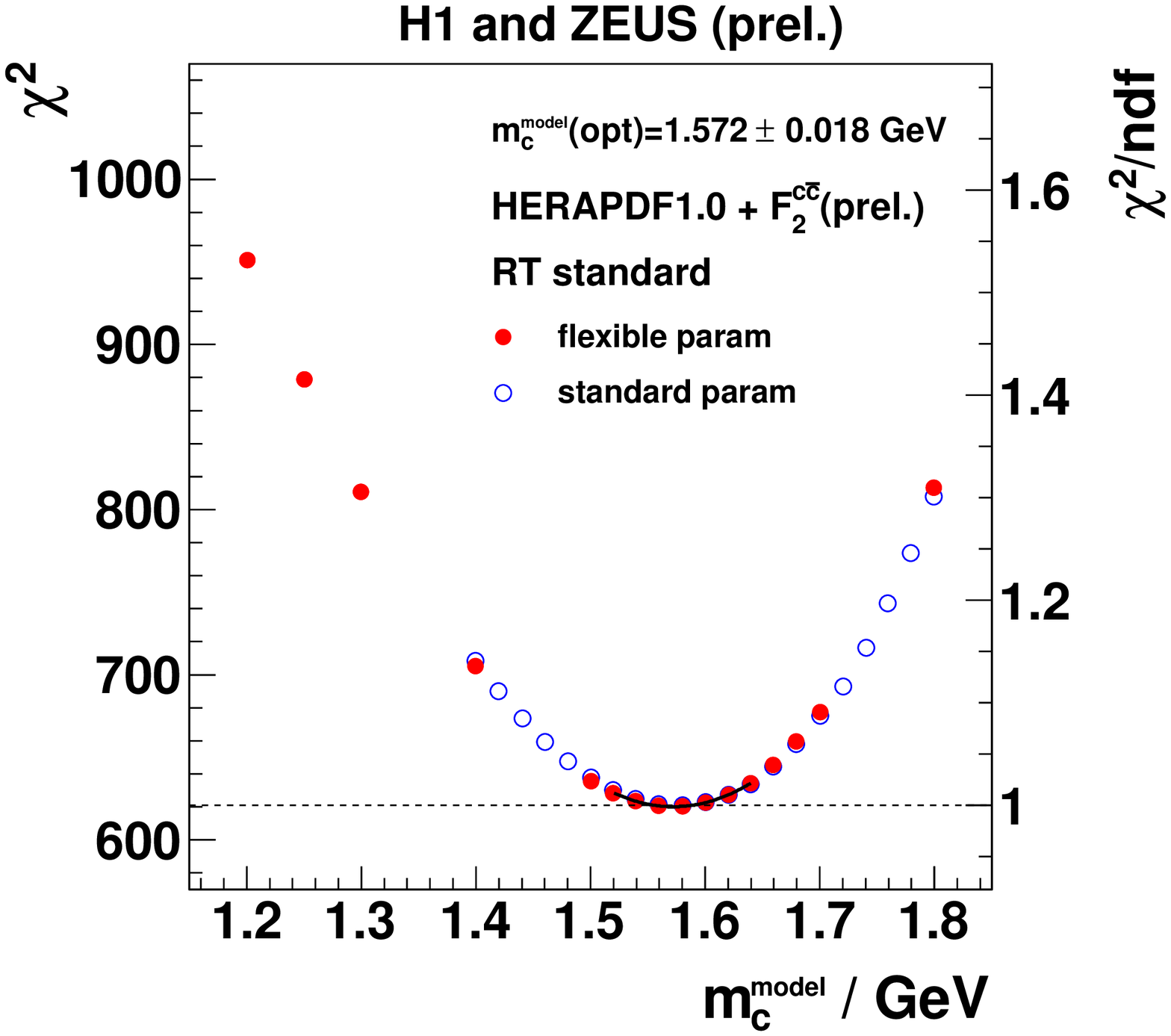 ,width=0.4\textwidth}}
\caption { $\chi^2$ scan vs the charm quark mass $m_c$ for the HERAPDF1.0 fit 
to just HERA-I inclusive data (left) and a fit which also inlcudes combined 
HERA $F^{c\overline{c}}_2$ data (right). Both of these fits use the RT-VFN heavy quark scheme
}
\label{fig:mcchiscan}
\end{figure}

The HERA combined charm data have been presented in Sec.~\ref{sec:charmdata}.
Fig.\ref{fig:f2cfit} shows the comparison of the HERA combined measurements 
of $F^{c\overline{c}}_2$ with the predictions of the HERAPDF1.0 fit.
These data can of course be included in the fit. There are $41$ jet data points
and, for the preliminary combination, these are provided with 
uncorrelated systematic errors and a single combined source of correlated 
error which was treated by the Offset method. The $\chi^2$ for the inclusive 
data is hardly changed by the addition of the charm data but for the $\chi^2$
for the charm data is very sensitive to the charm mass and the scheme used 
for heavy flavour treatment~\cite{hqscan}.
It is found that in order to obtain a good fit using the standard 
Thorne-Roberts variable Flavour Number Scheme (RT-VFN)~\cite{} it
is necessary to increase the standard value of the charm mass. 
Fig~\ref{fig:mcchiscan} shows a scan of the $\chi^2$ of the HERAPDF1.0 fit to 
the inclusive 
HERA-I data vs the charm quark mass parameter entering into the standard RT-VFN
scheme. In the same figure a scan for a similar fit to the inclusive
 HERA-I data plus 
the combined $F^{c\overline{c}}_2$ data is shown. 
The sensitivity of the charm data to 
the charm quark mass parameter is clear.

\begin{figure}[tbp]
\vspace{-0.5cm} 
%\vspace*{5pt}
\centerline{
\epsfig{figure=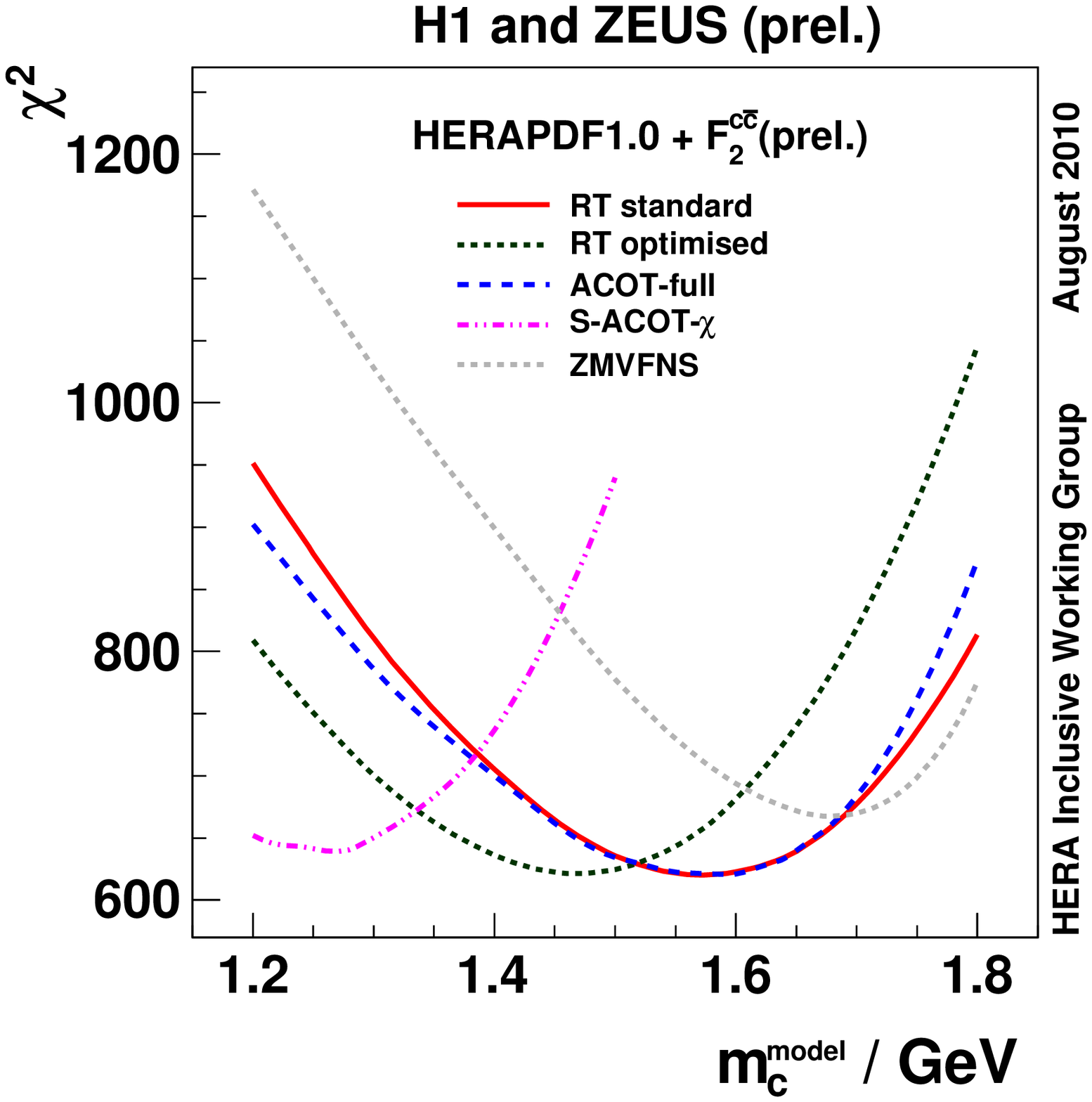 ,width=0.4\textwidth}
\epsfig{figure=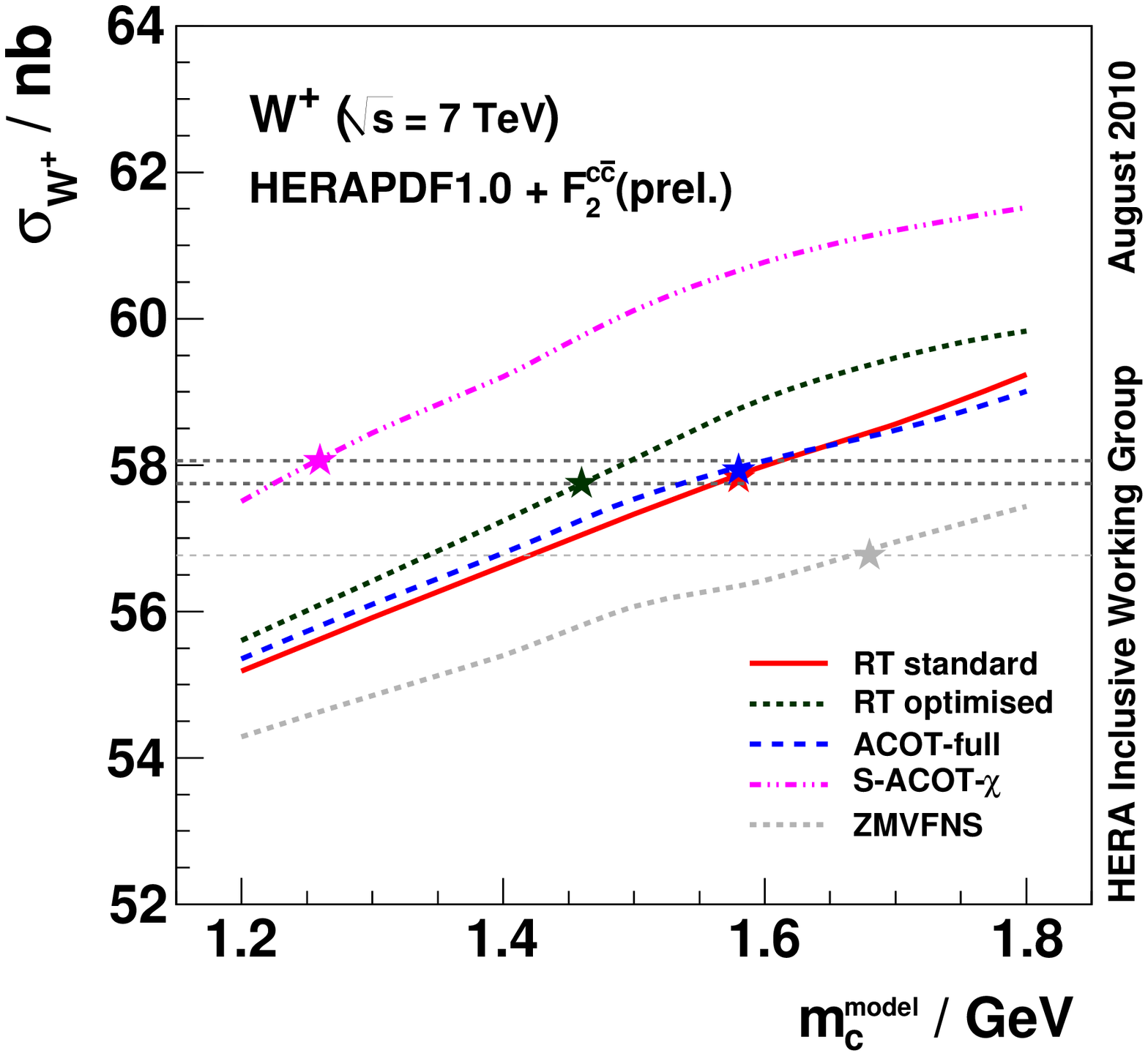 ,width=0.4\textwidth}}
\caption {Left: $\chi^2$ scan vs the charm quark mass $m_c$ for a fit to 
HERA-I inclusive and $F^{c\overline{c}}_2$ data for various heavy quark 
schemes. Right: Predictions for the $W^+$ cross-section at the LHC (7TeV) 
for these schemes vs $m_c$. The value of the charm mass parameter which gives 
the minimum $\chi^2$ is marked by a star for each scheme.
}
\label{fig:allcharm_wp}
\end{figure}

\begin{figure}[tbp]
\vspace{-0.2cm} 
%\vspace*{5pt}
\centerline{
\epsfig{figure=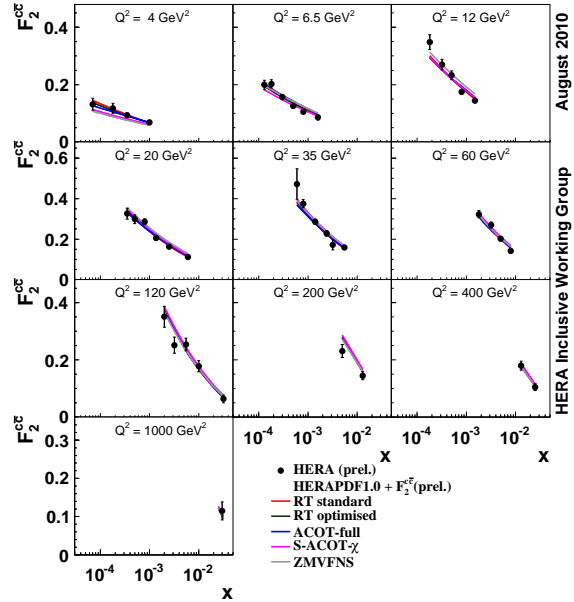 ,width=0.5\textwidth}}
\caption { 
$F^{c\overline{c}}_2$ data compared to the predictions of various 
heavy quark schemes, within the HERAPDF fit formalism.
}
\label{fig:f2cfitall} 

\end{figure}

However the Standard RT-VFN scheme is not the only possible heavy quark 
scheme. 
The fit to HERA-I inclusive plus $F^{c\overline{c}}_2$ data has been repeated 
for the Optimized RT-VFN scheme~\cite{rtopt}, the full ACOT scheme, 
the S-ACOT-$\chi$ scheme~\cite{sacot}
and the Zero-Mass Variable Flavour Number Scheme (ZM-VFN) in which 
light-quark coefficient functions are used for the heavy quarks, 
which are simply turned on at threshold $Q^2\sim m_c^2$. 
Fig.~\ref{fig:allcharm_wp} 
shows the $\chi^2$ scan for these different heavy quark schemes. It can be seen
that all schemes, bar the ZM-VFN, give acceptable fits, and that each scheme 
has its own preferred value of the charm quark mass. These values and the 
correspondin $\chi^2$ values are given in 
Table~\ref{tab:mc}. Fig~\ref{fig:f2cfitall} shows these fits compared to the 
charm data.

\begin{table}[tbp]
\centerline{
\begin{tabular}{|l|r|r|}
%\vspace{-1.0cm}
\hline
 Scheme& $m_c$(minimum)& $\chi^2$/ndp $F^{c\overline{c}}_2$ \\
\hline
 RT Standard & $1.58^{+0.02}_{-0.03}$  &  $42.0/41$\\
 RT Optimized  & $1.46^{+0.02}_{-0.04}$  & $46.5/41$\\
 ACOT-full & $1.58^{+0.03}_{-0.04}$ & $59.9/41$  \\
 S-ACOT $\chi$ & $1.26^{+0.02}_{0.04}$   & $68.5/41$\\
 ZM-VFN  & $1.68^{+0.06}_{-0.07}$ & $88.1/41$ \\
\hline
\end{tabular}
}
\caption{Charm mass parameters and $\chi^2$ values per number of data points 
($ndp$) for fits to 
$F^{c\overline{c}}_2$ data using various heavy quark schemes.
}
\label{tab:mc}
\end{table}

Predictions for $W^+, W^-, Z$ production at the LHC are sensitive to the value 
of the charm mass and to the heavy quark scheme used, as illustrated 
in Fig.~\ref{fig:allcharm_wp}. For any chosen value of the charm mass the 
spread of predictions for different schemes is $\sim 7\%$. However if each 
prediction is used at its own favoured value of the charm mass then this 
spread is reduced to $\sim2\%$ and, if the disfavoured ZM-VFN is excluded 
to$ \leqsim 1\%$.

\subsubsection{Including Low Energy run data in PDF fits}
\label{sec:lowEfit}

\begin{figure}[tbp]
\vspace{-0.5cm} 
%\vspace*{5pt}
\centerline{
\epsfig{figure=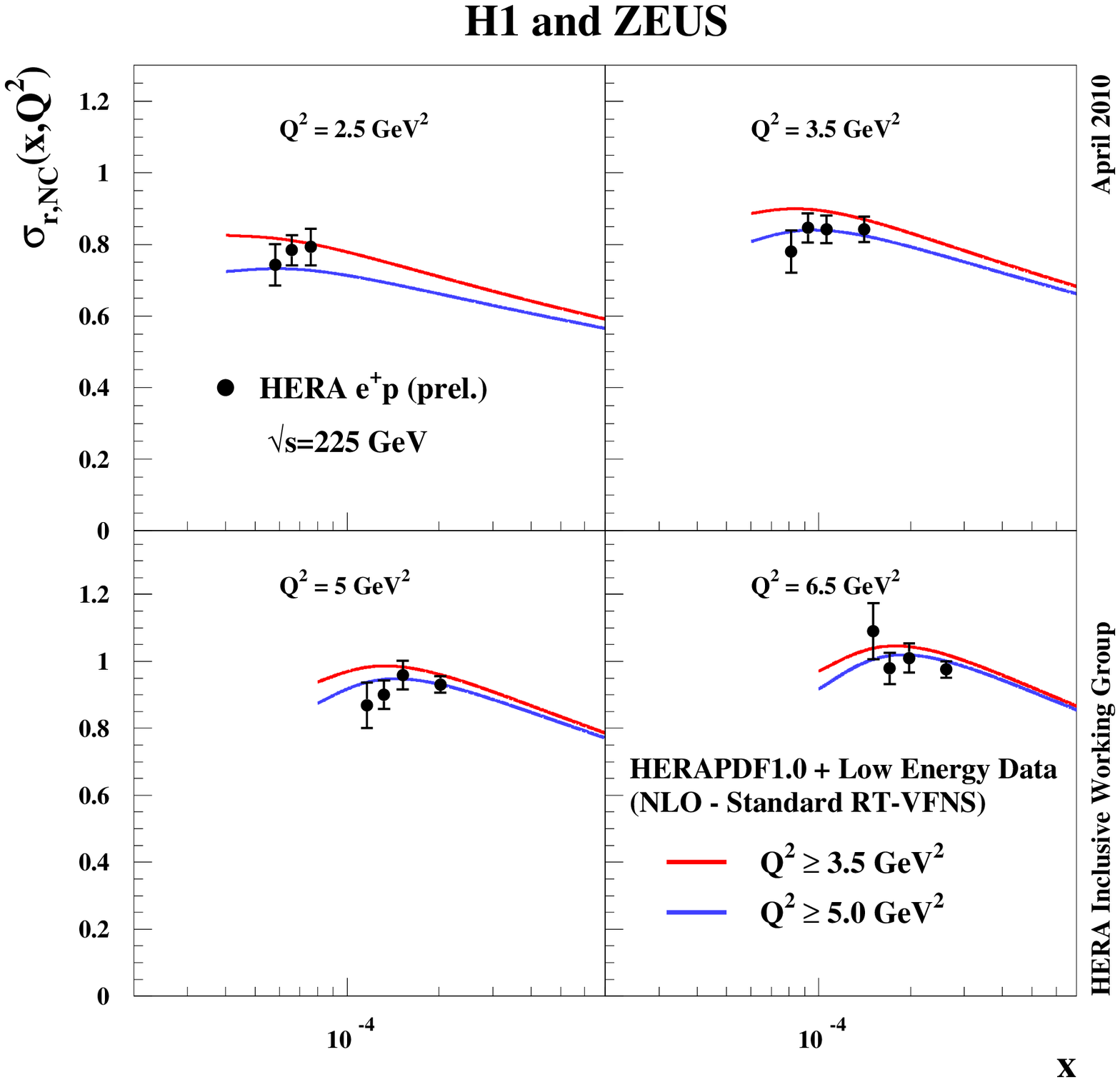 ,width=0.4\textwidth}
\epsfig{figure=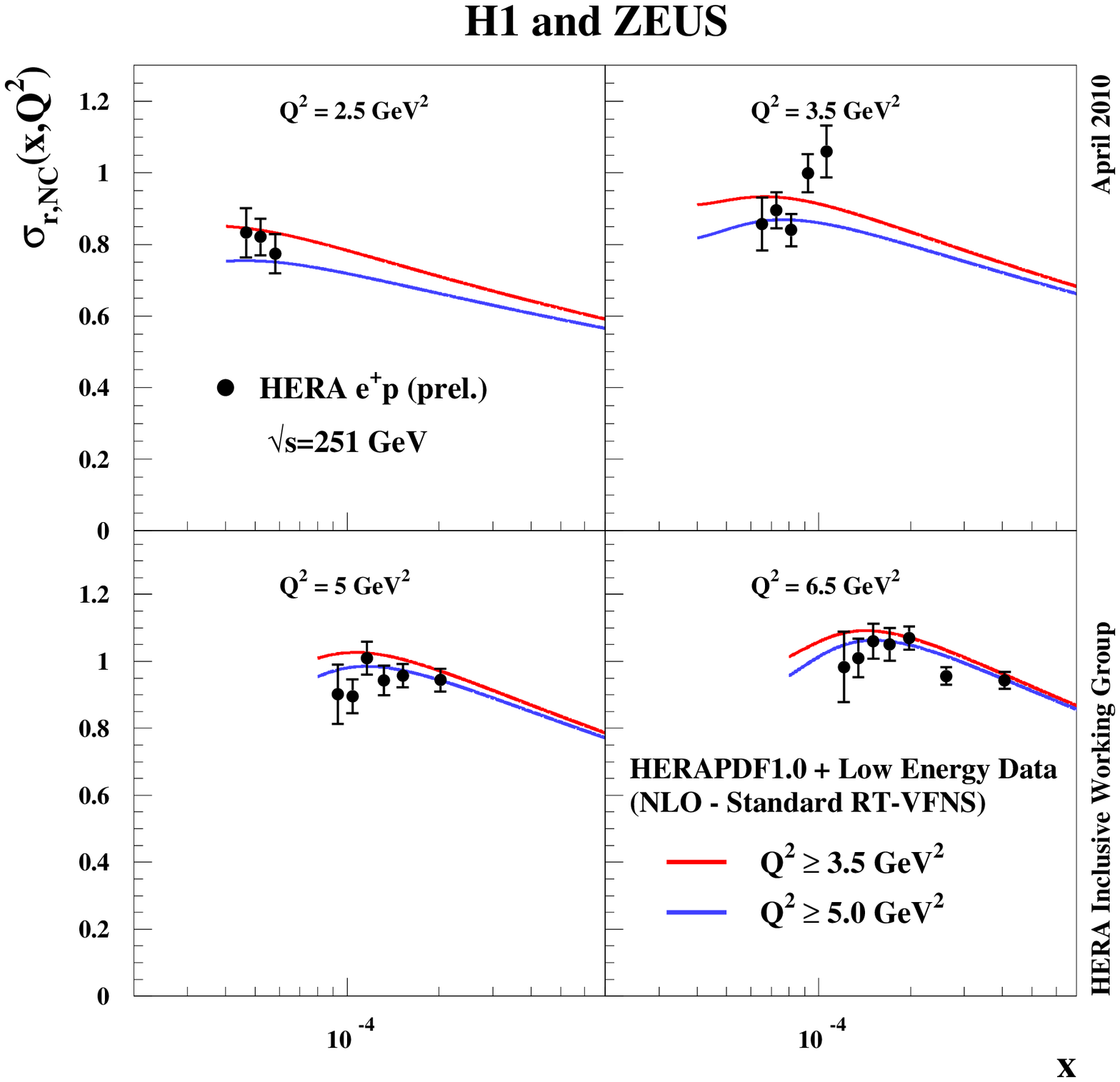 ,width=0.4\textwidth}}
\caption {The combined HERA data from running with proton beam energies 
$E_p=460$~GeV and $E_p=575$~GeV is shown for a few low-$Q^2$ bins, compared to
predictions from PDF fits to these data and the combined HERA-I high energy 
data. Predictions are shown for data subject to two different minimum  
$Q^2$ cuts: $3.5~$GeV$^2$ and $5.0~$GeV$^2$.
}
\label{fig:460575}
\end{figure}

\begin{figure}[tbp]
\vspace{-0.5cm} 
%\vspace*{5pt}
\centerline{
\epsfig{figure=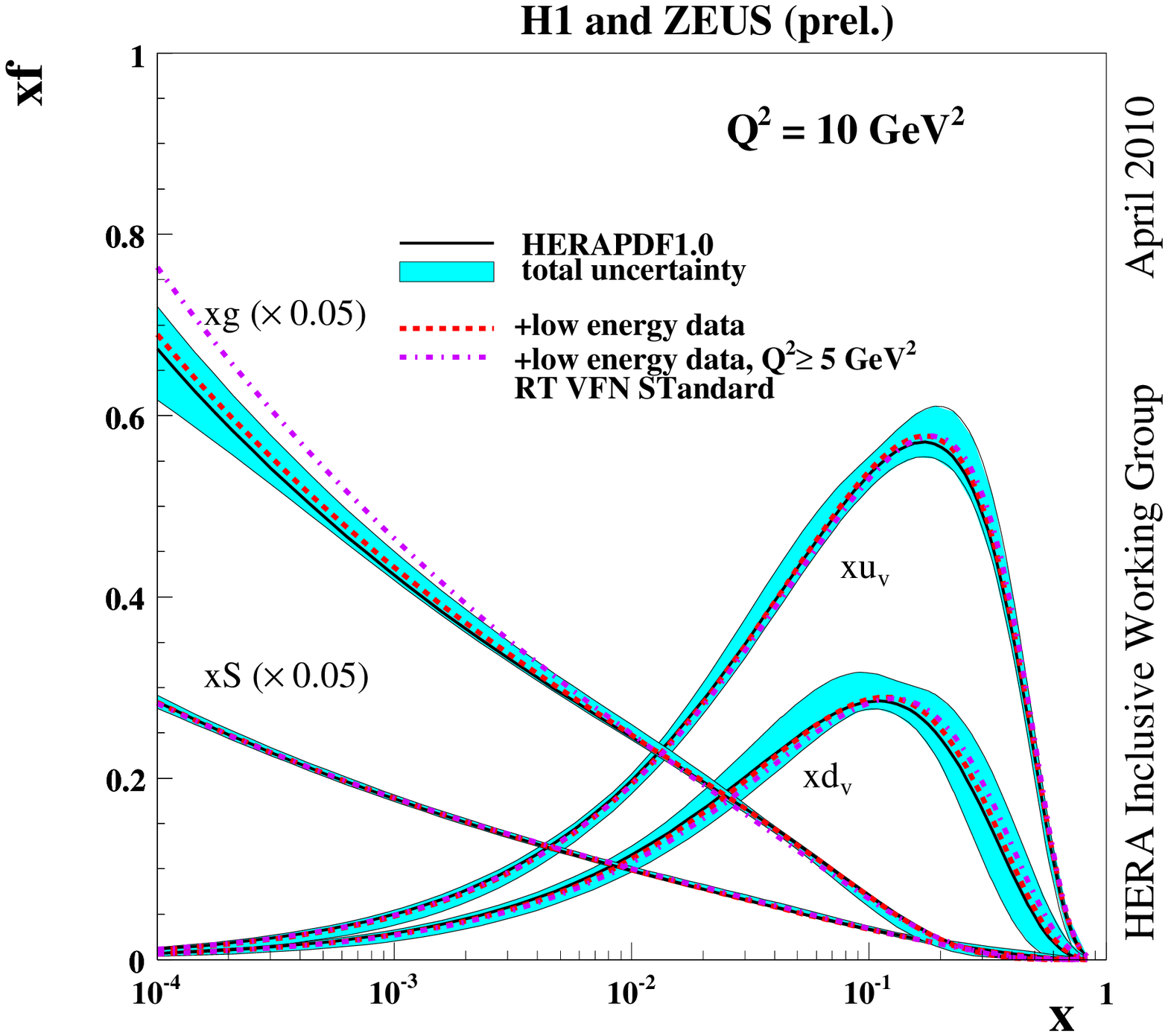 ,width=0.4\textwidth}
\epsfig{figure=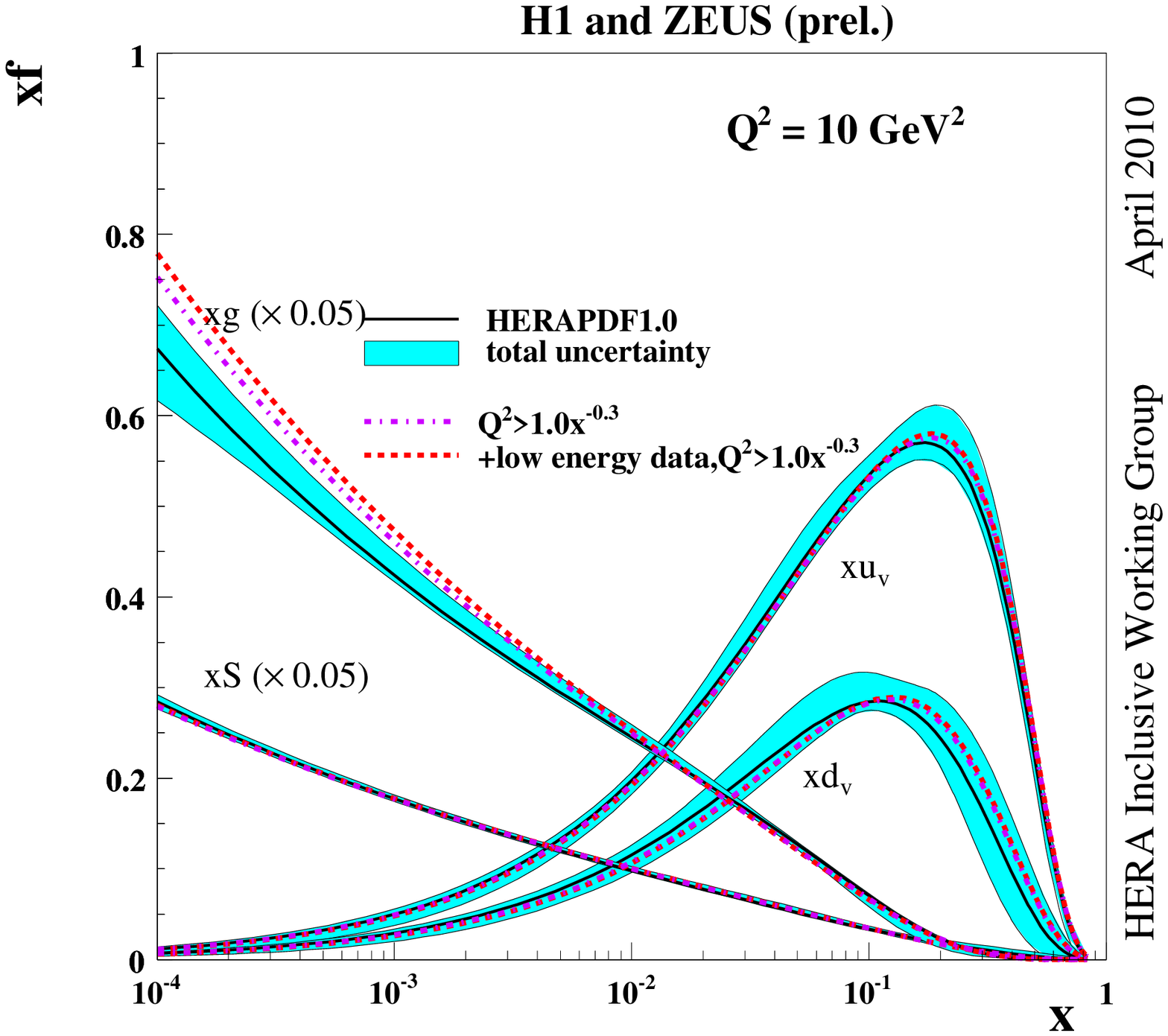,width=0.4\textwidth}}
\caption { The parton distribution functions from 
HERAPDF1.0, $xu_v,xd_v,xS=2x(\bar{U}+\bar{D}),xg$, at
$Q^2 = 10$~GeV$^2$.
The total uncertainties are shown. The gluon and sea %%@
distributions are scaled down by a factor $20$. (Left) The lines overlayed 
show the results of fits to the HERA-1 data plus
the low energy running data with the standard minimum $Q^2$ cut of  
$3.5~$GeV$^2$ and with a harder cut of $5.0~$GeV$^2$. (Right) 
The lines overlayed 
show the results of fits to the HERA-1 data and to the HERA-1 plus
the low energy running data, with the 'saturation inspired' 
cut of $Q^2 > 1.0 x^{-0.3}$GeV$^2$.
}
\label{fig:lowEpdf}
\end{figure}

The preliminary combined data from the low energy running described in 
Sec~\ref{sec:lowE} have been input to the HERAPDF fit together with the 
HERA-I combined high energy data. Thess data have $25$ sources of correlated 
systematic uncertainty from the individula experiments, and $3$ procedural 
sources of systematic uncertainty similarly to the high energy combination. 
These correlated errors are added 
in quadrature except for the $3$ procedural which are treated as fully 
correlated by the Hessian method.
There are $224$ combined data points on the NC $e^+ p$ cross section from the 
low energy proton beam running and 
when they are fit together with the $592$ combined data points from the HERA-I 
running the 
$\chi^2/ndf$ is $845.7/806$ for $10$ parameters. The partial $\chi^2/ndp$ are
$588/592$ for the high energy inclusive data and $257.6/224$ for the 
low energy inclusive
data. These data are sensitive to the minimum $Q^2$
cut imposed, as illustrated in Fig.~\ref{fig:460575}. A better fit is obtained 
with a larger, $Q^2>5$~GeV$^2$, cut. 
The partial $\chi^2/ndp$ after this cut are
$527.1/566$ and $200/215$ for the low energy data.
The data at low $x,Q^2$ access high $y$ and thus 
sensitive to the longitudinal structure function $F_L$. Because of the close
relationship of $F_L$ and the gluon PDF these data should affect the 
gluon PDF. This is illustrated in Fig.~\ref{fig:lowEpdf} which 
shows that variation of the $Q^2$-cut
affects the gluon PDF more for the fit including low energy data, since the 
result is outside the error bands which include this cut-variation for the 
HERAPDF1.0 fit.
Kinematically cutting out low $Q^2$ data also implies cutting out data at the 
lowest $x$ and the data are similarly sensitive to an $x>0.0005$ cut. Data at 
low $x$ may not be well fit by the DGLAP formalism since this is missing 
$ln(1/x)$ resummation terms and possible non-linear effects. 
Fig~\ref{fig:lowEpdf} also illustrates sensitivity to 
a 'saturation' inspired cut of $Q^2 > 1.0x^{-0.3}~$GeV$^2$.
However, one cannot claim that any break-down of the DGLAP formalism has yet 
been observed, since if the HERAPDF1.0 formalism is generalised to 
the extended parametrisation with $14$ parameters,
then the increased uncertainty in the low-$x$ 
gluon, illustrated in Fig.~\ref{fig:herapdf1.5f}, covers the sensitivity of 
the low energy data to the low $x,Q^2$ cuts.

\subsubsection{HERAPDF1.5}
\label{sec:herapdf1.5}

\begin{figure}[tbp]
\vspace{-0.2cm} 
%\vspace*{5pt}
\centerline{
\epsfig{figure=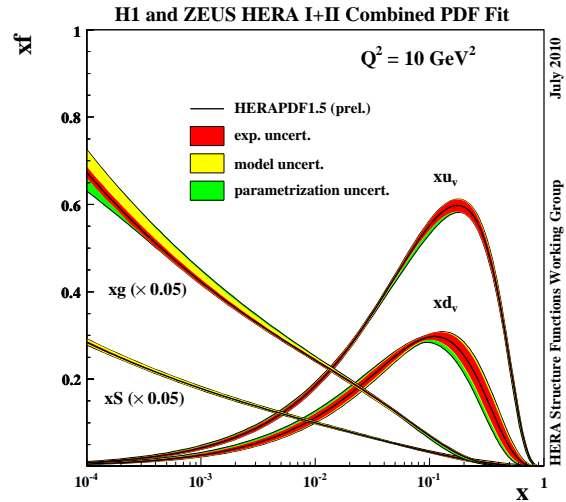 ,width=0.5\textwidth}}
\caption {The parton distribution functions from 
HERAPDF1.5, $xu_v,xd_v,xS=2x(\bar{U}+\bar{D}),xg$, at
$Q^2 = 10$~GeV$^2$. The experimental, model and parametrisation 
uncertainties are shown separately. The gluon and sea %%@
distributions are scaled down by a factor $20$. 
}
\label{fig:herapdf15}
\end{figure}

\begin{figure}[tbp]
\vspace{-0.2cm} 
%\vspace*{5pt}
\centerline{
\epsfig{figure=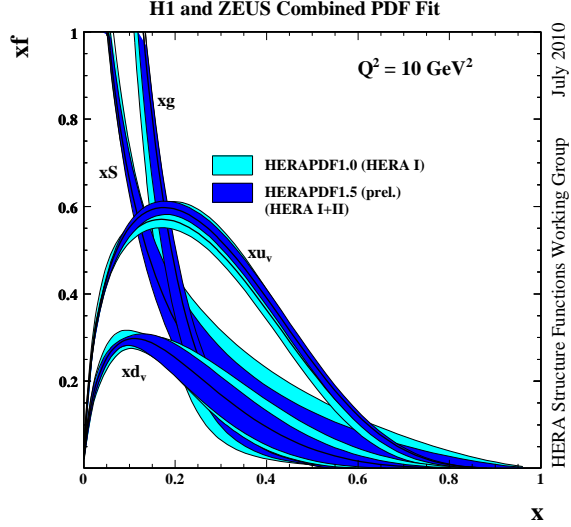,width=0.5\textwidth}}
\caption {The parton distribution functions from 
HERAPDF1.5, $xu_v,xd_v,xS=2x(\bar{U}+\bar{D}),xg$, at
$Q^2 = 10$~GeV$^2$, overlayed on the parton distributions from HERAPDF1.0.
The total uncertainties for each PDF are shown. A linear scale in $x$ is used 
to emphasize the reduction in uncertainties for HERAPDF1.5 at high $x$.
}
\label{fig:herapdf1510}
\end{figure}

The HERAPDF1.5~NLO fit uses the same formalism as HERAPDF1.0 but includes
preliminary HERA-I+II data.
The $\chi^2$ per degree of freedom for the HERAPDF1.5~NLO central fit is 
$760/664$, where the increased $\chi^2$ reflects the greater accuracy of the 
HERA-I+II combination. This fit has already been compared to the data in 
Figs.~\ref{fig:ncepem2}-~\ref{fig:dataCCm2}. 
The improvement to the PDFs is illustrated in Fig.~\ref{fig:herapdf15},
which shows the HERAPDF1.5 in a format 
such that it may be directly compared with 
HERAPDF1.0 in Fig.~\ref{fig:summary}. Fig.~\ref{fig:herapdf1510} shows the 
HERAPDF1.5 overlayed on HERAPDF1.0 on a linear $x$ scale, such that the 
improvement at high $x$ may be clearly seen. 

\begin{figure}[tbp]
\vspace{-0.5cm} 
%\vspace*{5pt}
\centerline{
\epsfig{figure=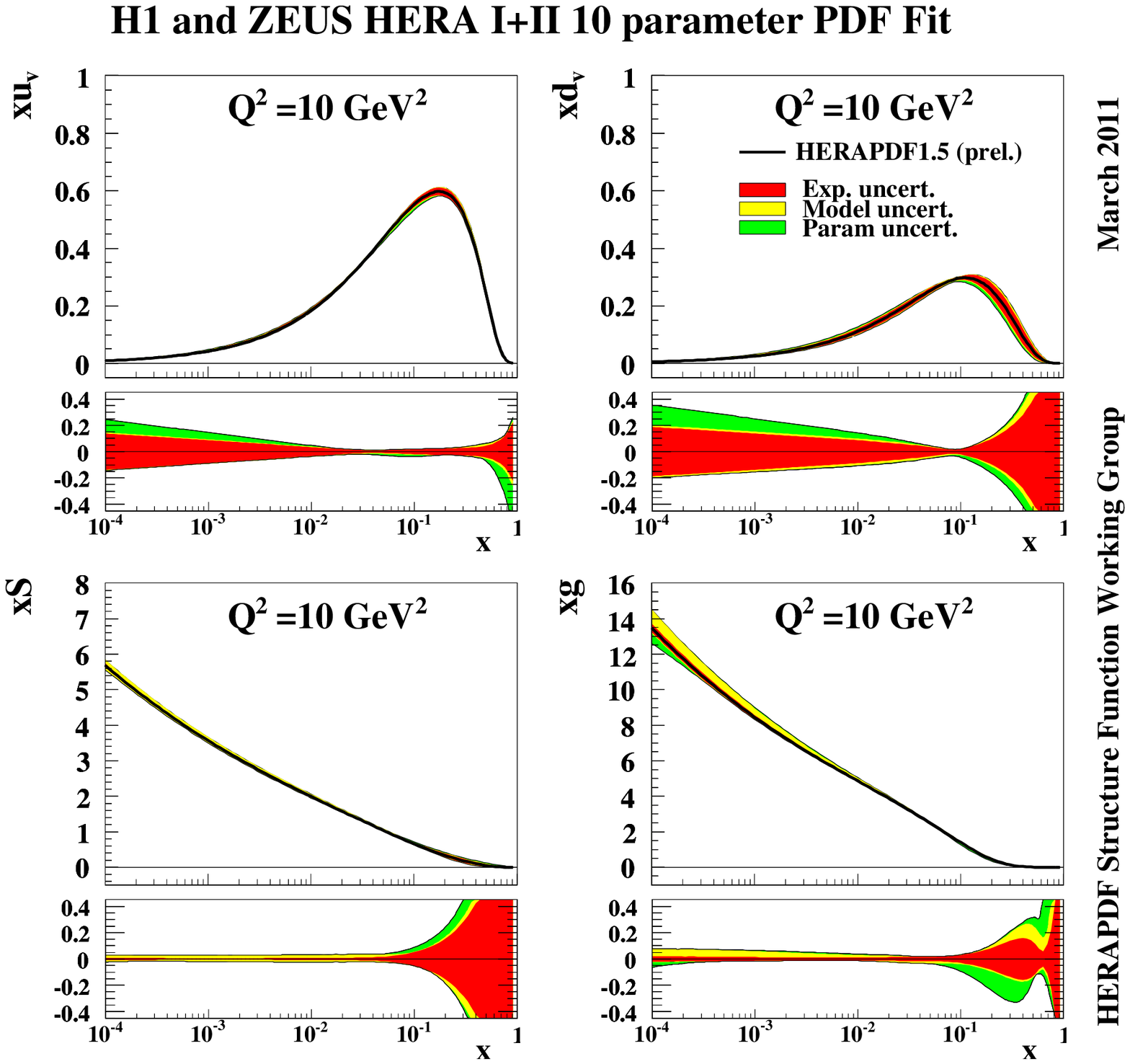 ,width=0.4\textheight}}
\vspace*{-0.2cm}
\centerline{
\epsfig{figure=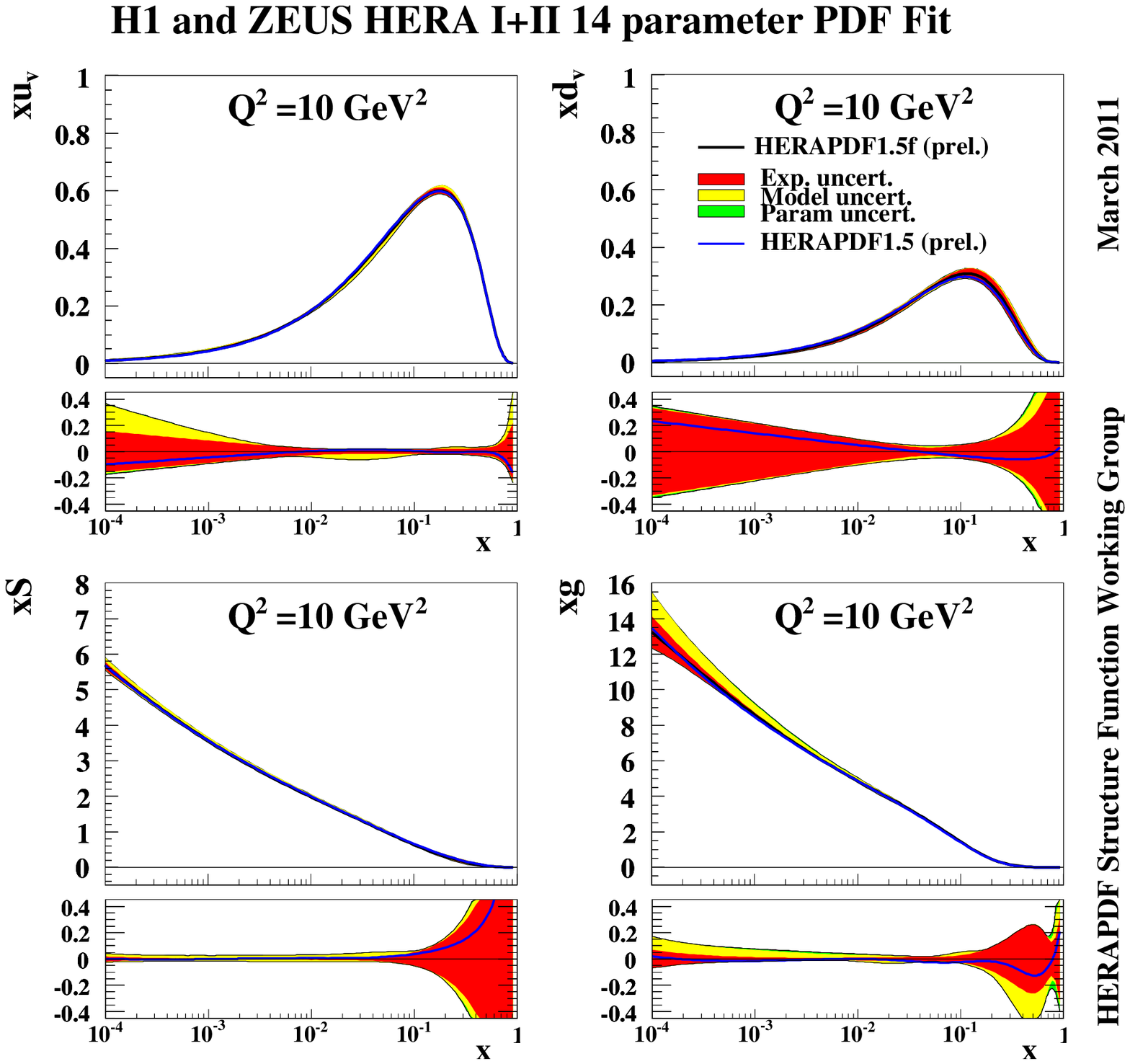 ,width=0.4\textheight}}
\caption { 
The parton distribution functions from 
HERAPDF1.5 and HERAPDF1.5f, $xu_v,xd_v,xS=2x(\bar{U}+\bar{D}),xg$, at
$Q^2 = 10$~GeV$^2$. Fractional uncertainty bands are shown 
below each PDF. 
The experimental, model and parametrisation 
uncertainties are shown separately. 
}
\label{fig:herapdf1.5f}
\end{figure}

The fit formalism was also extended
to included more PDF parameters, 
as already described in Sec.~\ref{sec:pdfchap}.
This fit, called HERAPDF1.5f NLO fit, has $\chi^2$ per degree of freedom 
$730/664$, where the improvement to the $\chi^2$ is mostly due to the 
treatment of the three procedural systematic errors by the 
Hessian rather than Offset method. There is a small decrease in $\chi^2$, 
$\Delta\chi^2=-5$ due to the increase of the number of parameters from $10$ 
to $14$. The PDFs of the HERAPDF1.5f and 1.5 NLO fits are 
compared in Fig.~\ref{fig:herapdf1.5f}, 
where one can see that the extra freedom in the parametrisation does not 
change the central values of the PDFs significantly. The total size of the PDF 
uncertainties are also not changed significantly, although some of the 
parametrisation 
uncertainty in HERAPDF1.5 is now included in the 
experimental uncertainty in HERAPDF1.5f.
The most significant change to the uncertainties is a modest 
increase in the uncertainty of the low-$x$ gluon. This covers the sensitivity 
to low-$x, Q^2$ cuts found in the low energy data combination, 
see Sec.~\ref{sec:lowEfit}.

\begin{figure}[tbp]
\vspace{-0.2cm} 
%\vspace*{5pt}
\centerline{
\epsfig{figure=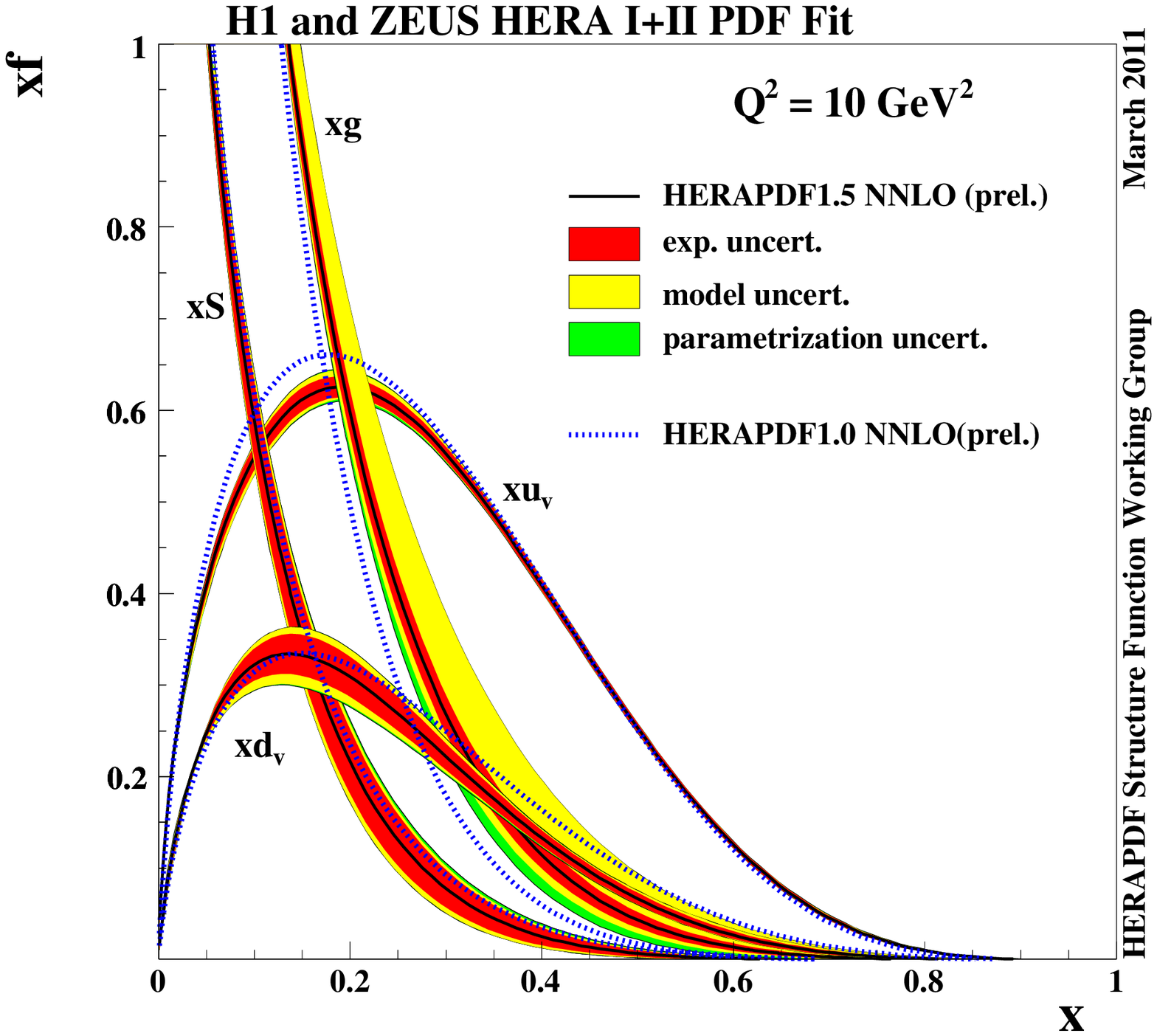,width=0.5\textwidth}}
\caption {The parton distribution functions from 
HERAPDF1.5 NNLO, $xu_v,xd_v,xS=2x(\bar{U}+\bar{D}),xg$, at
$Q^2 = 10$~GeV$^2$, compared to HERAPDF1.0 NNLO.
A linear scale in $x$ is used 
to emphasize the differences at high $x$.
}
\label{fig:herapdf15nnlovs10}
\end{figure}

The HERAPDF1.5 NNLO fit was performed on the same preliminary combined 
HERA I+II data. The $\chi^2$ per degree of freedom for 
for the HERAPDF1.5~NNLO central fit it is $740/664$. For this NNLO 
fit the addition
of extra parameters made a significant difference to the $\chi^2$, 
The change from a 10 to 14 parameter fit, results in a change of 
$\Delta\chi^2=-32$ with the 
largest difference coming from the addition of the term which allows 
freedom in the low $x$ gluon. Fig.~\ref{fig:herapdf15nnlovs10} compares the 
HERAPDF1.5~NNLO fit to HERAPDF1.0~NNLO which was an NNLO version of the 
HERAPDF1.0 using just $10$ parameters and fitting just HERA-I data. One can 
see that the extra parameters give somewhat different shapes to the valence 
quarks and a much harder high-$x$ gluon PDF. 

\begin{figure}[tbp]
\vspace{-0.2cm} 
%\vspace*{5pt}
\centerline{
\epsfig{figure=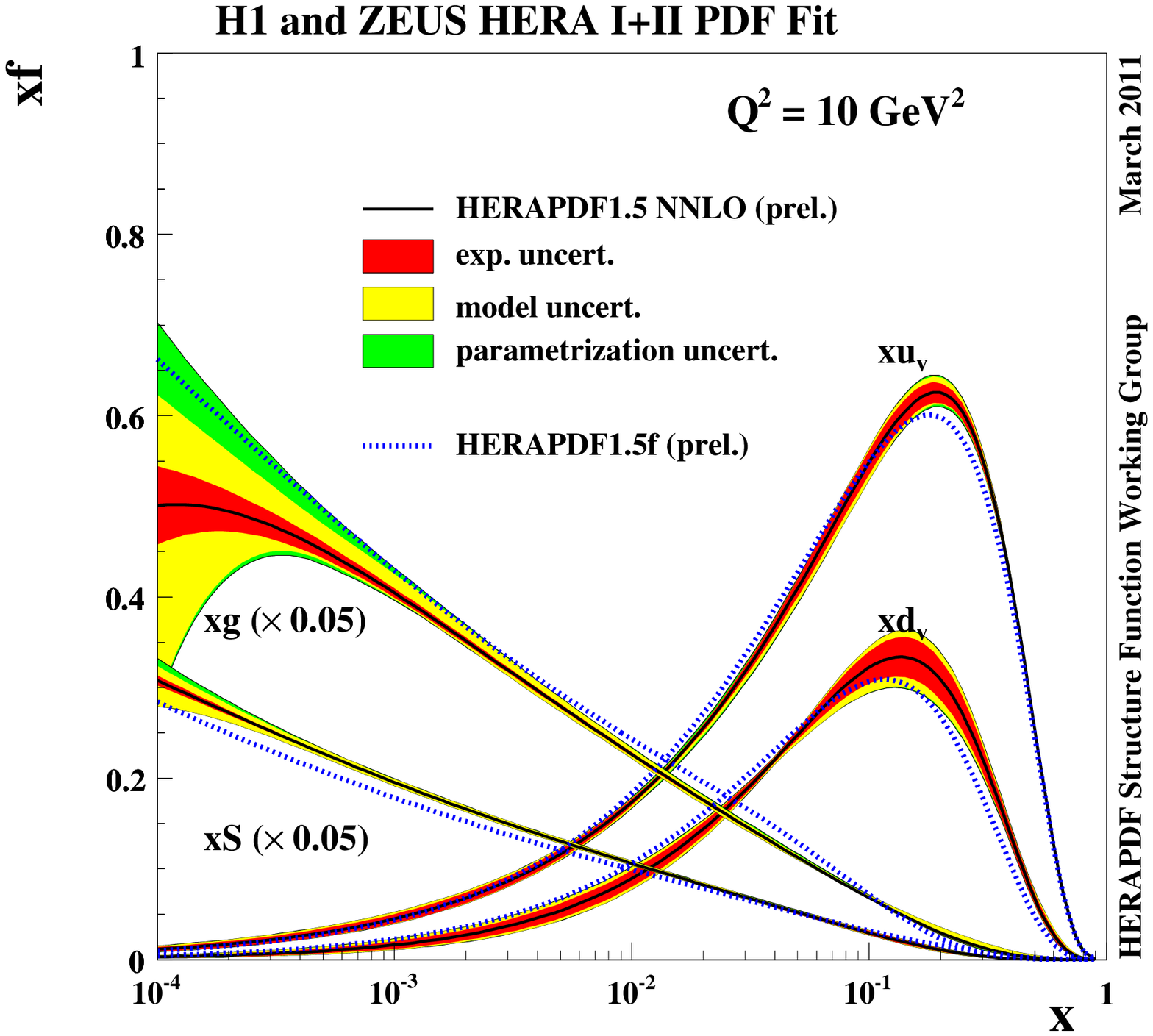,width=0.5\textwidth}}
\caption {The parton distribution functions from 
HERAPDF1.5NNLO, $xu_v,xd_v,xS=2x(\bar{U}+\bar{D}),xg$, at
$Q^2 = 10$~GeV$^2$, compared to the HERAPDF1.5f NLO fit.
 The gluon and sea %%@
distributions are scaled down by a factor $20$. 
}
\label{fig:herapdf15nnlovsnlo}
\end{figure}

Fig.~\ref{fig:herapdf15nnlovsnlo}
compares the HERAPDF1.5NNLO fit to the corresponding NLO fit HERAPDF1.5f. 
These fits have the same number of parameters.
The change from NLO to NNLO gives a somehat steeper sea and softer gluon at 
low $x$ consistent with the different rates of evolution at NNLO. The 
most striking difference is the greater level of uncertainty at low $x$ for 
the NNLO fit. This is mostly due to sensitivity to the low $Q^2$ cut on the 
data. One might have expected that an NNLO fit would fit low $x,Q^2$ data 
better than an NLO fit, however this would seem not to be the case, 
see also ref.~\cite{caola}.

\subsubsection{Including jet data in PDF fits: HERAPDF1.6}
\label{sec:jets}
The gluon PDF contributes only indirectly to the 
inclusive DIS cross sections. However, 
the QCD processes that give rise to scaling violations in the 
inclusive cross sections, namely the QCD-Compton (QCDC) and %%@
boson-gluon-fusion
(BGF) processes, can also be 
observed as events with distinct jets in the final 
state provided that the energy and momentum transfer are large enough.  The 
cross section for QCDC scattering depends on $\alpha_s(M_Z)$ and the quark 
PDFs. The cross section for the
BGF process depends on $\alpha_s(M_Z)$ and the gluon PDF. These two processes
are dominant in different kinematic regions. Thus jet cross 
sections give new information about the PDFs. For the inclusive data, 
the correlation between 
$\alpha_s(M_Z)$ and the gluon PDF limits the accuracy with which either can be 
determined. The jet data bring new information which helps to reduce the 
overall correlation.

In the HERAPDF1.6 NLO PDF fit the jet data sets presented in 
Sec.~\ref{sec:jetdata} are fitted together with the preliminary HERA I+II 
combined inclusive data. These data sets have $12$ correlated systematic 
errors, which are treated as fully correlated by the Hessian method.
The predictions for the jet cross sections have been calculated to NLO in 
QCD using the NLOjet++ program~\cite{nlojet} and have been input to the fit
by the FASTNLO interface~\cite{fastnlo}. The calculation of the NLO jet cross 
sections is too slow to be used iteratively in a fit. 
Thus NLOjet++ is used to compute LO and NLO weights 
which are independent of $\alpha_s$ and the 
PDFs. The FASTNLO program then calculates the NLO QCD cross sections,
 by convoluting these weights with the PDFs and $\alpha_s$. 
The predictions must be multiplied by 
hadronisation corrections before they can be used to fit the data. 
These were determined by using  Monte Carlo (MC) programmes, 
which model parton hadronisation to estimate the ratio of 
the hadron- to parton-level cross sections for each bin.
The hadronisation corrections are generally within a few 
percent of unity. The predictions for jet production were also corrected for 
$Z^0$ contributions.

\begin{figure}[tbp]
\vspace{-0.5cm} 
%\vspace*{5pt}
\centerline{
\epsfig{figure=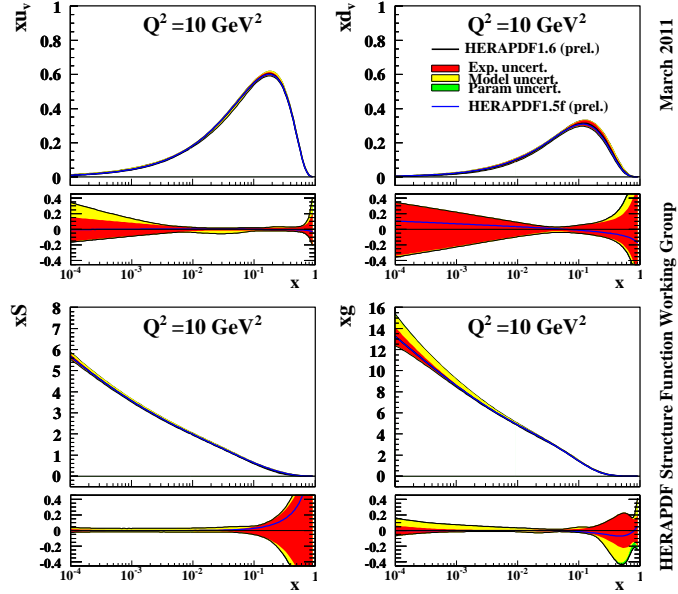 ,width=0.4\textheight}}
\caption { 
The parton distribution functions from 
HERAPDF1.6, $xu_v,xd_v,xS=2x(\bar{U}+\bar{D}),xg$, at
$Q^2 = 10$~GeV$^2$. Fractional uncertainty bands are shown 
below each PDF. 
The experimental, model and parametrisation 
uncertainties are shown separately. the central fit of 
HERAPDF1.5f is shown as a blue line
}
\label{fig:herapdf1.6}
\end{figure}

The fit is done with the same settings as for the HERAPDf1.5f fit. 
The $\chi^2/ndf$ for the fit is $812/766$, for a fit to $674$ inclusive 
data points and $106$ jet data points with $14$ parameters. 
The partial $\chi^2$ of the data sets is $730/674$ for the inclusive data and
$82/106$ for the jet data.  
Fig.~\ref{fig:herapdf1.6} shows the parton distributions and their 
uncertainties for the HERAPDF1.6 fit. HERAPDf1.5f is also shown on this plot
as a blue line. The fit with jets has rather similar central PDFs values to 
the fit without jets, apart from having a somewhat less hard high-$x$ sea. 
The uncertainties
are also similar to those of HERAPDf1.5f, with a slightly reduced uncertainty 
on the high-$x$ gluon. The quality of the fit to the jet data  
establishes that NLO QCD is able simultaneously to describe both 
inclusive cross sections and jet cross sections, thereby
providing a compelling demonstration of QCD factorisation.

\begin{figure}[htb]
%\vspace{-2.0cm}
\begin{center} 
\includegraphics[width=0.4\textwidth]{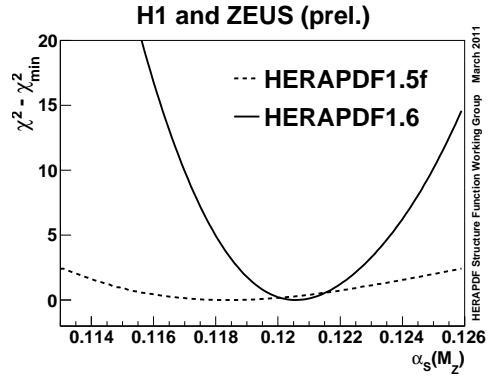} 
\caption {The difference between $\chi^2$ and its minimum value for the 
HERAPDF1.5f and HERAPDf1.6 fits as a function of $\alpha_s(M_Z)$
}
\end{center}
\label{fig:chiscan}
\end{figure}

The standard value of $\alpha_s(M_Z)$ used in the fits has been 
$\alpha_s(M_Z)=0.1176$. 
The correlation between $\alpha_s(M_Z)$ and the gluon 
PDF is too strong to make an accurate detrmination of $\alpha_s(M_Z)$ 
using purely inclusive data, but the jet data are sensitive to $\alpha_s(M_Z)$
such that one may let it be a free parameter of the fit. The value of 
$\alpha_s(M_Z)$ which results is
$
\alpha_S(M_Z) = 0.1202 \pm 0.0013(exp) \pm 0.0007(model/param) \pm 0.0012 (had) +0.0045/-0.0036(scale).
$
We estimate the model and parametrisation uncertainties for $\alpha_S(M_Z)$ 
in the same way 
as for the PDFs and we also add the uncertainties in the hadronisation 
corrections applied to the jets. The scale uncertainties are estimated by 
varying the renormalisation and factorisation scales chosen in the jet 
publications by a factor of two up and down. The dominant contribution to the 
uncertainty comes 
from the jet renormalisation scale variation. 
Fig.~\ref{fig:chiscan}  shows a $\chi^2$ scan vs $\alpha_S(M_Z)$ for the fits 
with and without jets, illustrating how much better $\alpha_S(M_Z)$ is 
determined when jet data are included. The model and parametrisation errors 
are also much better controlled.

\begin{figure}[htb]
%\vspace{-2.0cm} 
\begin{tabular}{cc}
\includegraphics[width=0.45\textwidth]{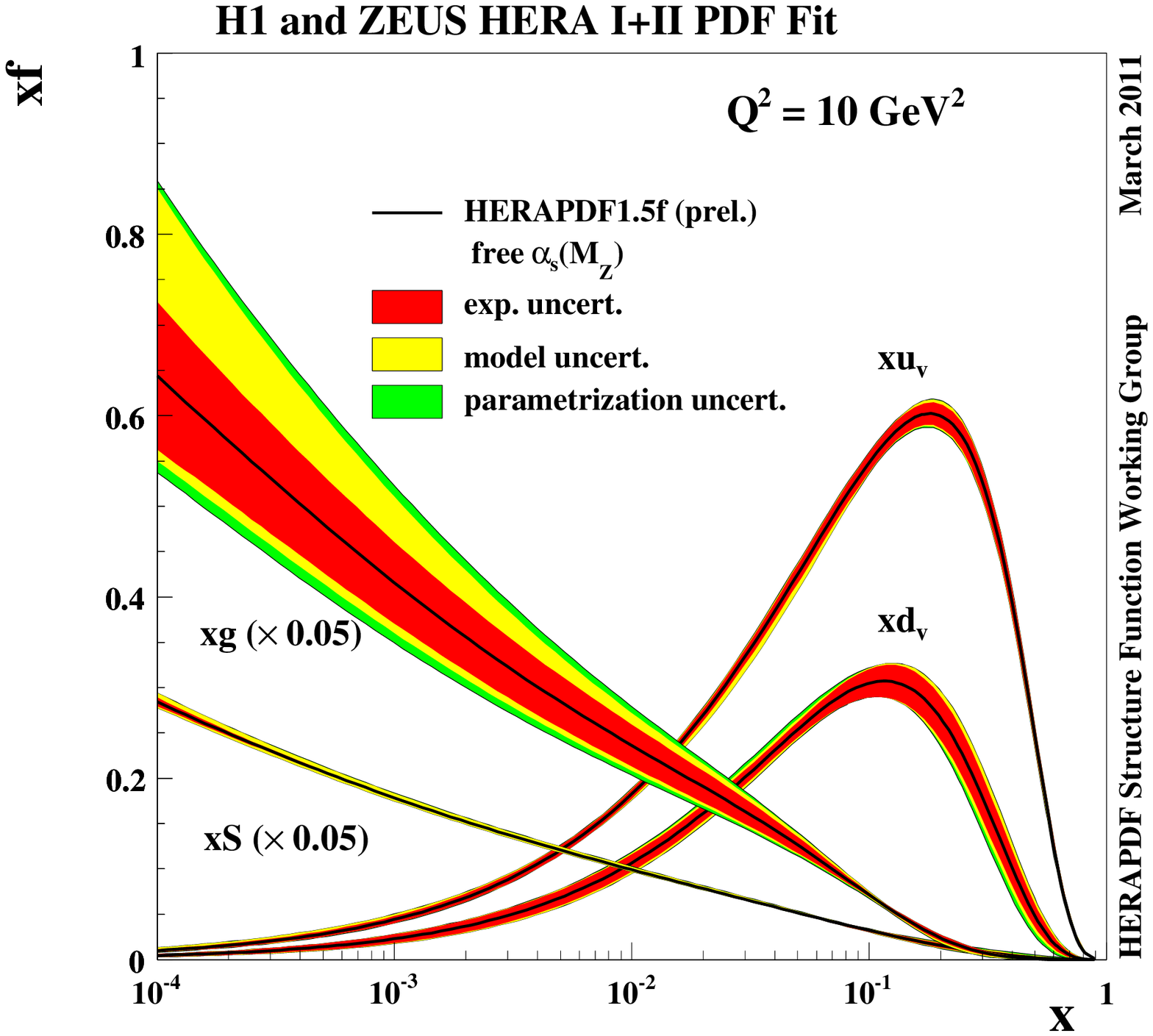} &
\includegraphics[width=0.45\textwidth]{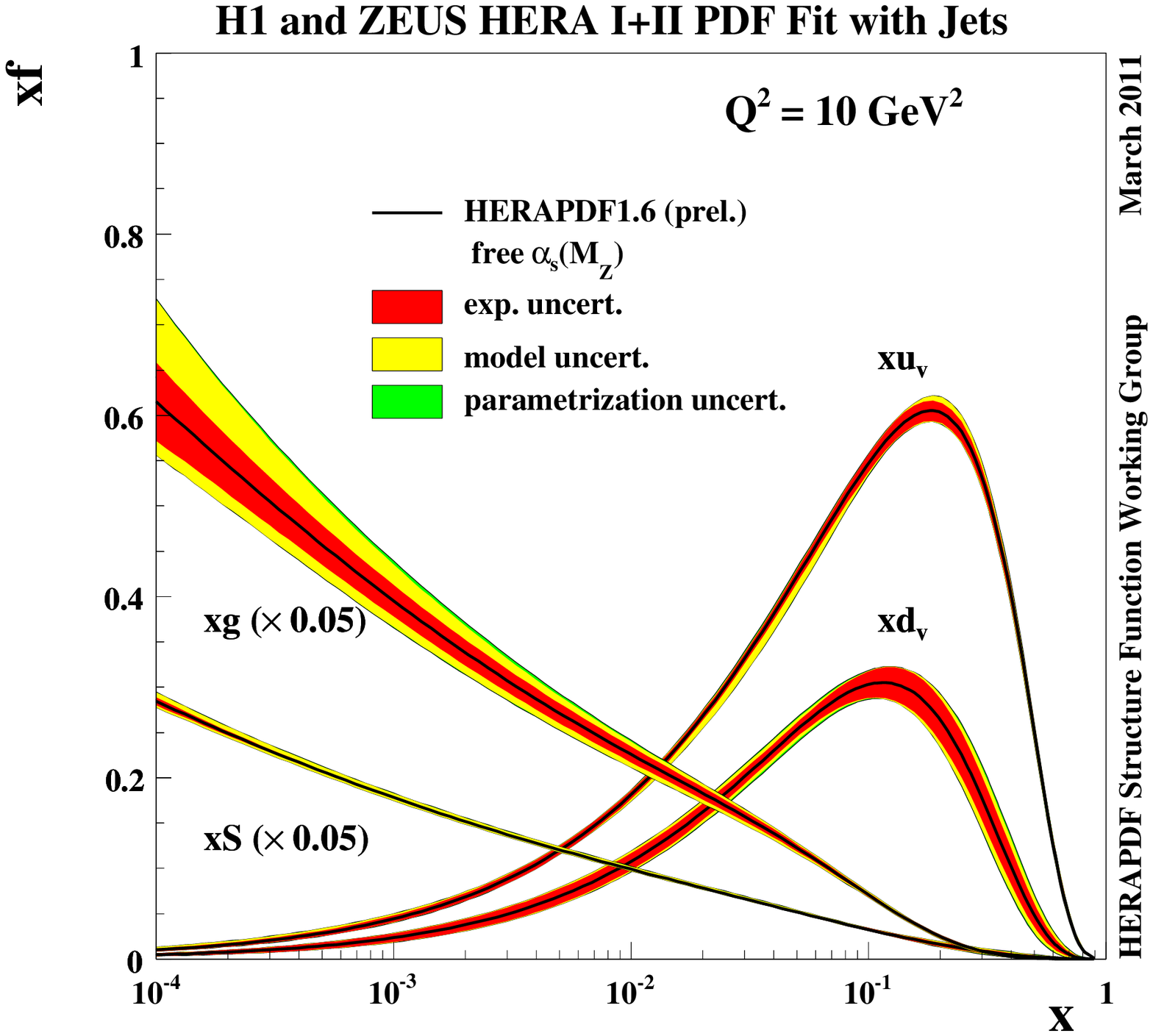}
\end{tabular}
\caption {The parton distribution functions
 $xu_v,xd_v,xS=2x(\bar{U}+\bar{D}),xg$, at $Q^2 = 10$~GeV$^2$, from 
HERAPDF1.5f and HERAPDf1.6, both with $\alpha_S(M_Z)$ 
treated as a free parameter of the fit.
The experimental, model and parametrisation 
uncertainties are shown separately. The gluon and sea %%@
distributions are scaled down by a factor $20$.
}
\label{fig:jetnojetalph}
\end{figure}

The $\chi^2$ for the HERAPDF1.6 fit 
with free $\alpha_S(M_Z)$ is 807.6 for 765 degrees of freedom. The 
partial-$\chi^2$
 for the inclusive data has barely changed but the partial-$\chi^2$ 
for the jet data decreases to 77.6 for 106 data points. 
Fig.~\ref{fig:jetnojetalph} shows the 
summary plots of the PDFs for HERAPDF1.5f and HERAPDF1.6, 
each with $\alpha_S(M_Z)$ left 
free in the fit. It can be seen that without jet data the uncertainty on 
the gluon PDF at low $x$ is large due to the strong 
correlation between the low-$x$ shape of
the gluon PDF and $\alpha_S(M_Z)$.  However once jet data are 
included the extra information on gluon induced processes reduces this 
correlation and the resulting 
uncertainty on the gluon PDF is not much larger than it 
is for fits with $\alpha_S(M_Z)$ fixed.

Direct 
photoproduction dijet cross sections have also been used in PDF fits to 
constrain the gluon as for example
in the ZEUS-jets analysis of ZEUS inclusive %%@
cross-section data and jet data~\cite{zeusjets}
However, such data have net yet been used in the HERAPDF fits because 
the cross-section predictions for photoproduced jets are sensitive to the 
choice of the input photon PDFs. In order to minimise sensitivity to this 
choice, the analysis can be 
restricted to use only the `direct' photoproduction 
cross sections. These are defined by the cut $x^{\rm obs}_\gamma > 0.75$, 
where $x^{\rm obs}_\gamma$ is a measure of the fraction of the photon's 
momentum that enters into the hard scatter. This is a direction for further 
study

\subsubsection{Bringing it all together: HERAPDF1.7}

Finally an NLO fit has been made bringing together all the data sets: HERA I+II
combined high energy data, combined low energy running data, 
$F^{c\overline{c}}_2$ data and jet data. This fit is called 
HERAPDF1.7~\cite{herapdf17}. The 
charm data are fit in the optimized version of the RT heavy quark scheme, 
with its preferred 
value of $m_c=1.5$. The value of $\alpha_s(M_Z)=0.119$ is fixed. This value 
gives the best fit to all the data in this fit, with the jet data 
dominating the sensitivity. Other settings are as for the HERAPDF1.6 fit 
except that the parameter $D_{uv}$ was found to be consistent with zero and 
hence only $13$ parameters have been used for the HERAPDF1.7 fit.
The correlated systematic uncertainties 
of the data sets are treated as for the individual 
fits: for the inclusive combined data sets at both low and high energy only the
procedural errors are treated as correlated by the Hessian method. 
For the $F^{c\overline{c}}_2$ data one source is treated as correlated by the 
offset method and for the jet data all $12$ sources are treated as correlated
by the Hessian method.

\begin{figure}[tbp]
\vspace{-0.5cm} 
%\vspace*{5pt}
\centerline{
\epsfig{figure=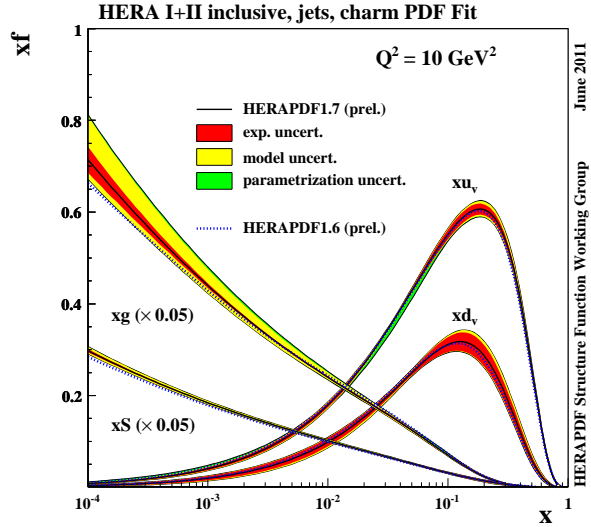,width=0.35\textheight}}
\caption { 
The parton distribution functions from 
HERAPDF1.7, $xu_v,xd_v,xS=2x(\bar{U}+\bar{D}),xg$, at
$Q^2 = 10$~GeV$^2$. 
The experimental, model and parametrisation 
uncertainties are shown separately. The sea and gluon
distributions are scaled down by a factor $20$.
}
\label{fig:herapdf1.7}
\end{figure}

The overall $\chi^2/ndf$ is  $1097.6/1032$ with partial $\chi^2/ndp$ of:
$44.1/41$ for $F^{c\overline{c}}_2$ data; $226.6/224$ for low energy data; 
$80.6/106$ for jet data and $746/674$ for HERA-I+II high energy data.
The data are all very compatible. The results of this combined fit are 
illustrated in Fig.~\ref{fig:herapdf1.7}

\subsection{Comparison of HERAPDF to other PDFs}
\label{compare}

\begin{figure}[tbp]
\vspace{-1.0cm} 
\centerline{\psfig{figure=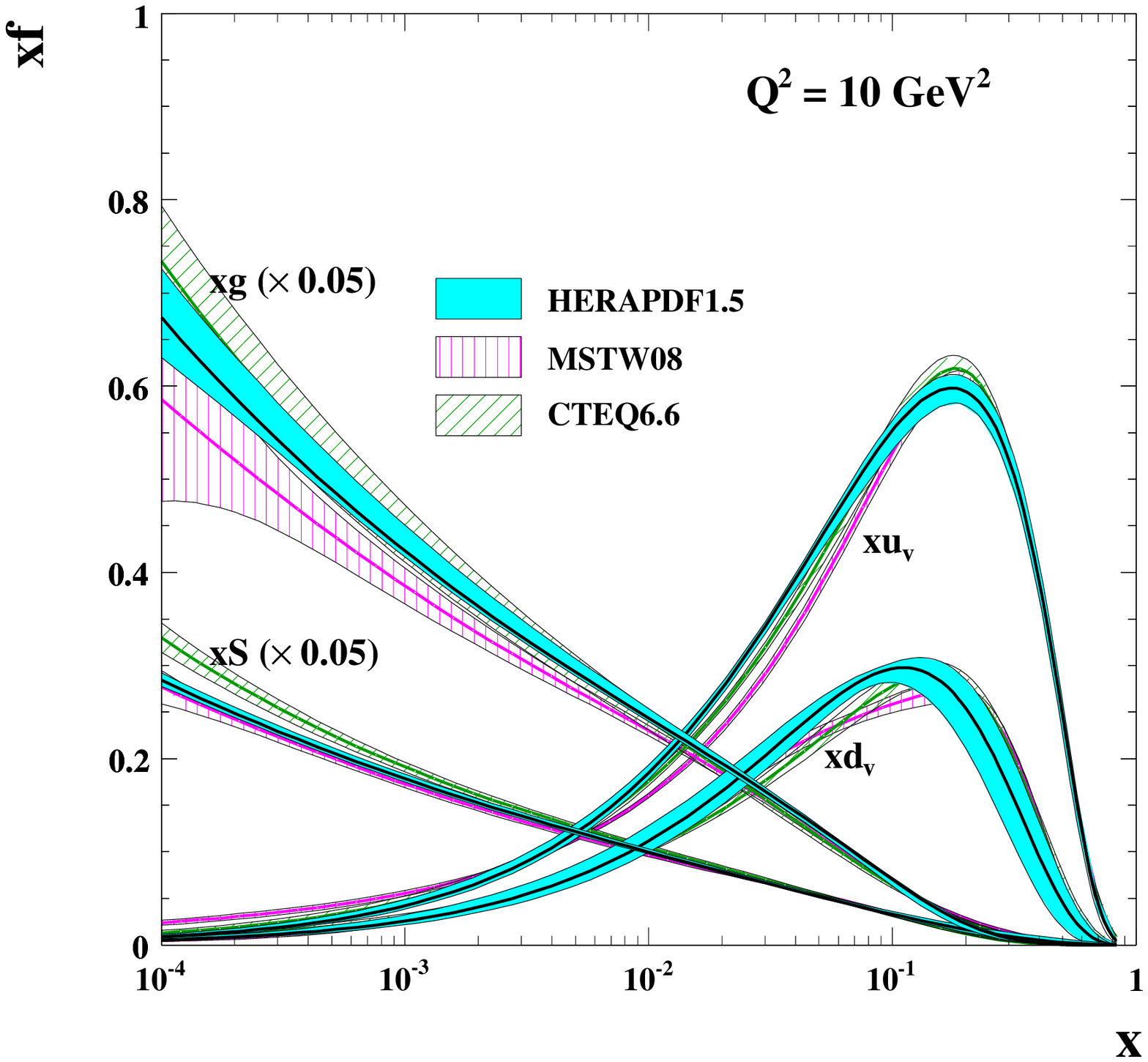,width=0.45\textwidth}~~ 
\psfig{figure=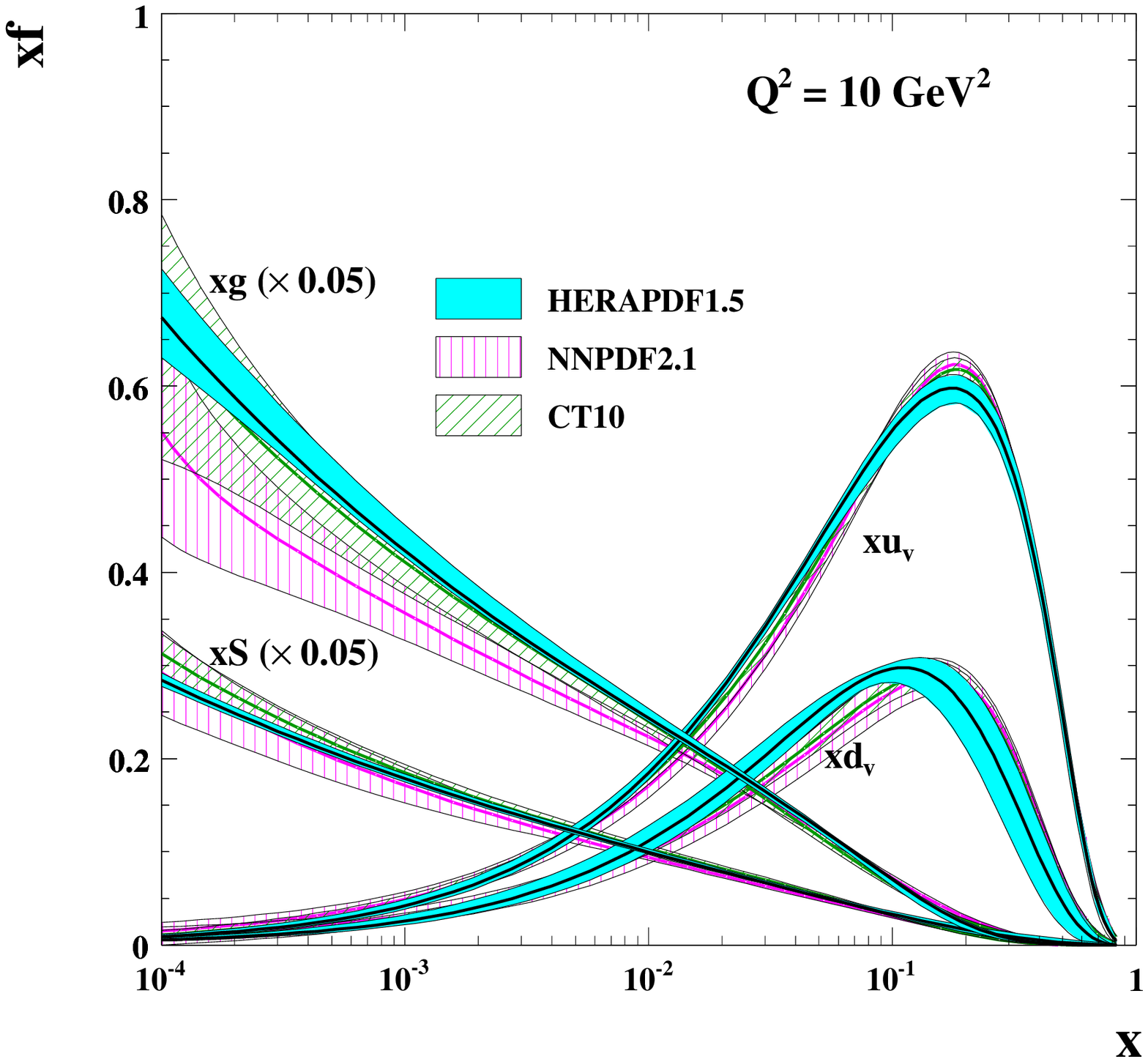,width=0.45\textwidth}}
\centerline{\psfig{figure=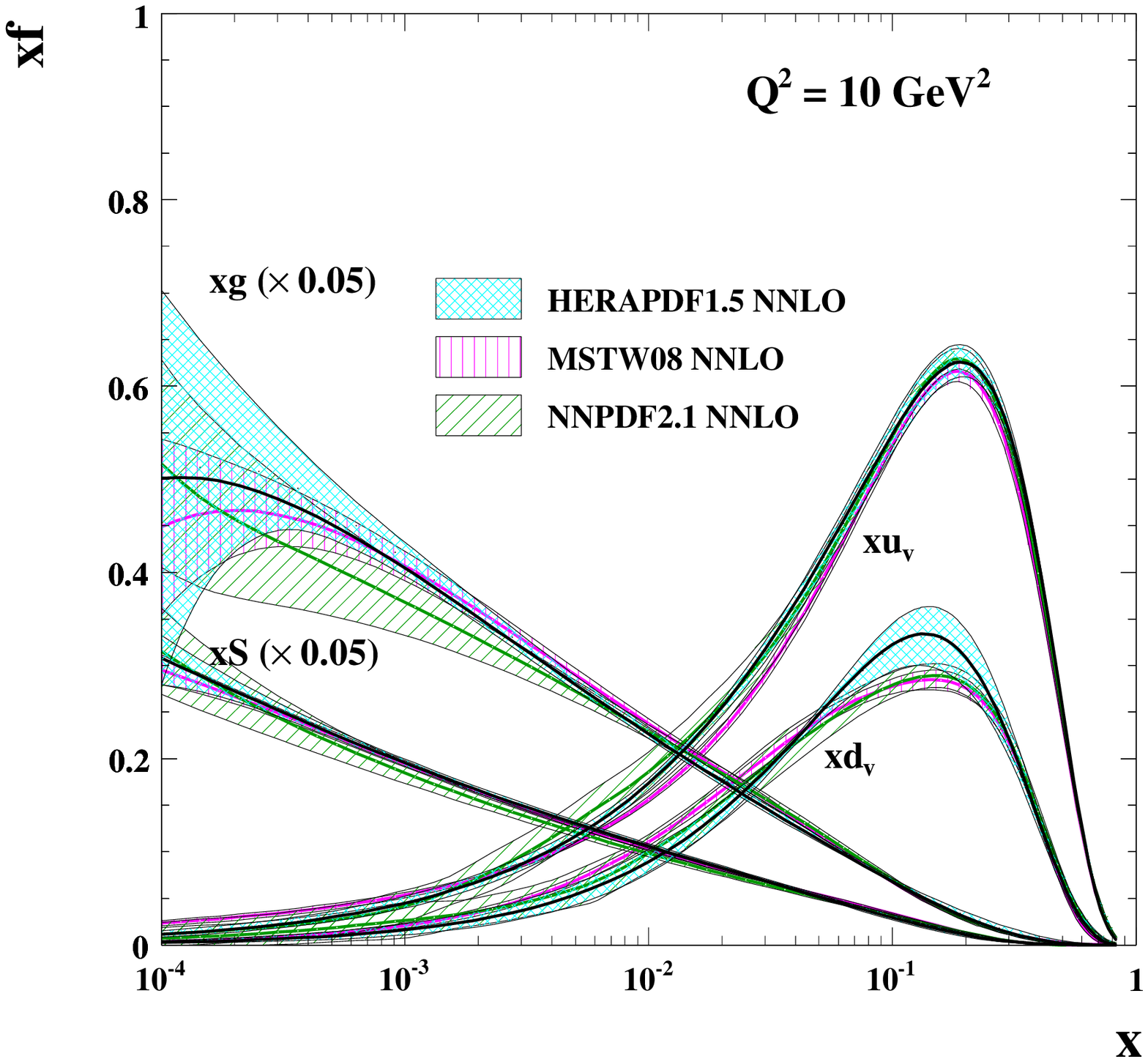,width=0.45\textwidth}~~ 
\psfig{figure=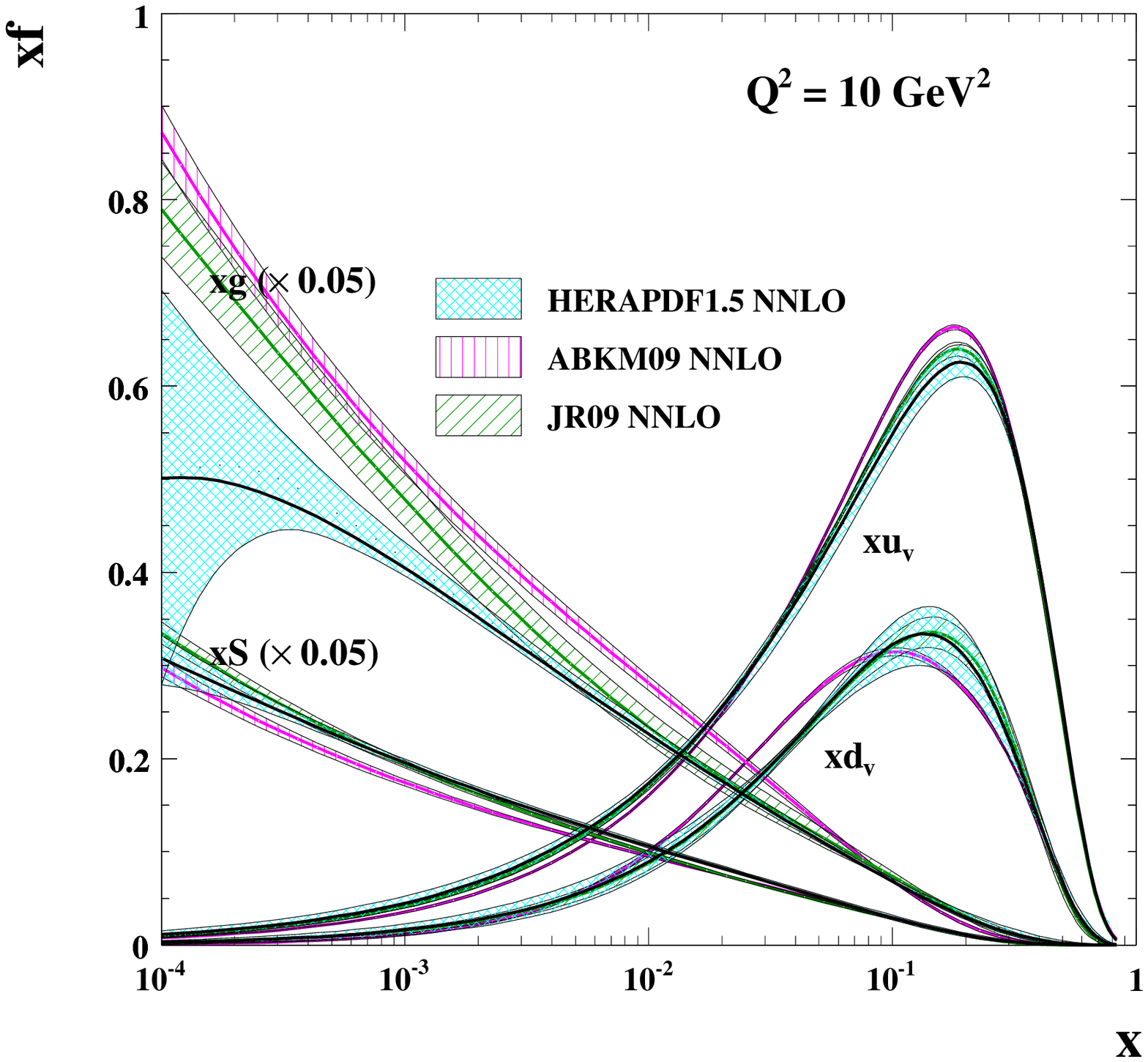,width=0.45\textwidth}}
\caption {HERAPDF1.5 compared to other PDFs. Top: NLO PDFs from MSTW08, CTEQ66,
NNPDF2.1, CT10. Bottom: NNLO PDfs from MSTW08, NNPDF2.1, ABKM09, JR09
}
\label{fig:pdfs}
\end{figure}
Fig.~\ref{fig:pdfs} compares HERAPDF1.5 to MSTW08~\cite{Martin:2009iq}, 
CTEQ6.6~\cite{cteq66}, CT10~\cite{ct10}, NNPDF2.1~\cite{nnpdf21}, 
ABKM09~\cite{abkm09}, JR09~\cite{jr09} at $Q^2=10~$GeV$^2$. 
All PDFs are shown with $68\%$ CL uncertainties. The top row compares 
NLO PDFs and the bottom row compares NNLO PDFs.  These PDF sets have been 
chosen for comparison because they have been selected for benchmarking by 
the PDF4LHC group~\cite{Alekhin:2011sk} (though CT10 and NNPDF2.1 are updates 
of the benchmarked PDFs).
All PDFs are broadly compatible but 
there are differences of detail which can have 
important consequences for predictions of LHC cross sections.

A concise way to compare predictions for various LHC 
cross sections is to compare parton-parton 
luminosities for quark-antiquark and gluon-gluon interactions.               
\begin{equation}
\frac{\partial L_{gg}}{\partial\hat{s}} = \frac{1}{s} \int_\tau^1 \frac{dx_1}{x_1} f_g(x_1,\hat{s}) f_g(x_2,\hat{s})
\end{equation}
\begin{equation}
\frac{\partial L_{\Sigma(\bar{q}q)}}{\partial\hat{s}} = \frac{1}{s} \int_\tau^1 \frac{dx_1}{x_1}~ \Sigma_{q=d,u,s,c,b}\left[ f_q(x_1,\hat{s}) f_{\bar{q}}(x_2,\hat{s}) + f_{\bar{q}}(x_1,\hat{s}) f_{q}(x_2,\hat{s})\right] 
\end{equation}
where $s$ is the centre of mass energy squared of the proton-proton collision 
and 
$x_1$ and $x_2$ are the fractional momenta of the partons in each proton, such 
that the centre of mass energy squared of the parton-parton collision is, $\hat{s}= \tau s$, where $\tau=x_1 x_2$.

\begin{figure}[tbp]
\vspace{-1.0cm} 
\centerline{\psfig{figure=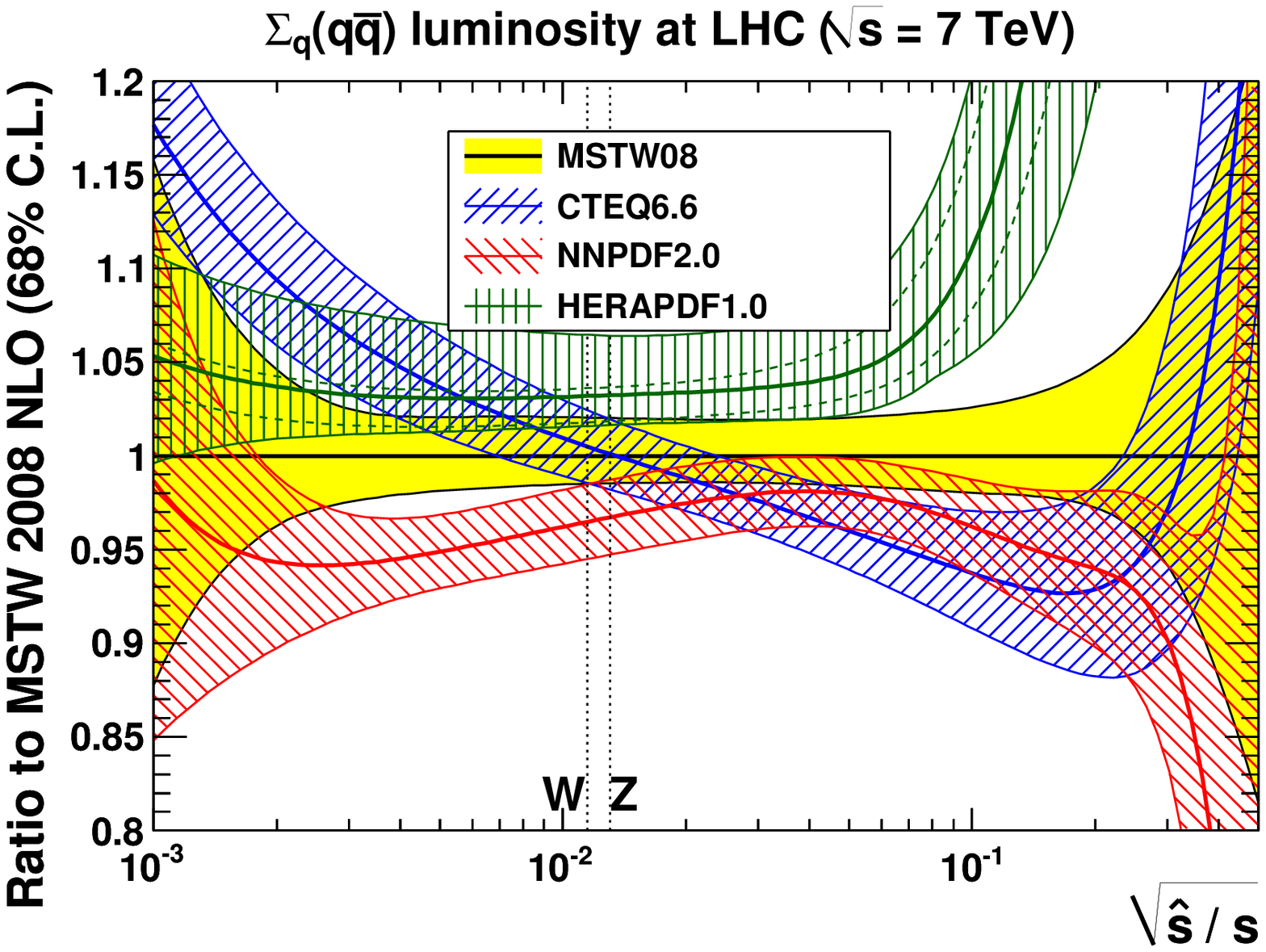,width=0.30\textwidth}~~ 
\psfig{figure=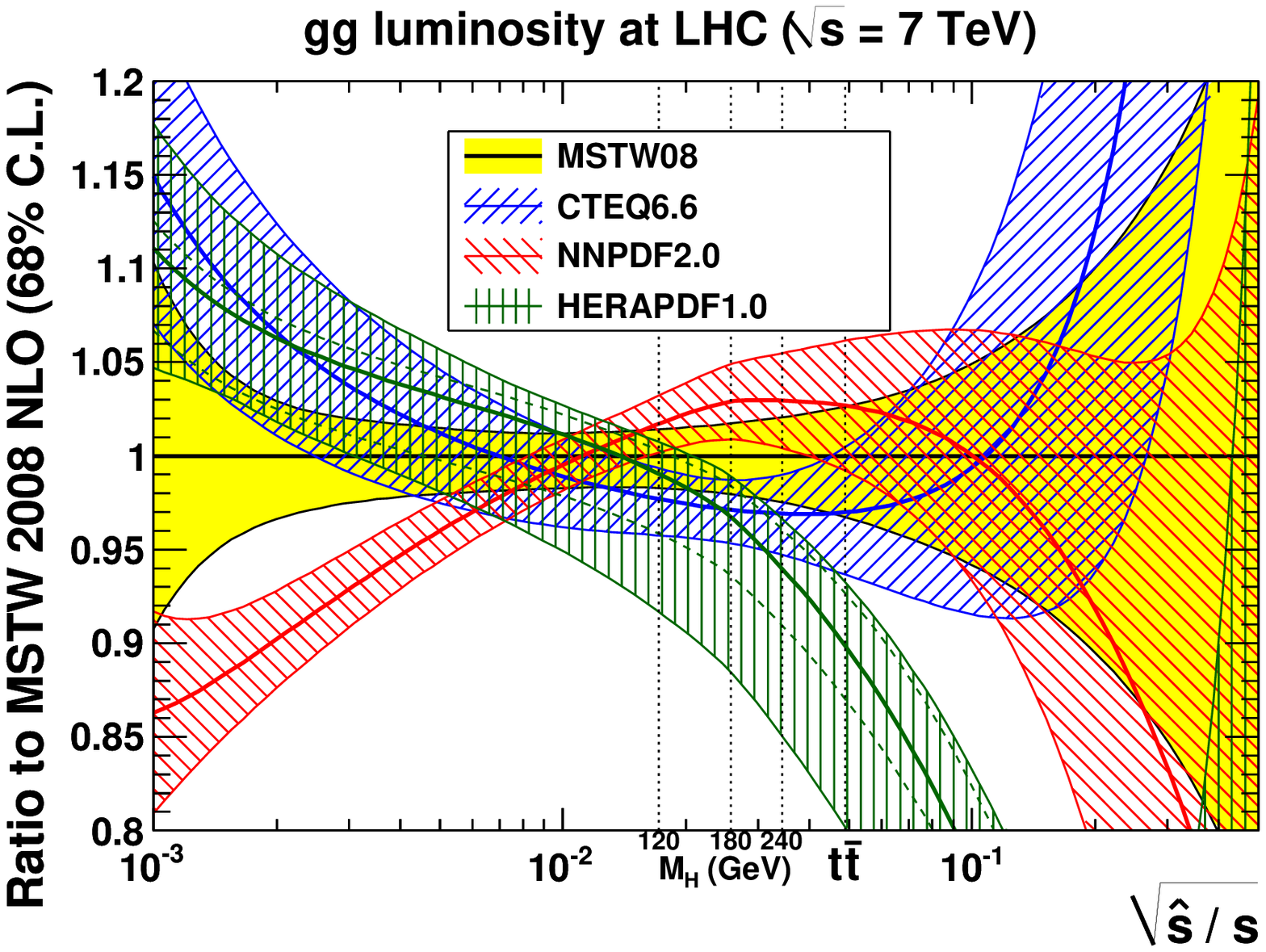,width=0.30\textwidth}}
\centerline{\psfig{figure=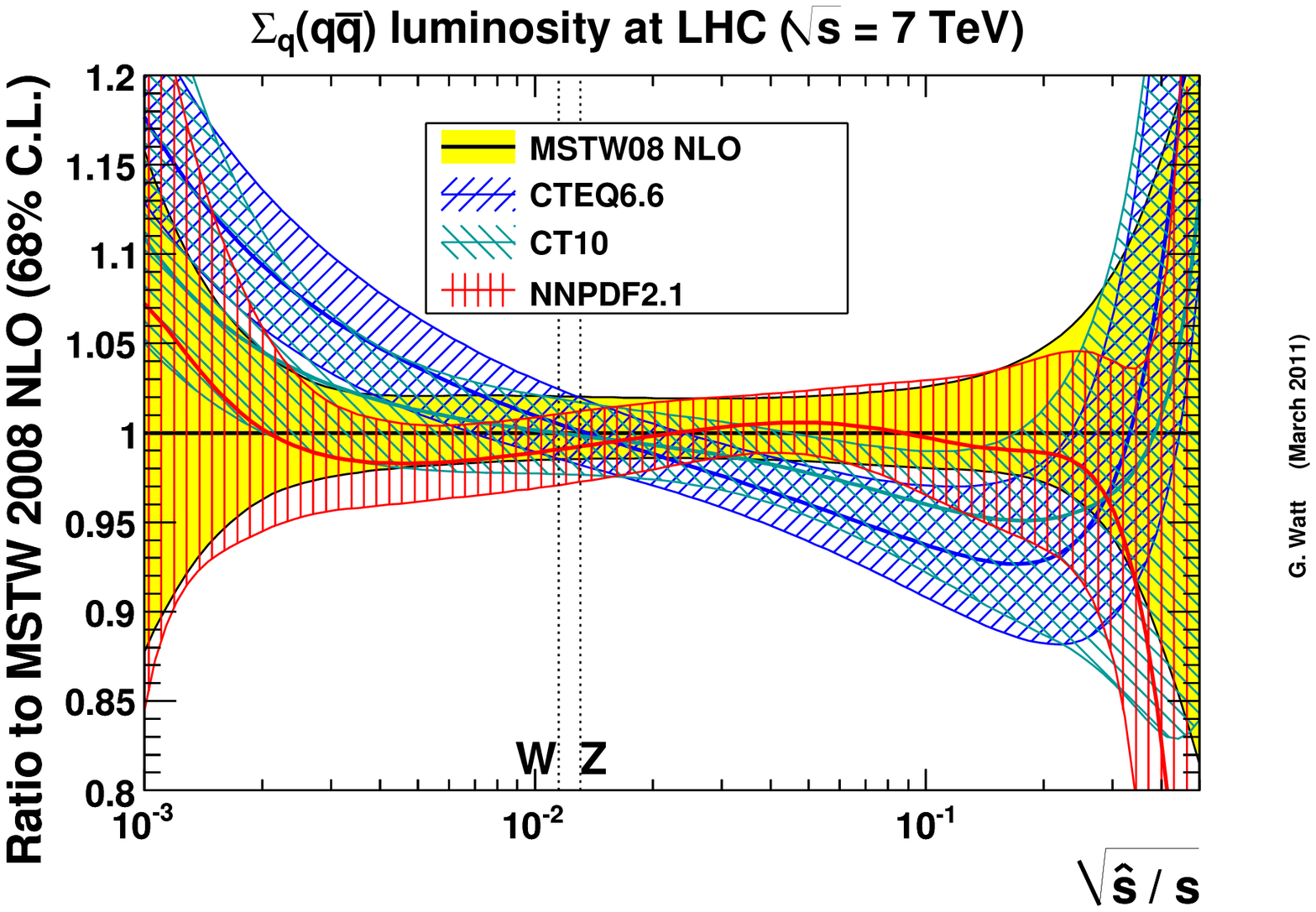,width=0.30\textwidth}~~ 
\psfig{figure=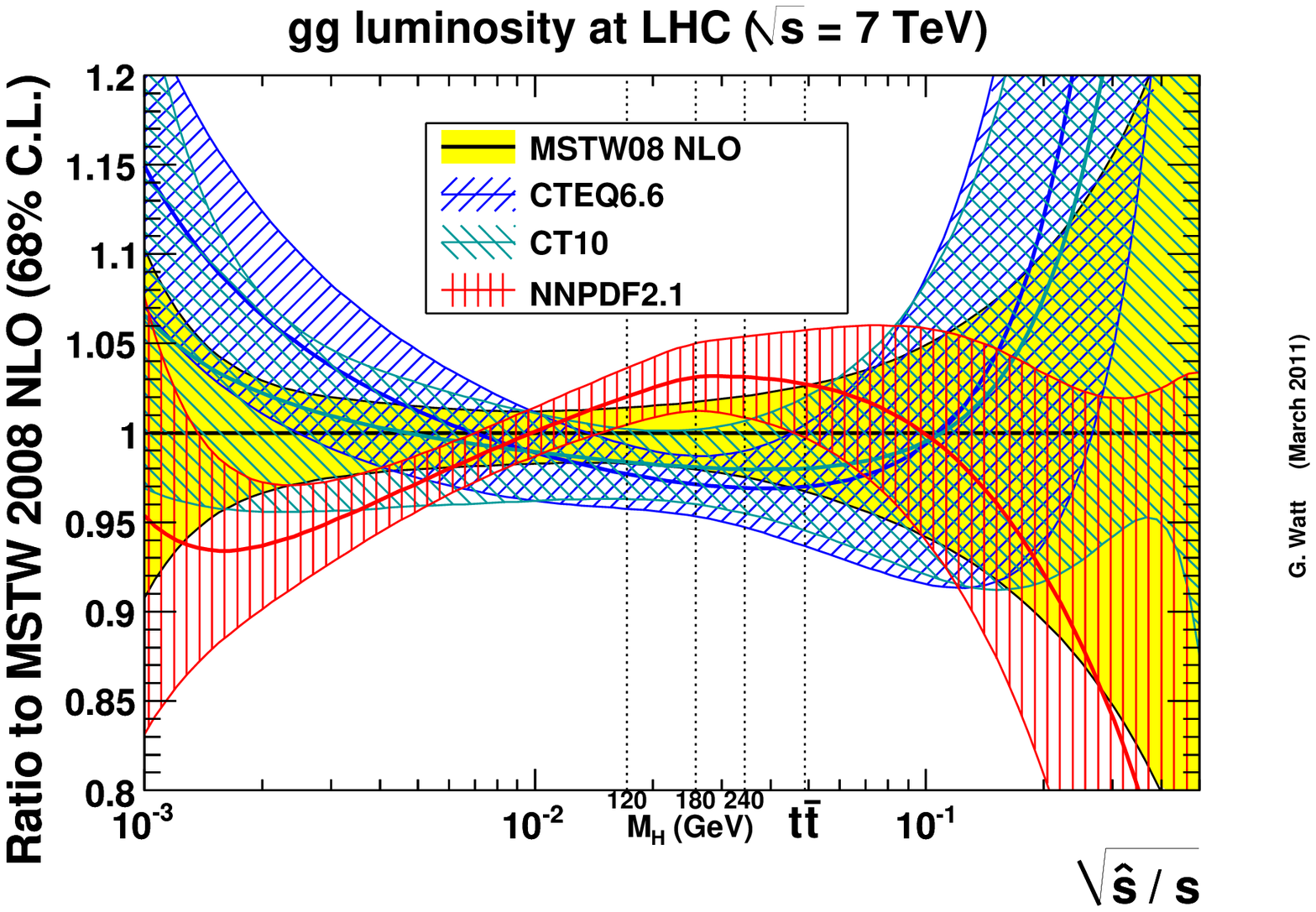,width=0.30\textwidth}}
\centerline{\psfig{figure=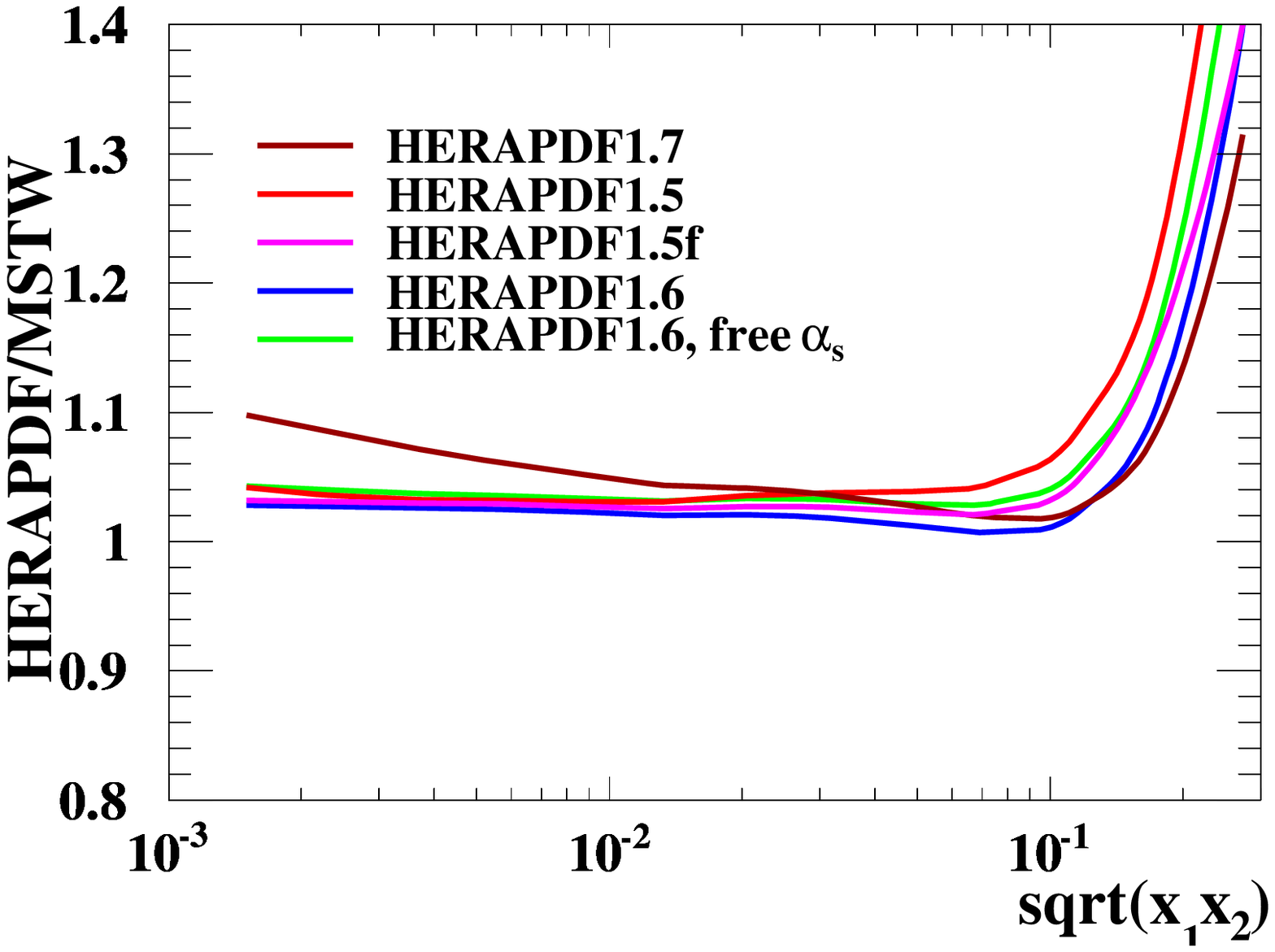,width=0.30\textwidth}~~ 
\psfig{figure=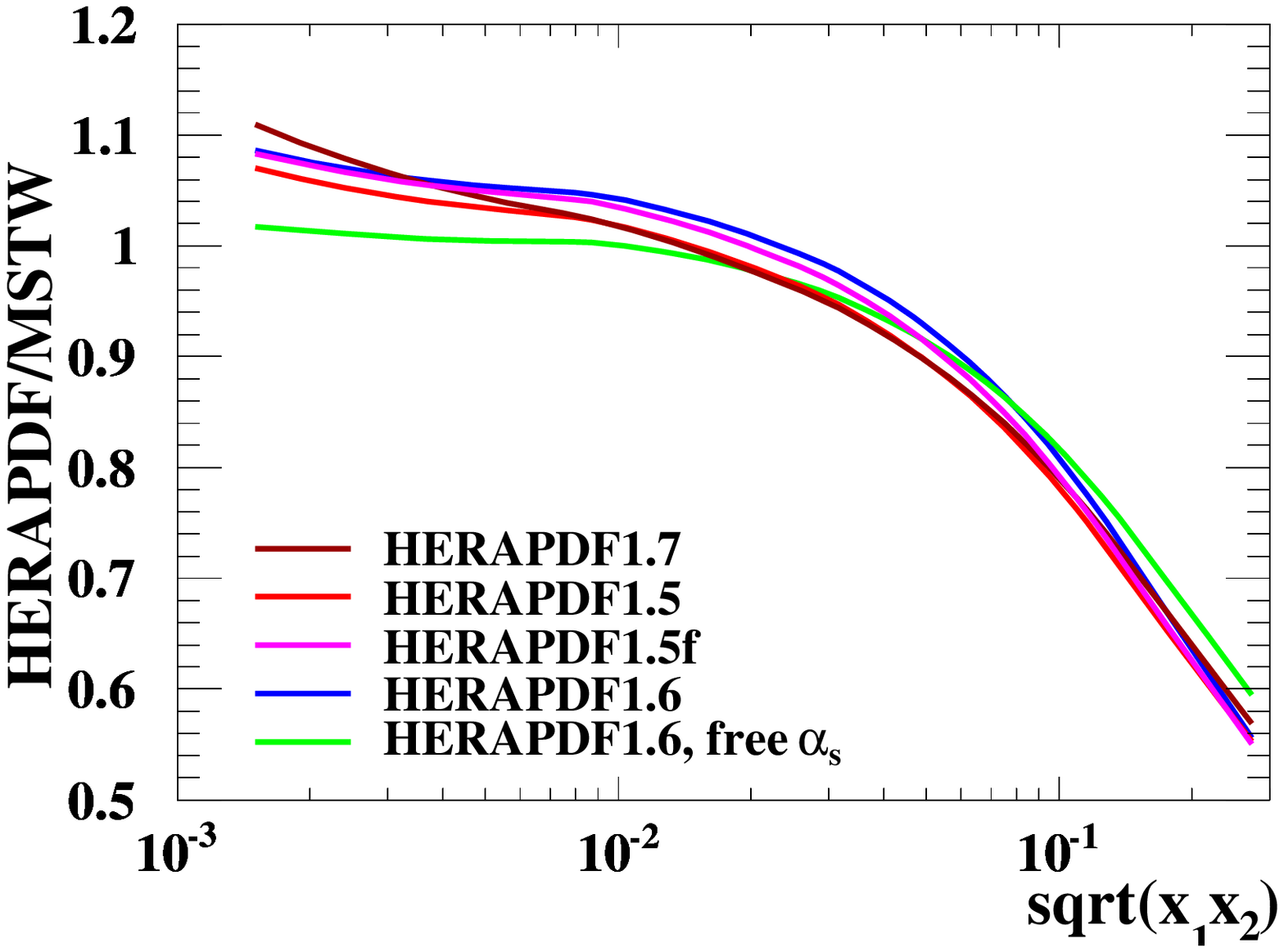,width=0.30\textwidth}}
\caption{Left hand side: the $q-\bar{q}$ luminosity in ratio to that of 
MSTW2008 for various PDFs. Right hand side: the same for the $g-g$ 
luminosities.
Upper row, PDFs as benchmarked by the PDF4LHC group:
 middle row, updates for 
CT10 and NNPDF2.1 issued in 2011: bottom row, the HERAPDF NLO updates 
described in this review.
}
\label{fig:lumi}
\end{figure}

\begin{figure}[tbp]
\vspace{-1.0cm} 
\centerline{\psfig{figure=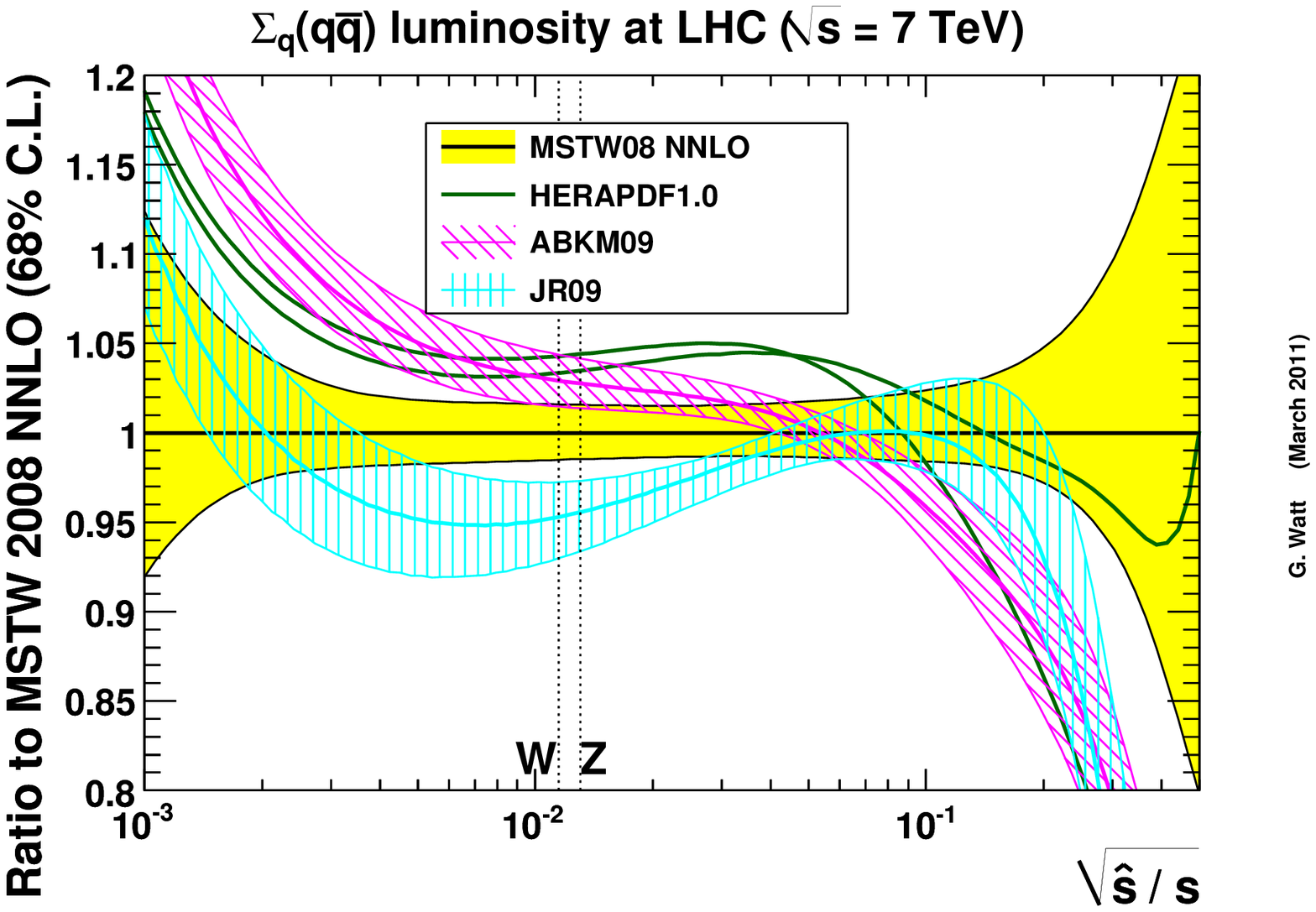,width=0.30\textwidth}~~ 
\psfig{figure=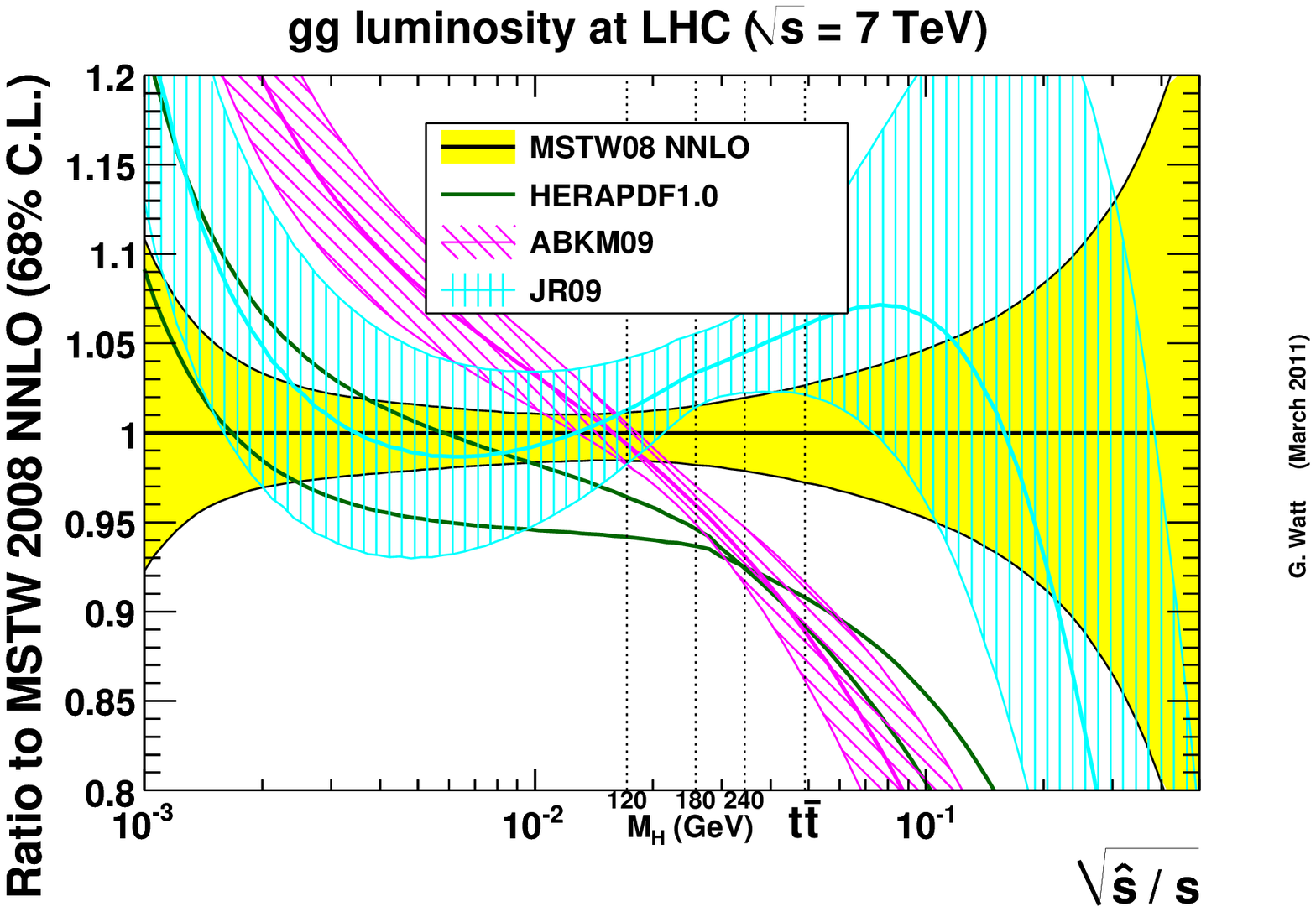,width=0.30\textwidth}}
\centerline{\psfig{figure=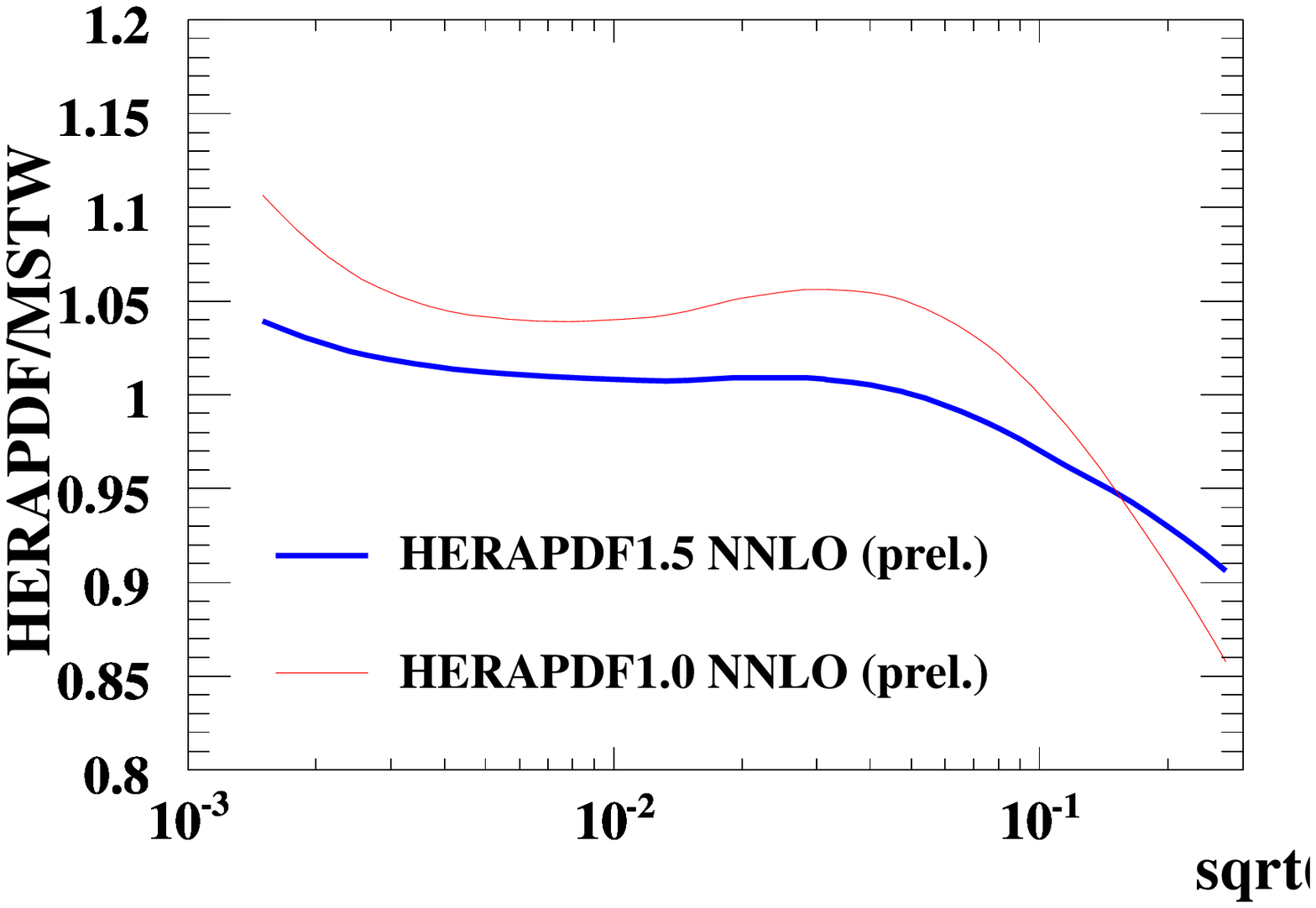,width=0.30\textwidth}~~ 
\psfig{figure=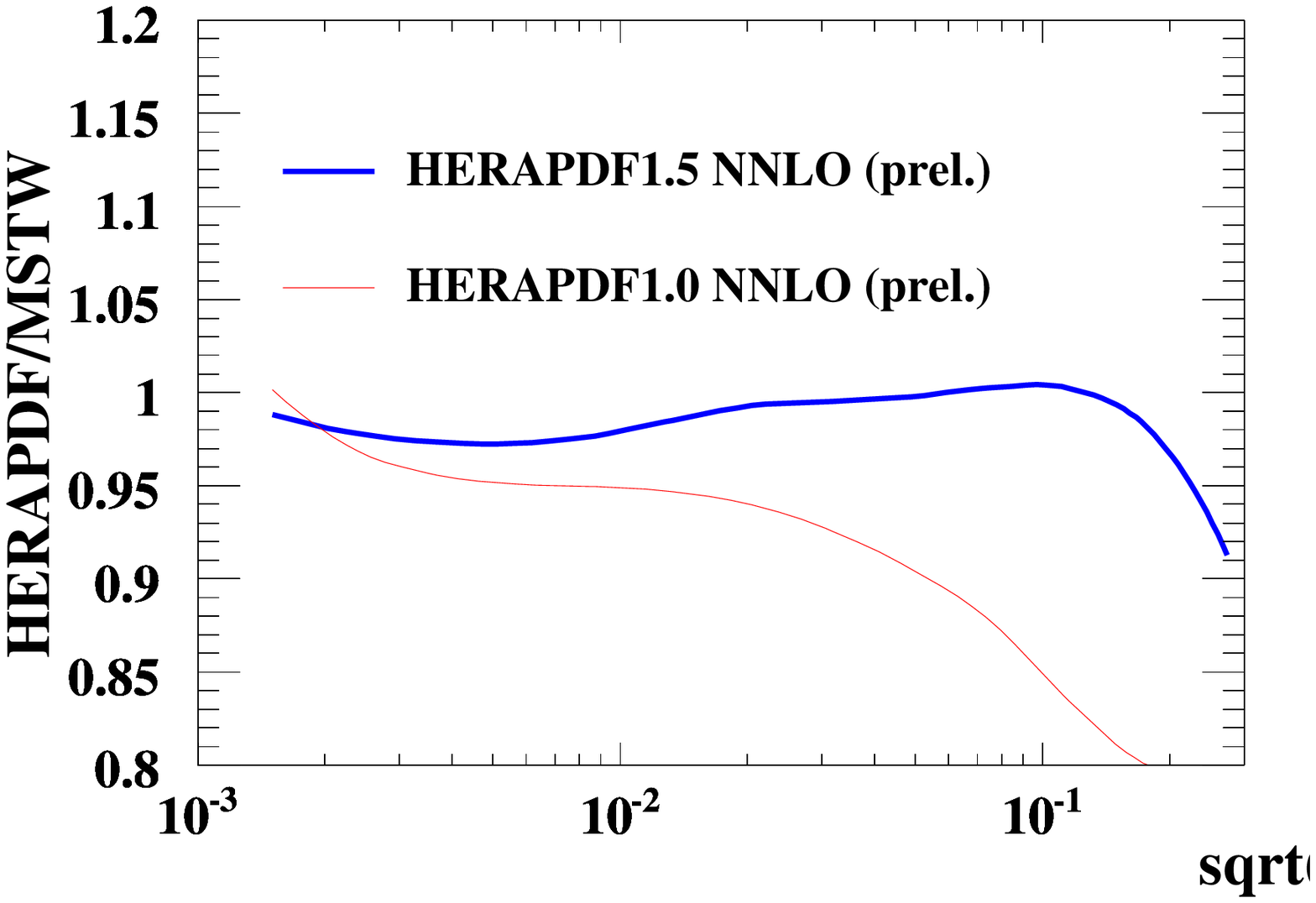,width=0.30\textwidth}}
\caption {Left hand side: the $q-\bar{q}$ luminosity in ratio to that of 
MSTW2008 for various PDFs. Right hand side: the same for the $g-g$ 
luminosities.
Upper row, NNLO PDFs as benchmarked by the PDF4LHC group ({\it Plots from G.Watt http://projects.hepforge.org/mstwpdf/pdf4lhc/}): bottom row, 
the HERAPDF1.5 NNLO PDFs described 
in this review compared to HERAPDF1.0 NNLO
}
\label{fig:luminnlo}
\end{figure}

Fig.~\ref{fig:lumi} shows $q-\bar{q}$ and $g-g$
luminosities for $p-p$ interactions at the LHC \footnote{Plots on top and middle rows from G.Watt http://projects.hepforge.org/mstwpdf/pdf4lhc} with 
$\surd{s} = 7~$TeV, 
in ratio to those of the MSTW2008 PDF, 
for PDFs issued by CTEQ, NNPDF and HERAPDF. 
This figure also shows the corresponding luminosity plots for the 
HERAPDF1.5, 1.6, 1.7 NLO updates described in this review.
Fig.~\ref{fig:luminnlo} shows similar luminosity comparison plots for 
NNLO PDFs from MSTW, ABKM, JR and HERAPDF.

There are several reasons why the PDF predictions differ. It
 is beyond the scope of the 
current review to describe all the other PDFs in detail. However a few 
remarks can be made
on the main differences between HERAPDF and other PDFs.
Firstly they are 
based on different data sets and different choices of cuts on these data sets
and this is closely related to the differing way in 
which the PDF uncertainties are estimated since the use of many different data 
sets has led to the use of increased $\chi^2$ tolerances for some of the PDF 
sets. Secondly, different choices of  PDF parametrisation are made and this
impacts on the size of the uncertainties. Thirdly, PDFs use different central 
values of $\alpha_s(M_Z)$ and this affects the shape of the PDFs, 
particularly the gluon PDF. 
Fourthly, the PDF analyses differ in the schemes 
used to account for heavy quark production and different heavy quark masses.

\subsubsection{Correlated systematic uncertainties and $\chi^2$ tolerance.} 

Most modern data used in PDF fits are statistically very precise such that 
systematic errors dominate. Thus the correct treatment of correlated 
systematic errors becomes very important. 
In PDF fits done prior to the year 2000 point-to-point correlated 
systematic errors were not specifically treated. They were
added in quadrature to the uncorrelated errors. This can lead to biassed 
results. The correct treatment of correlated systematic errors is discussed in
Ref~\cite{offhesse}. The consensus amongst PDF fitters is that
the uncertainty due to correlated systematic errors %%@
should be included in the theoretical prediction such that
\[ 
F_i(p,s) = F_i^{\rm NLOQCD}(p) + 
\sum_{\lambda} s_{\lambda} \Delta^{\rm sys}_{i\lambda}
\]
where $p$ are the PDF parameters, $s_\lambda$ 
represent independent (nuisance) variables for each source of
 systematic uncertainty and $\Delta^{\rm sys}_{i\lambda}$ represents the 
one standard deviation correlated systematic error on data point $i$ due to correlated error source $\lambda$. 
A representative form of the $\chi^2$ is then given by 
\begin{equation}
\chi^2 = \sum_i \frac{\left[ F_i(p,s)-F_i(\rm meas) \right]^2}{\sigma_i^2} %%@
+ \sum_\lambda s^2_\lambda 
\label{eq:chi2}
\end{equation}
where $\sigma_i$ is the uncorrelated error on each data point.
Thus the nuisance parameters are fitted together with the PDF parameters.
This method of treatment of correlated systematic has been termed the Hessian 
method. An alternative is the Offset method in which $s_\lambda=0$ for the 
central fit but the nuisance parameters are varied when determining the 
error on the PDF parameters~\cite{offhesse}. 

In the PDF fits of CTEQ/CT, MSTW and GJR/JR the Hessian method
 is used with %%@
increased $\chi^2$ tolerances such that a $68\%(90\%)$CL is not set by a 
variation of
$\Delta\chi^2 = 1(2.73)$ but by a larger variation. 
The reason for the use of such increased $\chi^2$ %%@
tolerances arises when using many different input data sets which are not 
all completely consistent.  
The tolerances are set so as to ensure that all the separate data %%@
sets are fit to within their $68\%(90\%)$CL. 
The tolerance %%@
can differ according to the parameter being fitted (or more exactly %%@
according to an eigenvector combination of parameters, 
see Sec.~\ref{sec:eigen})  but as a rough %%@
guide the CTEQ $90\%$CL tolerance is $\Delta\chi^2 \sim 100$
and the MSTW $90\%$CL tolerance is $\Delta\chi^2 \sim30$ for the MSTW2008 
analysis\footnote{Note that MSTW provide both $68\%$ CL
and $90\%$CL uncertainties, whereas for CTEQ/CT a factor of $1./\surd{2.73}$ 
must be applied to the $90\%$CL uncertainties to 
obtain $68\%$CL uncertainties.}. 
The GJR/JR analyses use $\Delta\chi^2 \sim 50$. 
The use of these increased $\chi^2$ tolerances has caused great controversy.
For example, Pumplin~\cite{pumplin} argues that a
$\chi^2$ tolerance of at most $\sim 10$ can be justified on the grounds of 
data incompatibility and that the %%@
more inflated values implicitly account for parametrisation %%@
variations. 

The ABKM group does not use increased tolerances and that is why their 
PDF uncertainties are generally smaller than those of other groups. HERAPDF 
also does not use an increased tolerance but considers additional 
model and parametrisation uncertainties, see Sec.~\ref{sec:pdfchap}. 
The NNPDF group use a completely different method of estimating 
PDF uncertainties, see Sec~\ref{sec:paramunc}. 

For the HERAPDF the Hessian procedure has already been applied to the data %%@
combination to set the best values for the systematic shifts and the %%@
combination procedure itself results in greatly reduced systematic errors, %%@
such that there is no longer a significant difference in the PDF %%@
uncertainties obtained using the Offset and Hessian methods of treating %%@
systematic uncertainties. The good $\chi^2$ for the combination fit also %%@
establishes that the resulting data set is very consistent, see %%@
Sec.~\ref{sec:datacomb}, such that in the HERAPDF the conventional tolerance
$\Delta\chi^2 = 1$ 
is appropriate for setting $68\%$CL uncertainties.

\subsubsection{Diagonalisation and Eigenvector PDF sets}
\label{sec:eigen}

In either the Hessian or the Offset method, the Hessian matrices and 
covariance matrices are not, in general,
diagonal. The variation of $\chi^2$ w.r.t. some parameters is much more rapid
than that of others, but because the parameters are correlated to each other
the effect of each parameter is not clear. When evaluating uncertainties 
on physical observables it
can be an advantage to use an eigenvector basis of PDFs, which provide an 
optimized representation of parameter space in the neighbourhood of the 
minimum.
The eigenvalues of the covariance matrix represent the squares of the errors 
on the combination of parameters which gives the corresponding eigenvector.

An eigenvector basis of PDFs is the usual way of summarizing the results 
of a PDF analysis including its 
error estimates. Two sets of PDF parameters
must be supplied for each eigenvalue, representing
displacement up and down along its eigenvector direction by the $\chi^2$
tolerance. The symmetric error on a quantity $F$ which is a function of the PDF
parameters is then simply given by
\[
<\sigma_F^2>=  \sum_j \left[\frac{F(p_j^+) - F(p_j^-)}{2}\right]^2
\]
where $F(p_j^+)$, $F(p_j^-)$ are the values of $F$ evaluated up and down
along eigenvector $j$. Asymmetric errors may be evaluated by the prescription:
\[
<\sigma_F^2(+)>=  \sum_j \left[max(F(p_j^+) - F(p_j^0),F(p_j^-) - F(p_j^0),0)\right]^2
\]
\[
<\sigma_F^2(-)>=  \sum_j \left[max(F(p_j^0) - F(p_j^-),F(p_j^0) - F(p_j^-),0)\right]^2
\]
where $F(p_j^0)$, is the central value of $F$.

The PDFs from the HERAPDF are made public in this format via the LHAPDF
(http://lhapdf.hepforge.org) interface. As well as the eigenvector sets, 
which give the experimental uncertainty of the HERAPDF, 
further sets are provided to 
cover the model and parametrisation variations. These should be combined with 
the experimental errors as specified in Sec.~\ref{sec:results}.
Further PDF sets are also provided
for a range of fixed $\alpha_s(M_Z)$ values, 
so that uncertainty due to $\alpha_s(M_Z)$ variation may also be evaluated.

The LHAPDF library is also the repository for the PDF sets from other PDF fitting groups.

\subsubsection{Choice of data sets and kinematic cuts}

The CTEQ6.6, MSTW2008, ABKM09 and JR09 PDF analyses do 
not use the recently combined inclusive cross section data from 
HERA-I~\cite{h1zeuscomb} which are up to three times more accurate than the 
separate H1 and ZEUS data sets used by previous PDF analsyses. 
These combined HERA data are shifted in normalisation by $\sim 3\%$ with 
respect to the previous HERA data, and this explains the higher 
luminosity of the HERAPDF at low $\tau$. 

Conversely the HERAPDF analysis uses only HERA data, whereas the CTEQ, MSTW 
and NNPDF analyses are 'global' PDFs which also use: older fixed target data, 
both from DIS and from Drell-Yan production; Tevatron $W,Z$ cross section data 
and jet production data. The ABKM and JR PDFs each use some but not all of 
these non-HERA data sets. 

The use of a single consistent data set with a clear statistical interpretation
of uncertainty limits was one of the primary motivations behind the HERAPDF. 
However there are other reasons why the use of some of the 
other data sets may lead to 
further uncertainties. Firstly, the neutrino-$Fe$ fixed-target scattering data 
from CCFR and NuTeV, which is often used to help to determine the valence 
densities, needs 
corrections for nuclear effects (the 'EMC effect'). 
Although such nuclear corrections are made in the global PDF analyses, they are not perfectly determined 
and the uncertainty due to these corrections is not fully
accounted~\cite{cjstuff}. 
More recently similar critisms have been made of the use 
deuterium target data (either in DIS or Drell-Yan). 
Accardi et al~\cite{jlabstuff} have reconsidered deuterium 
corrections for the fixed target data. They find large uncertainties in these 
corrections and this results in greater (unaccounted for) uncertainty in 
the high-$x$ $d-$quark for fits where the deuterium data is the principal 
source of information on the $d-$quark. (For the HERAPDF the information on 
$d-$quark comes from CC $e^+ p$ scattering).

Fixed proton target data do not suffer from these problems, but the kinematic
reach of such data does extend into the high-$x$, low-$Q^2$ region, where 
the theoretical interpretation of the data requires consideration of
 target mass 
corrections and higher twist terms. Most PDF analyses make a 
$W^2\geqsim 15$GeV$^2$ cut to avoid this region 
(for the HERAPDF this is unnecessary since all HERA data is at 
large $W$). The ABKM analysis choses to include the low $W$ data
 and model the higher twist terms. The high-$x$ region is also receiving 
attention from the CTEQ-JLAb group~\cite{keppel}.

A further problem in the use of older fixed target 
data is that results were often presented
and used in terms of $F_2$ rather than reduced cross sections.
ABM~\cite{abmf2} have examined the use 
of NMC $F_2$ data in the global fits. The NMC 
extraction of $F_2$ relied on assumptions 
on the value of $F_L$ which are not consistent with modern QCD calculations.
ABM find that using NMC published values of $F_2$, rather than the NMC 
cross section data, raises their extracted 
values of $\alpha_S$ erroneously.

The HERAPDF avoids bias from erroneous assumptions about heavy target 
corrections, 
deuterium corrections, higher twist corrections and $F_L$ corrections,  
by using only HERA pure proton target 
cross section data, but a price is paid in terms of the uncertainties of the 
 high-$x$ parton distributions, which are generally larger than those of the 
other groups.

It is also notable that the HERAPDFs have a harder high-$x$ sea 
and a softer high-$x$ gluon PDF at NLO. It has been suggested that this may be 
because Tevatron jet data are not included in the HERAPDF fit. However the 
story is not quite so simple.

Global fits use Tevatron high-$E_T$ jet production data to help to 
pin down the 
high-$x$ gluon. The HERAPDF analysis uses HERA-jet data for the same 
purpose, although the HERA jet data do not extend to as high $x$ values as the 
Tevatron jet data. These Tevatron jet data have very large 
correlated systematic uncertainties compared to HERA jet data 
such that much trust must be put in the evaluation of systematic 
uncertainties. Tevatron Run-I jet 
data suggested a hard high-$x$ gluon, but Run-II data soften this. The MSTW 
analysis uses only 
Run-II data whereas the CT/CTEQ analyses use both Run-I and Run-II data. 
These choices can explain the harder gluon luminosities of the CT PDFs 
at high-$x$. 
Watt and Thorne~\cite{wattthorne} obtain poor $\chi^2$ when comparing the 
Tevatron jet data to the  
HERAPDF1.0, 1.5 predictions. However their fits only 
compare to the central predictions of the HERAPDF. A more valid 
comparison would account for the HERAPDF error bands. If the 
Tevatron jet data are input to the HERAPDF1.5 fit a much better $\chi^2$
($\chi^2/ndp = 1.48$ for CDF and $1.35$ for $D0$ jets) is obtained. 
Significantly, the resulting PDFs do not lie outside the HERADF1.5 error bands
(although they do imply a harder high-$x$ gluon- on the upper edge of the error
 band). The reason that the HERAPDF can give a reasonable description of 
Tevatron jet data, while still having a relatively soft high-$x$ gluon PDF, is 
that high-$E_T$ jets result not only 
from the high-$x$ gluon but also from high-$x$ quarks and HERAPDF has a rather 
hard high-$x$ quark PDF.

The ABKM analysis also choses not to use Tevatron jet data, partly because 
new physics effects may be hidden in the data, biassing the PDFs. 
Consequently, ABKM has a soft high-$x$ gluon luminosity. Nevertheless,
ABM gives a good description of Tevatron jet data~\cite{abmjets}. 
A further issue regarding the use of Tevatron jet 
data concerns their use together 
with deuterium fixed-target data.  The greater uncertainty in 
the high-$x$ $d-$quark, due to uncertain deuterium corrections, 
will feed into the high-$x$ gluon PDF, since the $d-g$ process provides a 
substantial part of the Tevatron jet
cross section. However this larger uncertainty is usually not 
accounted for~\cite{jlabstuff}.

\subsubsection{Parametrisation and model uncertainty}
\label{sec:paramunc}
HERAPDF central fits have a relatively small number of parameters $\sim 14$.
However, parametrisation uncertainty is estimated by making fits with 
additional parameters freed, or with a change of the choice of the starting 
scale, $Q^2_0$, which is equivalent to a re-parametrisation. The comparison of 
HERAPDF1.5, which uses $10$ free parameters and HERAPDF1.5f which uses $14$ 
free parameters in Fig.~\ref{fig:herapdf1.5f}, shows that this procedure for 
accounting for parametrisation uncertainty largely accounts for the 
uncertainty introduced when the the extra parameters are freed in the central 
fit.
 
The HERAPDF results in a similar central value and uncertainty estimates to 
those of the global PDFs in many kinematic regions. In the case of the central 
values this is because the HERA data dominate the global input data. 
In the case of the uncertainty 
estimates it is partly due to the fact that the HERAPDF experimental 
uncertainties are augmented by estimates of the model and parametrisation 
variations, which are not accounted in the CT and MSTW analyses. 
This lends support to the idea that the 
increased $\chi^2$ tolerances of MSTW and CT partly cover some of these 
additional model and parametrisation uncertainties. 

The NNPDF global analysis uses a completely different approach both to PDF 
parametrisation and to the determination of PDF uncertainties. 
All errors (statistical, systematic and normalisation) 
as given by experimental collaborations are represented by 
Monte Carlo replica sets of artificial data. 
A neural net is used to learn the shape of these replicas rather than 
using a fixed parametrisation at the starting point. This can be regarded as 
equivalent to using a very large number of parameters.    
 The PDFs are not 
determined by a $\chi^2$ fit but by stopping the learning algorithm before 
overlearning occurs. The results are not presented in terms of eigenvectors 
of the fit but in terms of a set of replicas such that their mean 
gives the best estimate of the central PDF and the standard deviation from 
this mean gives the
$68\%$CL uncertainty estimate. It is remarkable that this entirely different 
procedure gives broadly similar central values and uncertainty estimates as 
those of the MStW and CTEQ global fits. To some extent this 
vindicates the standard 
procedure, in particular with regard to the use of increased $\chi^2$
 tolerances
 to set the $68\%$CL uncertainties.

\subsubsection{The value of $\alpha_s(M_Z)$}

Some groups (HERAPDF, CTEQ, NNPDF) adopt a fixed value of $\alpha_s(M_Z)$, 
inspired by the PDG value, and others (ABKM, GJR, MSTW) fit $\alpha_s(M_Z)$ 
simultaneously with the PDF parameters and use their best fit value.
All groups bar GJR use values 
$\sim 0.118-0.120$ at NLO but there is a definite low($0.113$)- high($0.117$) 
split at NNLO. HERAPDF, CT(EQ), NNPDF and MSTW provide PDFs at different 
$\alpha_s(M_Z)$ values so that the effect of variation of $\alpha_s(M_Z)$ 
on cross section predictions can be evaluated.

MSTW obtain the highest value of $\alpha_S(M_Z)$, at both NLO and NNLO,
 and these high values have been 
atributed to the use of Tevatron jet data in their fits. However, ABM have 
tried inputting these jet data to their fit and have found that 
this has only a small effect on their extraction of a low value of
$\alpha_s(M_Z)$~\cite{abmjets}. There is also a 'folk-lore' 
that DIS data prefer 
lower values of $\alpha_s(M_Z)$. However both MSTW~\cite{Martin:2009bu} and NNPDF~\cite{Lionetti:2011pw} have performed DIS only 
fits in which they find that only the BCDMS data prefer low $\alpha_s(M_Z)$ 
values. The HERA data actually prefer quite high values as shown in 
Sec.~\ref{sec:jets}. The effect of this on the gluon-gluon luminosity 
may be seen in Fig.~\ref{fig:lumi} 
by comparing the HERAPDF1.6 curve, with fixed $\alpha_s(M_Z)=0.1176$, to that 
of the HERAPDF1.6 free $\alpha_s(M_Z)$ curve, 
which has $\alpha_s(M_Z)=0.1202$. 
The larger $\alpha_s(M_Z)$ value leads to a smaller low-$x$ gluon and a 
somewhat harder high-$x$ gluon such that the  
gluon-gluon luminosity is then in better agreement with that of MSTW2008, 
which also use a large $\alpha_s(M_Z)$ value.

\subsubsection{Heavy Quark Schemes}
The ABKM and GJR groups use 
Fixed-Flavour-Number (FFN) treatments, HERAPDF, CTEQ and MSTW use various 
General-Mass-Variable-Flavour-Number (GMVFN) treatments and 
NNPDF2.0~\cite{nnpdf20} used 
a Zero-Mass-Variable-Flavour-Number treatment(ZMVFN). These heavy quark schemes
are discussed in Ref.~\cite{bookch}.
The use of the zero-mass treatment explains why the NNPDF2.0 
luminosities lie lower than those of
 CTEQ, MSTW and HERAPDF at low $\tau$. 
This may be seen by comparing the top row of Fig.~\ref{fig:lumi} 
to the middle row where the NNPDF2.1
luminosity, which used a GMVFN, is seen to be in much better agreement 
with the other PDFs. This is because, when charm mass is accounted for, charm 
is suppressed at threshold and the light quark densities must be somewhat 
larger in order to describe the deep inelastic cross-section. 
However not all GMVFNs are the same.
Predictions for $F_2^c$ differ 
between schemes~\cite{Rojo:2010gv} and the choice of scale within a scheme can 
also affect predictions. 
The value of the charm and beauty masses also differ between the PDF analyses.
HERAPDF, NNPDF and MSTW now provide PDFs at different charm and beauty mass 
values so that the effect of this can be evaluated. In future the 
combined data on 
$F_2^{c\bar{c}}$, discussed in Sec.~\ref{sec:charm}, should help to reduce the 
uncertainty on PDFs coming from the choice of scheme and the value of the 
charm mass.

The heavy quark mass schemes described in Sec.~\ref{sec:charm} 
all use a charm quark mass parameter which should be the pole-mass.  
However the pole-mass has a strong dependence on the order of the 
perturbative calculation and 
may best be regarded as a parameter. 
It may be better to consider the \msbar running-mass. HERA data on 
$F_2^{c\bar{c}}$ has also been used for a determination of this 
mass~\cite{moch}
 
\subsection{Comparisons of HERAPDF predictions to Tevatron and LHC data}

\begin{figure}[tbp]
\vspace{-1.0cm} 
\centerline{\psfig{figure=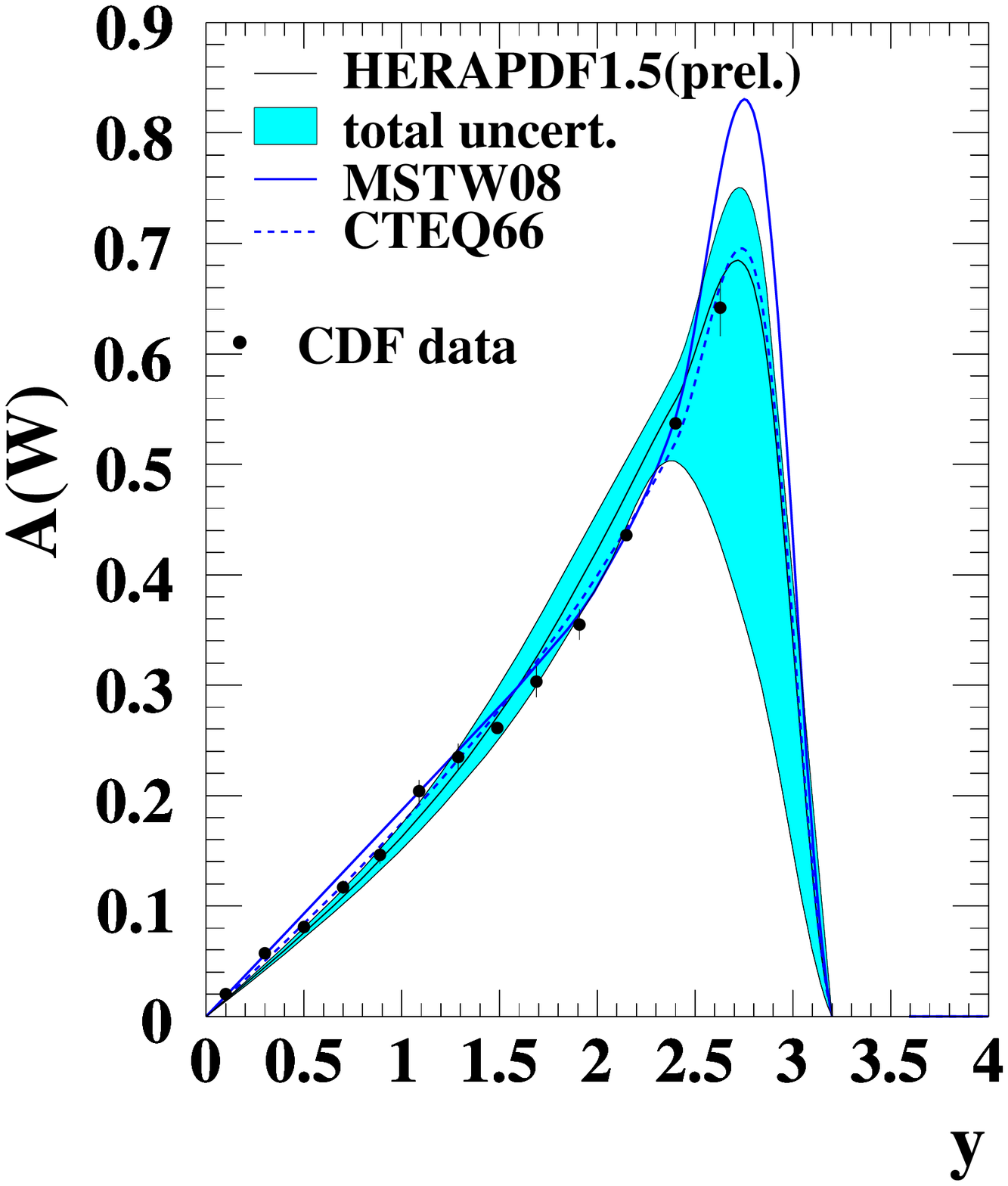,width=0.30\textwidth}~~ 
\psfig{figure=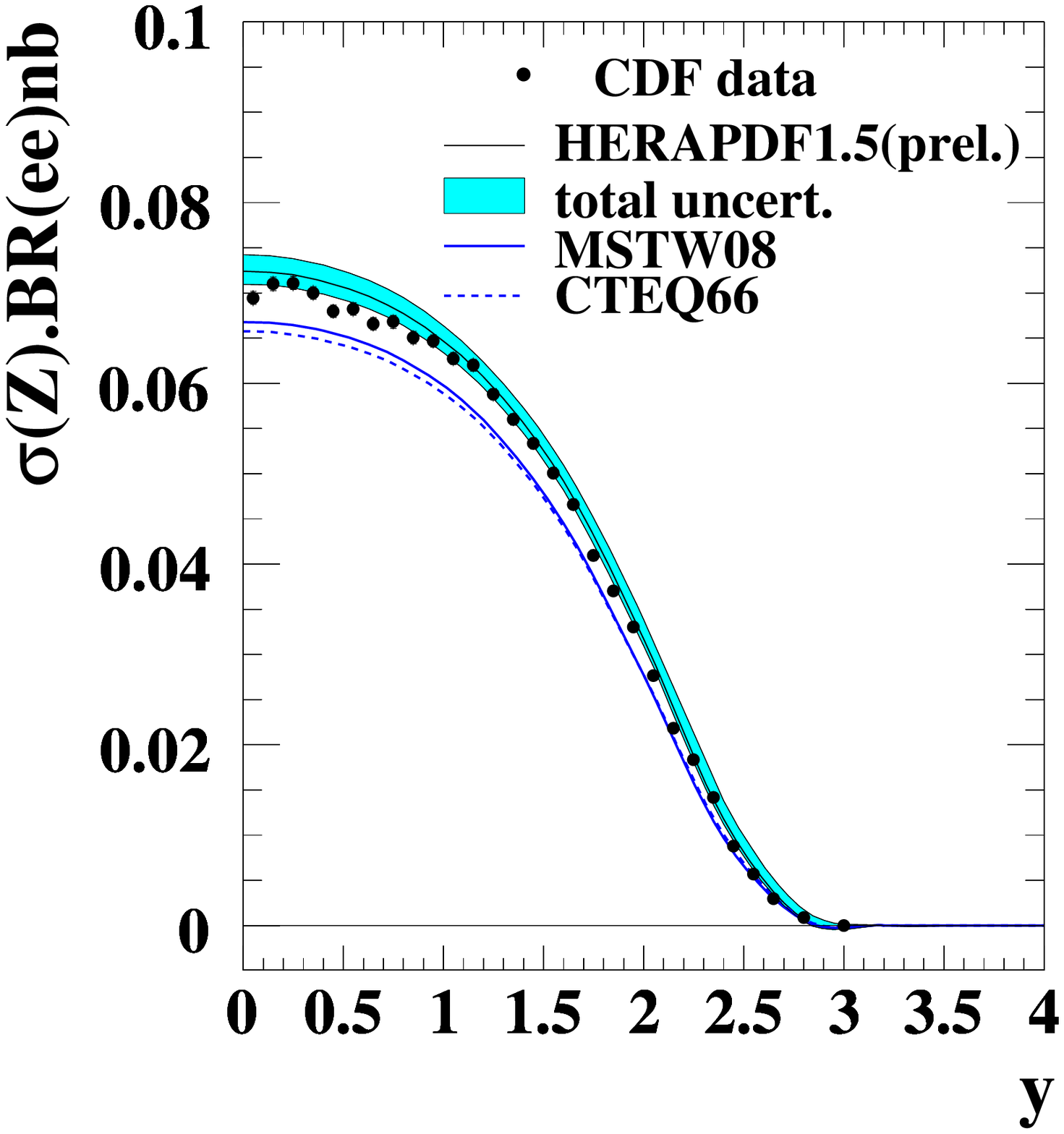,width=0.30\textwidth}}
\caption {Left hand side: data on the direct $W$-asymmetry from CDF; right 
hand side: data on the $Z0$ rapidity spectrun from CDF; compared 
to NLO predictions from CTEQ6.6, MSTW08 and HERAPDF1.5. The blue band indicates 
the uncertainties on the HERAPDF prediction.
}
\label{fig:tevWZ}
\end{figure}

Finally we present some representative 
comparisons of HERAPDF predictions to PDF sensitive 
data from the Tevatron and LHC colliders. Fig.~\ref{fig:tevWZ} presents 
comparisons to CDF data on the direct $W$-asymmetry~\cite{cdfwasym} and $Z0$ 
rapidity spectrum~\cite{cdfz0}.
These data are well described by the HERAPDF1.5 prediction\footnote{The predictions of the HERAPDF1.6 and 1.7 PDFs are very similar to that of HERAPDF1.5}.
A fit of the data to the central value of the prediction yields a $\chi^2$ of 
36 for $28$ data points for the $Z0$ data and of $41$ for $13$ data points for 
the asymmetry data. These descriptions are improved if the data is input to the
HERAPDF fit, to $\chi^2/ndp = 26/28$ for the $Z0$ data and $21/13$ for the 
asymmetry data\footnote{Note that the $\chi^2/ndp$ for these asymmetry data are as well described by the HERAPDF as they are by other PDFs which have used them, e.g. NNPDF.}. The resulting PDFs lie well within the HERAPDF1.5 error bands.
The HERAPDF uncertainty bands could be reduced by input of these data. 
This is a future project beyond the scope of the current review.
 
\begin{figure}[tbp]
\vspace{-1.0cm} 
\psfig{figure=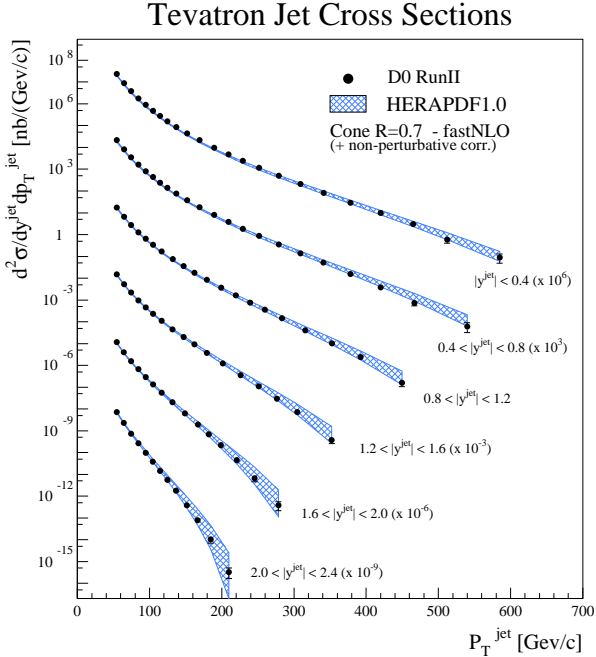,width=0.60\textwidth}
\caption {D0 data on inclsuive jet production compared to NLO 
predictions from HERAPDF1.0} 
\label{fig:tevjets}
\end{figure}

Fig.~\ref{fig:tevjets} presents 
comparisons of HERAPDF1.0 predictions to D0 data on the inclusive jet production ~\cite{d0jets}
Because of the large correlated systematics of these data it is not possible 
to assess the quality of the description by eye. If these data are input to 
the HERAPDF1.5 fit a $\chi^2/ndp=145/110$ can be obtained.
Similary if CDF inclusive jet production data~\cite{cdfjets} 
are input to the HERAPDF1.5 
NLO fit a $\chi^2/ndp=113/76$ is obtained. In both cases the resulting PDFs 
move to the  edge of the HERAPDF1.5 error band- tending to favour a harder 
high-$x$ gluon. For this reason, the HERAPDF1.6 $\alpha_s(M_Z)=0.1202$ fit,
which already has a harder gluon than the 1.5 fit, 
gives the best description of these data out of all the NLO HERAPDF sets.

The HERAPDF1.5NNLO PDF fit gives a better description of these
data than the NLO PDFs- the central PDF of HERAPDF1.5NNLO yields a $\chi^2$ 
per data point of  
$\chi^2/ndp=72/76$. However this can only be approximate since the
theoretical description of the jet data itself contains only an approximate 
calculation for the NNLO jet cross-section.

\begin{figure}[tbp]
\vspace{-1.0cm} 
\centerline{\psfig{figure=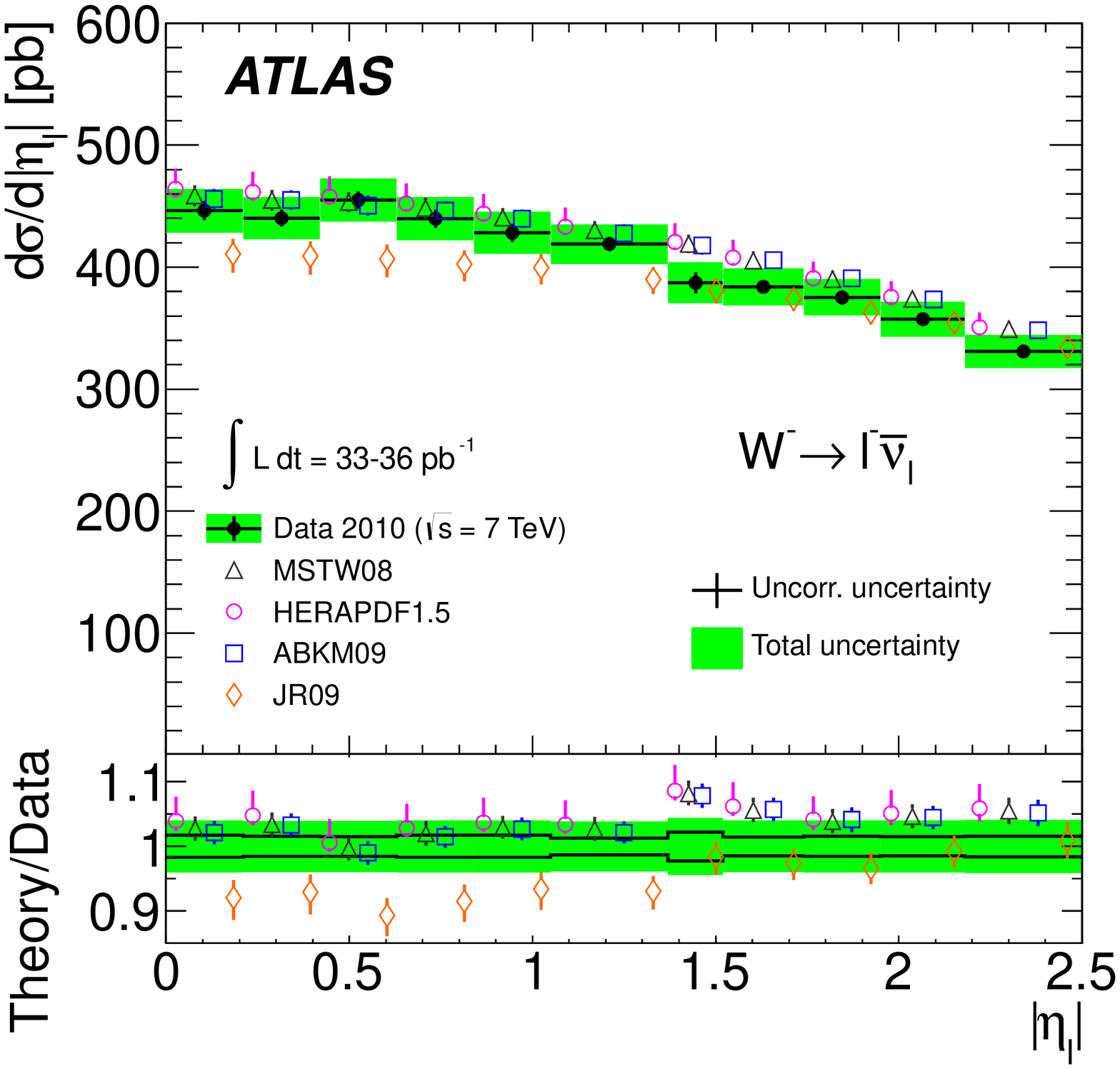,width=0.30\textwidth}~~ 
\psfig{figure=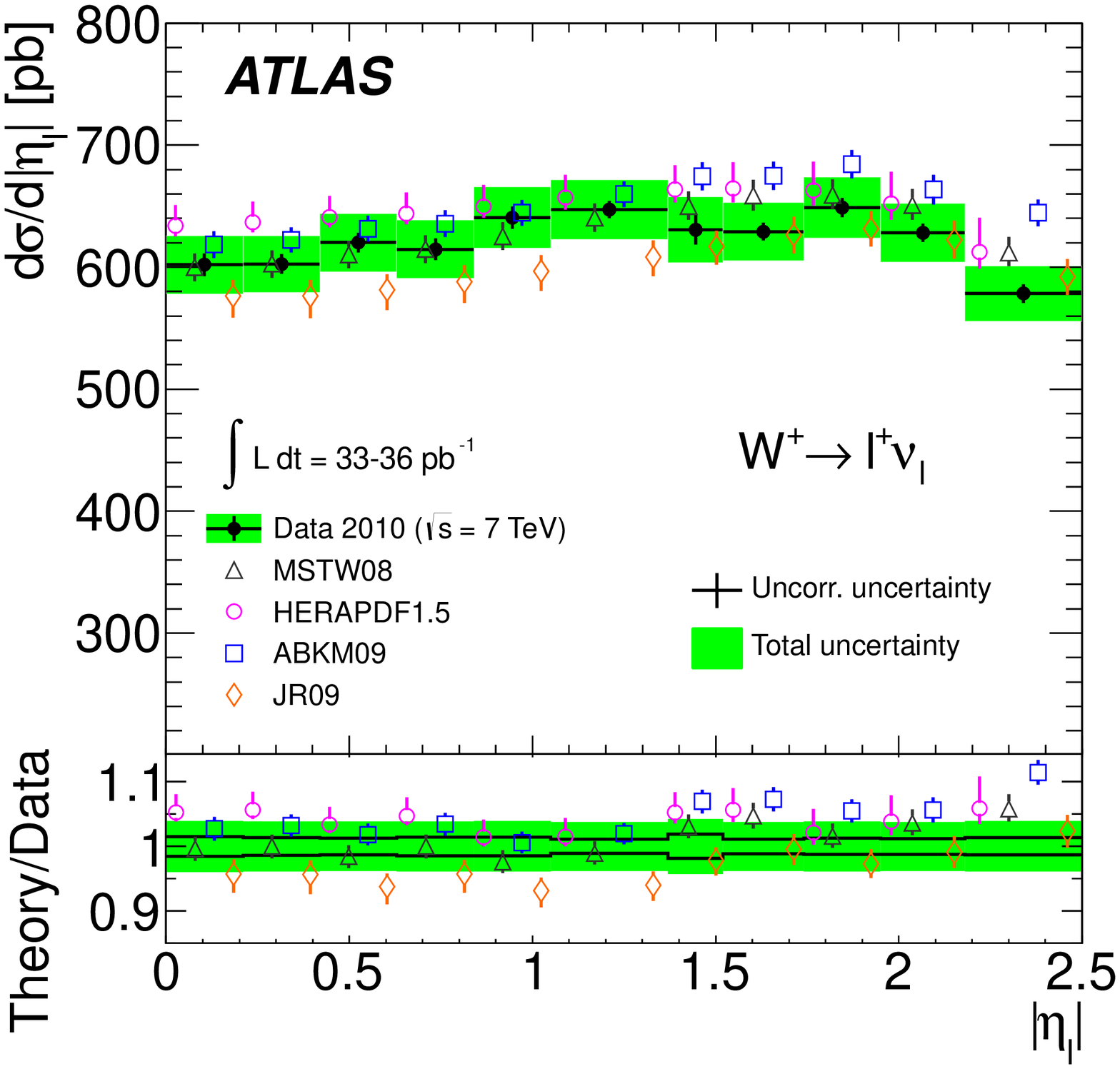,width=0.30\textwidth}}
\centerline{\psfig{figure=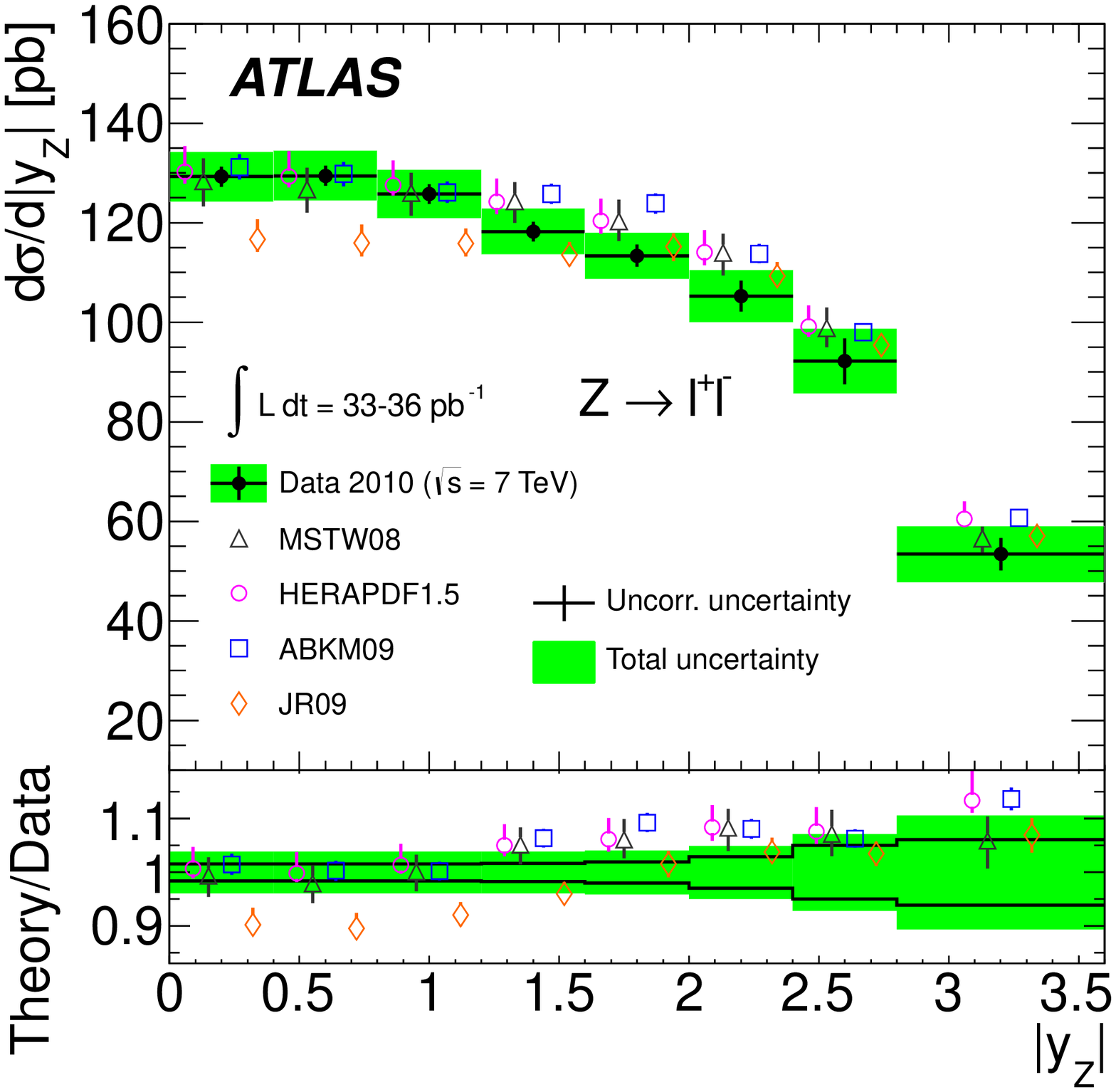,width=0.30\textwidth}~~ 
\psfig{figure=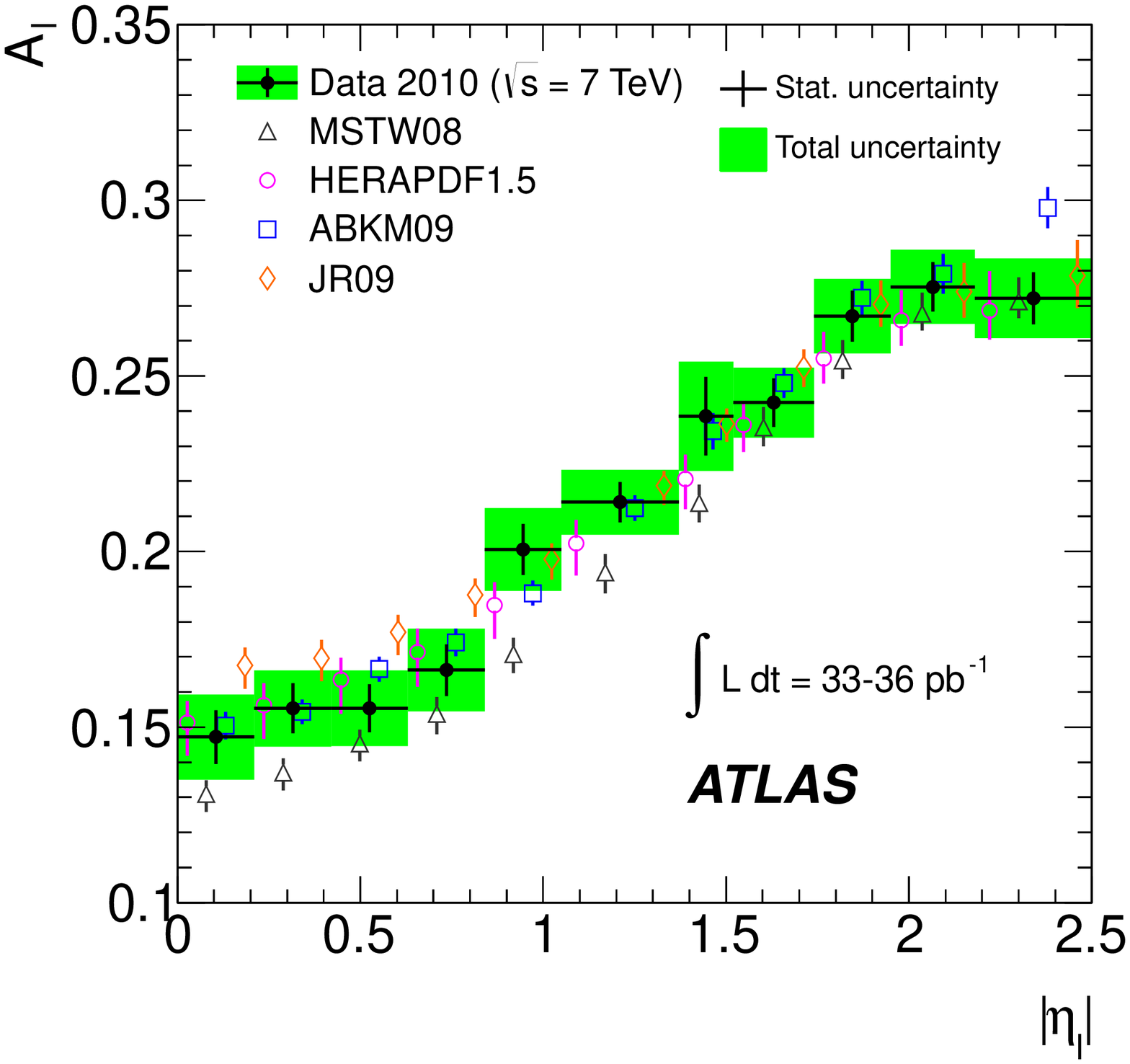,width=0.30\textwidth}}
\caption {Comparisons of ATLAS data on $W^+$ and $W^-$ decay lepton 
pseudorapidity spectra, $Z0$ rapidity spectra and $W$ decay lepton asymmetry 
data to NNLO predictions from MSTW08, HERAPDF1.5, ABKM09, JR09.
}
\label{fig:atlasWZ}
\end{figure}

\begin{figure}[tbp]
\vspace{-1.0cm} 
\psfig{figure=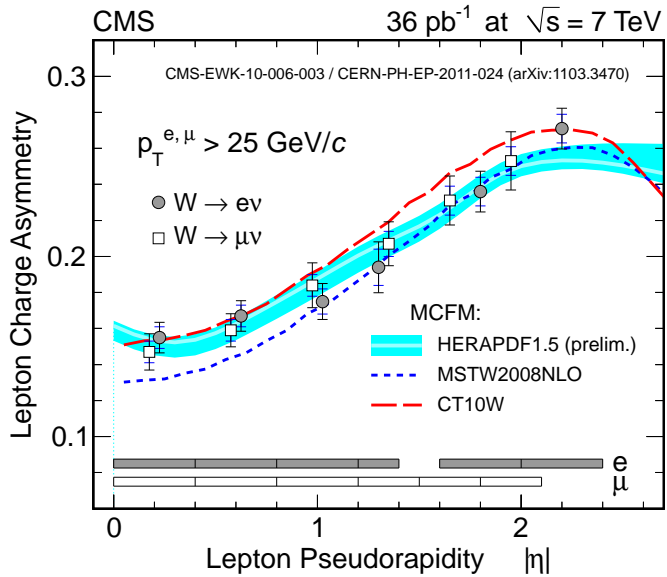,width=0.60\textwidth}
\caption {CMS data on $W$ decay lepton asymmetry compared to NLO 
predictions from HERAPDF1.5, MSTW08 and CT10W} 
\label{fig:cmsW}
\end{figure}

Fig.~\ref{fig:atlasWZ} presents 
comparisons of various PDFS, including HERAPDF1.5, 
to ATLAS data on the $W$-lepton decay pseudorapidity distributions 
and the $Z0$ rapidity distribution, as well as on the $W$-lepton 
asymmetry~\cite{atlasWZ} .
Fig.~\ref{fig:cmsW} presents 
comparisons of HERAPDF1.5 predictions to $234$pb$^{-1}$ of preliminary 
CMS 2011 data on the $W$ decay lepton asymmetry ~\cite{cmswasym}. 
These LHC $W$ and $Z$ cross section data 
are well described by the HERAPDF. However, a detailed study by the ATLAS 
Collaboration~\cite{atlastrange} using the ATLAS $W$ and $Z$ data and the 
HERA-I combined data has indicated a preference of the ATLAS data 
for unsuppressed strangeness at $x\sim 0.01$. 
Further discussion of this is beyond the scope of the present review.

\begin{figure}[tbp]
\vspace{-1.0cm} 
\centerline{\psfig{figure=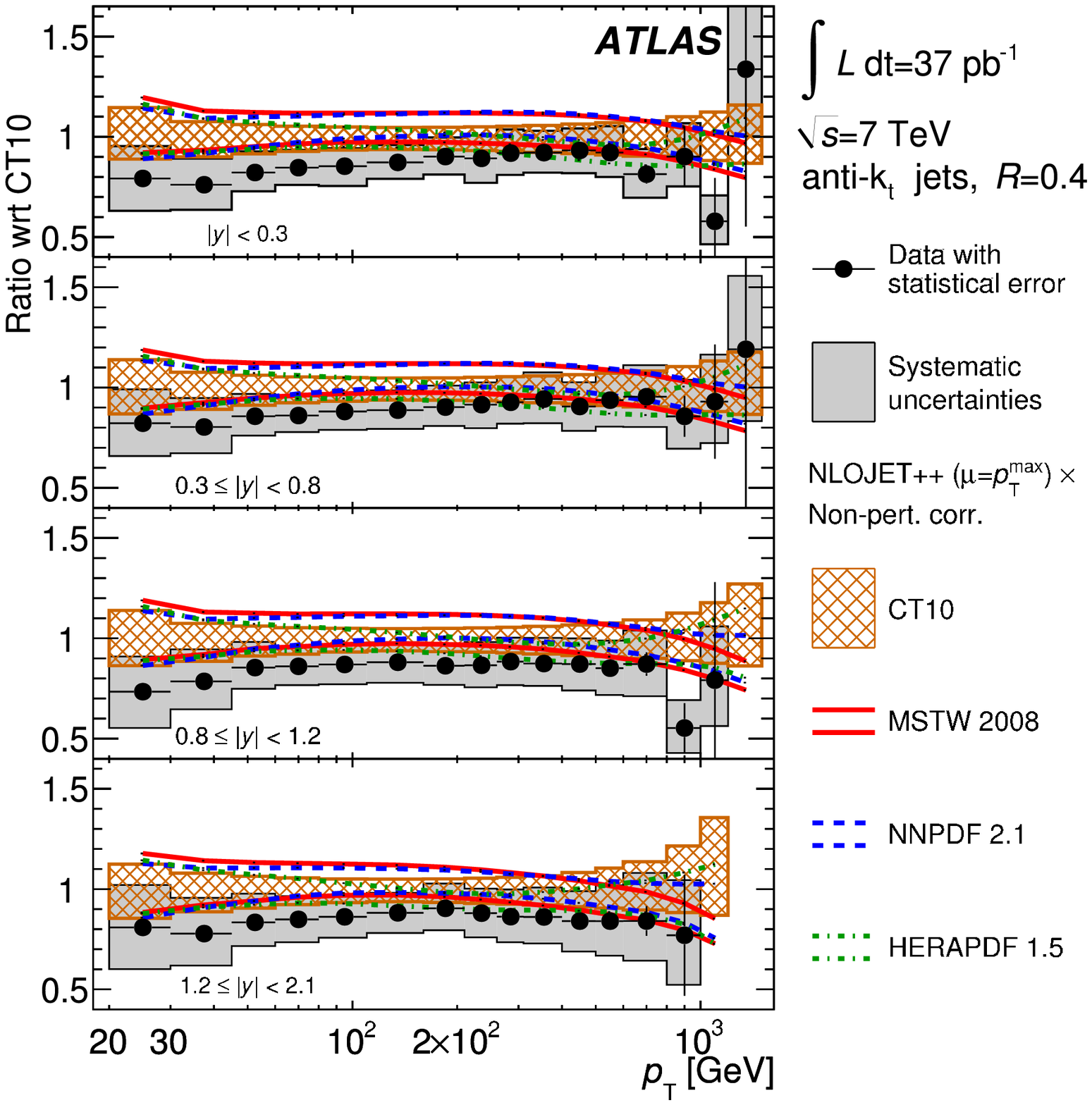,width=0.45\textwidth}~~ 
\psfig{figure=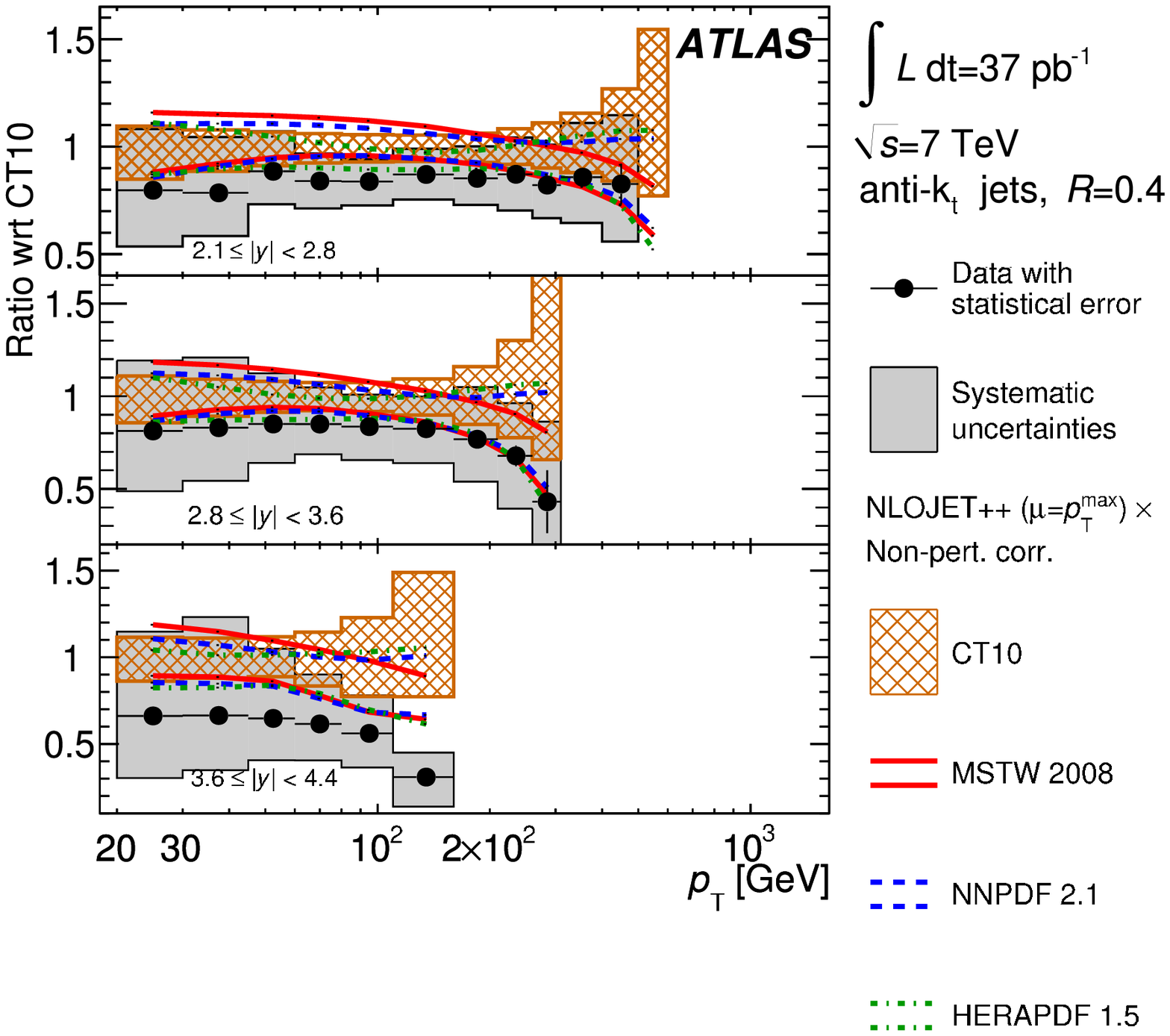,width=0.45\textwidth}}
\caption {ATLAS data on inclusive jet production in central and forward 
rapidity regions in ratio to the NLO predictions of CT10 and compared to NLO
predictions from HERAPDF1.5, NNPDF2.1 and MSTW08} 
\label{fig:atlasjets}
\end{figure}

\begin{figure}[tbp]
%\vspace{-1.0cm} 
\centerline{\psfig{figure=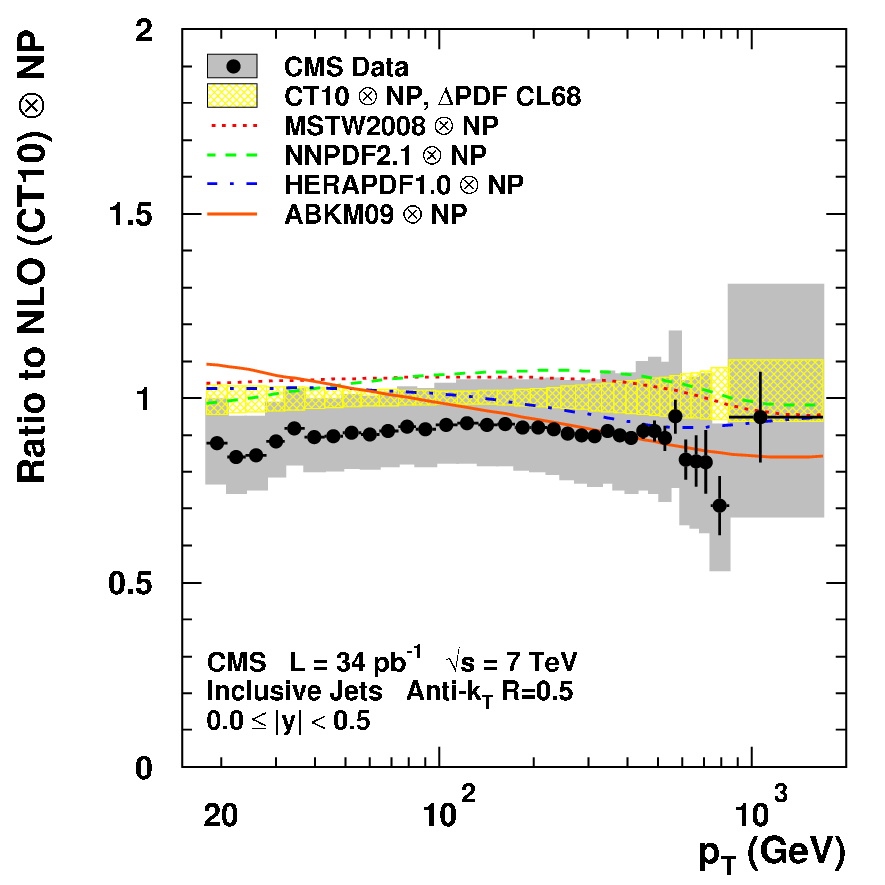,width=0.45\textwidth}~~ 
\psfig{figure=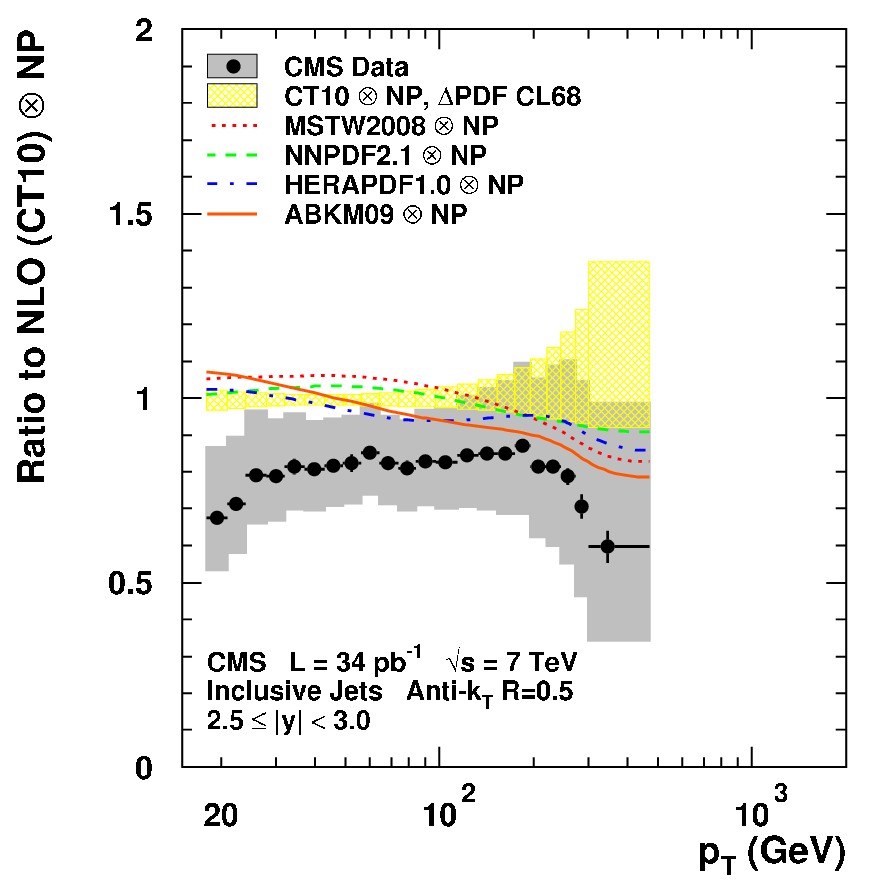,width=0.45\textwidth}}
\caption {CMS data on inclusive jet production in central and forward 
rapidity regions in ratio to the NLO predictions of CT10 and compared to NLO
predictions from HERAPDF1.0, ABKM09, MSTW08 and NNPDF2.1} 
\label{fig:cmsjets}
\end{figure}

Fig.~\ref{fig:atlasjets} presents 
comparisons of various PDF predictions, including HERAPDF1.5, to ATLAS data on the inclusive jet production ~\cite{atlasjets}.
Fig.~\ref{fig:cmsjets} presents 
comparisons of various PDF predictions, including HERAPDF1.5, to CMS data on the inclusive jet production ~\cite{cmsjets}.
Because of the large correlated systematics of these data 
it is not possible 
to assess the quality of the description by eye. The ATLAS jet data are 
published with information on these correlations and a $\chi^2$ per data point 
of $\sim 60/90$ can be obtained for each of the HERAPDFs, and the $\chi^2$ 
for the MSTW, CT and NNPDFs are similar. 
Thus the data are not yet very discriminating, 
however they indicate a preference for a somewhat less hard high-$x$ gluon 
than the Tevatron jet data.

\section{Summary}

Deep inelastic lepton-hadron 
scattering data from the HERA collider now dominate 
the world data on deep inelastic scattering since they cover an unprecedented 
kinematic range. 
The H1 and ZEUS experiments 
are combining their data in order to provide the most complete and 
accurate set of deep-inelastic data as the legacy of HERA. 
 
Data on 
inclusive cross-sections have been combined for the HERA-I phase of running 
and a preliminary combination has been made also using the HERA-II data. 
This latter exersize also includes the data run at lower proton beam 
energies in 2007. Combination of  $F_2^{c\bar{c}}$ data is underway, 
and combination of  $F_2^{b\bar{b}}$ dat and of jet data is foreseen.

The HERA collaborations have used these combined data to 
determine parton distribution functions (PDFs) in the proton. 
Because the HERA experiments investigated
 $e^+p$ and $e^-p$, charge current(CC) and neutral current (NC) scattering, 
the inclusive HERA data provide infromation on
 flavour separated up- and down-type quarks and 
antiquarks and on the gluon- from its role in the scaling violations of 
perturbative quantum-chromo-dynamics. The lower proton beam energy data 
provide further information on the gluon at small $x \leqsim 0.01$ 
since they allow a determination of 
the longitudinal structure function. The charm data provide additional 
information on heavy quark schemes and heavy quark mass values. The jet data 
(separate data from H1 and ZEUS at the time of writing) provide additional 
information on the gluon PDF in the $x$ range, $0.01 \leqsim x \leqsim 0.1$ and on $\alpha_s(M_Z)$.

The analysis of these data sets has resulted in the
the HERAPDF parton distribution functions. In this review we have described 
and compared these sets with each other and with PDF sets from other groups.
We have also demonstrated that the HERAPDF sets give successful descriptions 
of data on $W$ and $Z$ production and on jet production from the Tevatron 
and the LHC.
The currently recommended version 
of these PDFs, which are available on LHAPDF, are the HERAPDF1.5 NLO and 
NNLO sets.

%--------------------------------------------------------------------------%%@
----
%       Bibliography
%--------------------------------------------------------------------------%%@
----

\end{document}